\documentclass[aps,pra,superscriptaddress,twocolumn,a4paper,reprint,noeprint]{revtex4-1}
\usepackage[utf8]{inputenc}
\usepackage{graphicx}
\usepackage{amsmath}
\usepackage{amsfonts}
\usepackage{amssymb}
\usepackage[usenames,dvipsnames,svgnames,table]{xcolor}
\usepackage[USenglish]{babel}
\usepackage{hyperref}
\usepackage{grffile}

\def\Tr{\mathop{\textrm{Tr}}}
\def\RR{\mathop{\textrm{Re}}}
\def\II{\mathop{\textrm{Im}}}
\def\Pf{\mathop{\textrm{Pf}}}

\newcommand{\bsigma}{\mbox{\boldmath$\sigma$}}

\newcommand{\bad}{$\circ$}
\newcommand{\tdelta}{\tilde\delta}

\allowdisplaybreaks

\begin{document}

\title{Symmetry, nodal structure, and Bogoliubov Fermi surfaces for nonlocal pairing}

\author{Carsten Timm}
\email{carsten.timm@tu-dresden.de}
\affiliation{Institute of Theoretical Physics, Technische Universit\"at Dresden, 01062 Dresden, Germany}
\affiliation{W\"urzburg-Dresden Cluster of Excellence ct.qmat, Technische 
Universit\"at Dresden, 01062~Dresden, Germany}

\author{Ankita Bhattacharya}
\affiliation{Institute of Theoretical Physics, Technische Universit\"at Dresden, 01062 Dresden, Germany}

\date{July 26, 2021}

\begin{abstract}
Multiband effects can lead to fundamentally different electronic behavior of solids, as exemplified by the possible emergence of Fermi surfaces of Bogoliubov quasiparticles in centrosymmetric superconductors which break time-reversal symmetry. We extend the analysis of possible pairing symmetries, the corresponding nodal structure, and the Bogoliubov Fermi surfaces in two directions: We include nonlocal pairing and we consider internal degrees of freedom other than the effective angular momentum of length $j=3/2$ examined so far. Since our main focus is on the Bogoliubov Fermi surfaces we concentrate on even-parity pairing. The required symmetry analysis is illustrated for several examples, as a guide for the reader. We find that the inclusion of nonlocal pairing leads to a much larger range of possible pairing symmetries. For infinitesimal pairing strength, we find a simple yet powerful criterion for nodes in terms of a scalar product of form factors.
\end{abstract}

\maketitle

\section{Introduction}

In condensed matter physics, we are used to study the electronic properties of materials by considering one band at a time. Only when we are interested in excitations at higher energies, e.g., by electromagnetic waves, we include multiple bands. However, it is not always true that the low-energy and equilibrium properties of a solid can be understood based on the single-band paradigm. A case in point is the recent realization that in centrosymmetric multiband superconductors that break time-reversal symmetry (TRS) and have gap nodes, these nodes are generically two dimensional \cite{ABT17,BAM18}. We call these two-dimensional nodes \emph{Bogoliubov Fermi surfaces} (BFSs). This result has put meat on the bones of the proof \cite{KST14,ZSW16} that in such systems two-dimensional Fermi surfaces can be protected by a $\mathbb{Z}_2$ topological invariant. This invariant is related to the Pfaffian of the Bogoliubov--de Gennes (BdG) Hamiltonian, unitarily transformed into antisymmetric form~\cite{ABT17,BAM18}.

Another example is the belief that optical excitations across the gap are forbidden in clean superconductors, which is based on the single-band paradigm but does not hold for multiband superconductors, as recently shown by Ahn and Nagaosa \cite{AhN21}. One criterion for when optical excitations across the gap are allowed is the existence of BFSs---this holds both for centrosymmetric and for noncentrosymmetric superconductors.

Experimental signatures of BFSs \cite{LBT20} and their instability against spontaneous breaking of inversion symmetry either electronically \cite{OhM20,TIH20,HeL21} or by lattice distortions \cite{TBA21} have been considered by several groups. BFSs can be protected by multiple topological invariants \cite{BrS17,BAM18}, which imposes constraints on how they can merge and gap out for strong coupling. Recently, a classification of nodal structures for all magnetic space groups based on compatibility relations has been put forward \cite{OnS21}, also predicting BFSs. Furthermore, Herbut and Link \cite{HeL21} have revealed an analogy between the antisymmetric form of the BdG Hamiltonian---the existence of which is essential for the definition of the Pfaffian \cite{ABT17,BAM18}---and classical relativity. In this analogy, one can understand the BFSs from a condition of orthogonal fictitious electric and magnetic fields in momentum space \cite{HeL21}. The analogy is useful for studying the interaction-induced instability of the BFSs. However, it is restricted to four-dimensional internal Hilbert spaces.

Link and Herbut \cite{LiH20} have also presented a complementary study of BFSs in \emph{noncentrosymmetric} multiband superconductors with broken TRS and gap nodes. Although no $\mathbb{Z}_2$ topological invariant exists, BFSs are typically present since the breaking of TRS causes band shifts that are larger than the induced gaps \cite{LiH20}. This result is reminiscent of BFSs generated by band shifts induced by a superflow \cite{Vol03,AMR20}, which breaks TRS explicitly.

In the language of tight-binding models, the presence of multiple bands in the vicinity of the Fermi energy results from the existence of multiple (Wannier) orbitals per unit cell that appreciably contribute to the Bloch states at the Fermi energy. These orbitals can be located at the same site or, for structures with a basis, at different sites. We will refer to these orbital and site degrees of freedom, together with the electron spin, as \emph{internal degrees of freedom}. Superconducting pairing states can be nontrivial with respect to these internal degrees of freedom (internally anisotropic \cite{BAM18}). This allows nontrivial pairing states even for perfectly local pairing, which corresponds to a momentum-independent gap matrix, as we will discuss further below.

If the orbital degrees of freedom form a degenerate triplet, for example $p_x$, $p_y$, $p_z$ or $d_{yz}$, $d_{zx}$, $d_{xy}$ in a cubic crystal field, they can be combined with the spin to form states with effective angular momentum $j=1/2$ or $j=3/2$ \cite{BWW16,ABT17,TSA17,YXW17,SRV17,BAM18,OhM20,TIH20,TBA21,DPB21}. The latter leads to a natural description of the fourfold $\Gamma_8$ band-touching points in cubic crystals. While the results concerning the existence of BFSs and their $\mathbb{Z}_2$ topological invariant were general, they were mostly illustrated by the example of the $j=3/2$ model \cite{ABT17,BAM18,OhM20,TIH20,TBA21}. It is important to realized that the possibilities for internal degrees of freedom are much richer. Many superconductors that do not fit the $j=3/2$ description nevertheless have multiple bands close together and close to the Fermi energy.

Moreover, the examples studied so far were restricted to local pairing. However, nonlocal pairing is generically present and is often necessary to obtain any superconductivity if local pairing is excluded by a repulsive Hubbard interaction. We will see that nonlocal pairing typically allows for a large range of additional pairing states with symmetries that do not appear for local pairing.

In this paper, we extend the analysis of BFSs in centrosymmetric superconductors in two directions: We include nonlocal pairing and we consider internal degrees of freedom other than an effective angular momentum $j=3/2$. While we mainly discuss the physically most relevant case that the internal degrees of freedom include the spin and the time-reversal transformation squares to minus the identity operation, we also derive results for the converse case. We are mainly interested in the properties of BFSs in this more general setting and therefore concentrate on even-parity pairing. We provide details on the symmetry analysis to help readers perform such an analysis for specific systems of interest. Everything said here applies to three-dimensional crystals.

To be clear, we also specify what we do not consider: We do not address systems with a normal-state Fermi surface that is not topologically equivalent to a sphere enclosing the $\Gamma$ point, for example quasi-two-dimensional systems such as $\mathrm{Sr_2RuO_4}$ \cite{SMB20}. In such cases, directions in momentum space with symmetry-imposed nodes for infinitesimal pairing might not intersect with the normal-state Fermi surface so that these symmetry-imposed nodes would not be present. Furthermore, we do not consider accidental gap nodes, pairing states that combine different irreducible representations (irreps) of the point group, and new BFSs that emerge for strong coupling away from the normal-state Fermi surface. The description of such phenomena would only require straightforward extensions of the theory. Moreover, we do not address odd-frequency pairing \cite{LiB19}, which is expected to permit additional contributions to pairing states of given symmetry. Recently, Dutta \textit{et al.}\ \cite{DPB21} have shown that odd-frequency pairing amplitudes generically appear together with BFSs.


The remainder of this paper is organized as follows: In Sec.\ \ref{sec.theory}, we describe the symmetry analysis that produces all symmetry-allowed contributions to pairing of any symmetry for any crystallographic point group, together with the nodal structure for infinitesimal pairing strength and criteria for the inflation of nodes into BFSs beyond infinitesimal pairing. In Sec.\ \ref{sec.appl}, we apply this general framework to a number of model systems of increasing complexity. Finally, in Sec.\ \ref{sec.conclusions}, general insights gained by the preceding sections are discussed and an outlook on open questions is given. Several formal points are presented in Appendices.

\section{General analysis}
\label{sec.theory}

In this section, we describe the general symmetry analysis. We assume the normal state to satisfy inversion symmetry and TRS. The first condition implies that the only possible point groups are $C_i$, $C_{2h}$, $D_{2h}$, $D_{3d}$, $C_{4h}$, $D_{4h}$, $C_{6h}$, $D_{6h}$, $S_6 = C_{3i}$, $T_h$, and $O_h$. Of these groups, $C_i$, $C_{2h}$, and $D_{2h}$ have only one-dimensional irreps. $C_{4h}$, $C_{6h}$, and $S_6$ in addition have two-dimensional real irreps which decompose into one-dimensional complex irreps. The real irreps are relevant for the analysis of Hermitian irreducible tensor operators. Finally, $D_{3d}$, $D_{4h}$, $D_{6h}$, $T_h$, and $O_h$ also have multidimensional complex irreps. Pairing states described by multidimensional irreps are the most interesting for us since they lead to multicomponent order parameters, which naturally accommodate the breaking  of~TRS.

It is advantageous to consider the \emph{magnetic} point group of the crystal since this allows us to treat point-group symmetries and TRS on equal footing \cite{Wig32,Wig59,DiW62,Cra65,BrD68,DDJ08}. Since TRS is preserved in the normal state, the antiunitary time-reversal operator $\mathcal{T}$ is an element of the magnetic point group. Hence, we are dealing with a \emph{gray} group: If $G$ is the structural point group, then the gray point group is $M = G + \mathcal{T}G$, where $\mathcal{T}G$ contains all elements of $G$ multiplied by time reversal $\mathcal{T}$ (the elements of $G$ commute with $\mathcal{T}$~\cite{BrD68}).

The theory of \emph{complex} corepresentations of magnetic groups \cite{Wig32,Wig59,DiW62,Cra65} has been reviewed for example by Bradley and Davies \cite{BrD68}. However, this theory is not the appropriate one for our purposes since it is based on the notions of unitary equivalence of matrices and reducibility of corepresentations by unitary transformations. This leads to the result that two corepresentations that represent $\mathcal{T}$ by $+\openone$ and $-\openone$, respectively, can be equivalent. Since we need to distinguish operators that are even or odd under time reversal this is not the appropriate equivalence relation. We rather require \emph{real} corepresentations based on orthogonal equivalence, which leaves the properties under time reversal invariant, and reducibility by orthogonal transformations. By using Wigner's construction of corepresentations \cite{Wig59,BrD68} but restricting oneself to orthogonal transformations, it is fairly easy to see that for every irrep $\Gamma$ of $G$, the gray magnetic point group $M$ has two irreps $\Gamma_+$ and $\Gamma_-$, which are distinguished by a (further) index $\pm$ which indicates the sign under time reversal. Character tables of real corepresentations of the magnetic point groups are given by Erb and Hlinka~\cite{ErH20}.

One more piece of group theory is needed: Since we are studying single-fermion Hamiltonians, we have to consider double groups, i.e., a rotation by $2\pi$ does not give the group identity but only a rotation by $4\pi$ does. This leads to additional ``double-valued'' or spinor irreps \cite{DDJ08}. However, the double-valued irreps do not play any role in our analysis for the following reason: We will expand Hamiltonians into sums of basis matrices with momentum-dependent coefficients which are basis functions of irreps. The momentum components $k_x$, $k_y$, $k_z$ are basis functions of single-valued irreps, which implies that momentum-dependent functions can only be basis functions of single-valued irreps. Hence, double-valued irreps do not occur in the expansion.

We now introduce relevant notations. The superconductor is described by a BdG Hamiltonian of the form
\begin{equation}
\mathcal{H}(\mathbf{k}) = \begin{pmatrix}
    H_N(\mathbf{k}) & \Delta(\mathbf{k}) \\
    \Delta^\dagger(\mathbf{k}) & -H_N^T(-\mathbf{k})
  \end{pmatrix} ,
\label{1.HBdG.2}
\end{equation}
where $H_N(\mathbf{k})$ is the normal-state Hamiltonian and $\Delta(\mathbf{k})$ is the pairing potential. Both are operators on the $N$-dimensional Hilbert space of internal degrees of freedom and are given by $N\times N$ matrices for a particular basis. Antisymmetry of fermionic states implies that~\cite{Gor58,SiU91,CTS16}
\begin{equation}
\Delta^T(-\mathbf{k}) = -\Delta(\mathbf{k}) .
\label{1.Deltaanti.2}
\end{equation}
By construction, the BdG Hamiltonian satisfies particle-hole (charge-conjugation) symmetry $\mathcal{C}$, which is expressed as
\begin{equation}
\mathcal{U}_C\, \mathcal{H}^T(-\mathbf{k})\, \mathcal{U}_C^\dagger = - \mathcal{H}(\mathbf{k}) ,
\label{1.UCtrans.3}
\end{equation}
with $\mathcal{U}_C = \sigma_1 \otimes \openone$. We denote the Pauli matrices by $\sigma_1$, $\sigma_2$, $\sigma_3$ and the $2\times 2$ identity matrix by $\sigma_0$. Identity matrices in any dimension are denoted by $\openone$. Symmetries in the structural point group $G$ are expressed as
\begin{equation}
\mathcal{U}\, \mathcal{H}(R^{-1} \mathbf{k})\, \mathcal{U}^\dagger = \mathcal{H}(\mathbf{k}) ,
\label{1.genUtrans.3}
\end{equation}
where $R$ is an appropriate three-dimensional generalized rotation matrix and
\begin{equation}
\mathcal{U} = \begin{pmatrix}
    U & 0 \\ 0 & U^*
  \end{pmatrix}
\label{1.genUtrans.3a}
\end{equation}
is unitary. This form of $\mathcal{U}$ follows from particle-hole symmetry. The most important case of us is inversion symmetry or parity $P$, which is implemented by a unitary matrix $U_P$. We assume inversion symmetry of the normal state, i.e.,
\begin{equation}
U_P\, H_N(-\mathbf{k})\, U_P^\dagger = H_N(\mathbf{k}) .
\end{equation}
Finally, TRS takes the form
\begin{equation}
\mathcal{U}_T\, \mathcal{H}^T(-\mathbf{k})\, \mathcal{U}_T^\dagger = \mathcal{H}(\mathbf{k}) ,
\label{1.UTtrans.3}
\end{equation}
where
\begin{equation}
\mathcal{U}_T = \begin{pmatrix}
    U_T & 0 \\ 0 & U_T^*
  \end{pmatrix}
\end{equation}
is (the matrix form of) the unitary part of the antiunitary time-reversal operator. Time reversal $\mathcal{T}$ can square to plus or minus the identity. We denote the sign of $\mathcal{T}^2$ by $s_T = \pm 1$. Then $U_T U_T^* = s_T\, \openone$ and thus
\begin{equation}
U_T^T = s_T\, U_T .
\label{1.UTsymm.3}
\end{equation}
If the internal degrees of freedom include the electron spin we have $s_T = -1$ and $U_T$ is antisymmetric. The matrix $U_T$ is also unitary so that its dimension $N$ must be even since its spectrum consists of pairs $\pm e^{i\phi}$. The case $s_T = +1$ can only be realized if the spin does not occur explicitly, for example because electrons in one spin state are pushed to high energies by a strong magnetic field. Then $\mathcal{T}$ is not the physical TRS but an effective antiunitary symmetry. For $s_T = +1$, $U_T$ is symmetric and the dimension $N$ is not restricted. Beyond these considerations, the specific form of $U_T$ as well as the specific form of structural point-group transformations depend on the physical nature of the internal degrees of freedom.

It is useful to write the pairing matrix as
\begin{equation}
\Delta(\mathbf{k}) = D(\mathbf{k})\, U_T .
\label{2.DeltaD.2}
\end{equation}
Under a general unitary symmetry transformation, Eq.\ (\ref{1.genUtrans.3}), the pairing matrix transforms as
\begin{equation}
\Delta(\mathbf{k}) \mapsto U\, \Delta(R^ {-1}\mathbf{k})\, U^T ,
\end{equation}
i.e., not like a matrix. One easily sees that $D(\mathbf{k})$ transforms as
\begin{equation}
D(\mathbf{k}) \mapsto U\, D(R^{-1}\mathbf{k})\, U^\dagger ,
\end{equation}
i.e., like a matrix. Analogously, one finds that under time reversal, $D(\mathbf{k})$ transforms as
\begin{equation}
D(\mathbf{k}) \mapsto U_T\, D^*(-\mathbf{k})\, U_T^\dagger .
\label{2.DUTtrans.3}
\end{equation}
Hence, $D(\mathbf{k})$ transforms like $H_N(\mathbf{k})$ under the magnetic point group, noting that $H_N^T(\mathbf{k}) = H_N^*(\mathbf{k})$ because of Hermiticity.

The condition (\ref{1.Deltaanti.2}) from fermionic antisymmetry together with Eqs.\ (\ref{1.UTsymm.3}) and (\ref{2.DeltaD.2}) implies
\begin{equation}
U_T\, D^T(-\mathbf{k})\, U_T^\dagger = -s_T\, D(\mathbf{k}) .
\label{2.Dfanti.3}
\end{equation}
For the standard case of $s_T=-1$, this relation is similar to TRS but differs from it since $D(\mathbf{k})$ is generally not Hermitian so that $D^T(\mathbf{k})$ is not the same as $D^*(\mathbf{k})$.

The general steps of the symmetry analysis are now as follows:

(1) Construct a basis $\{h_\nu\}$ of Hermitian matrices on the space of the internal degrees of freedom so that the $h_\nu$ transform as irreducible tensor operators of the magnetic point group $M$. In the case of point groups with two-dimensional real irreps that decompose into two one-dimensional complex irreps, use the real irreps since this allows one to find Hermitian $h_\nu$; the irreducible tensor operators of the corresponding one-dimensional complex irreps are generally not Hermitian. We call the $h_\nu$ \emph{basis matrices} and normalize them in such a way that $\Tr h_\nu^2 = N$ (the identity matrix is then normalized). It is possible that not all irreps occur. If the dimension of the internal Hilbert space is $N$ there are $N^2$ basis matrices.

To find the appropriate basis matrices and their irreps, it is necessary to determine the explicit forms of the symmetry operators, i.e., of the unitary matrices $U_T$ for time reversal and $U_g$ for at least a set of generators $g$ of the structural point group $G$.

(2) Generate a list of all irreps of $M$ that possess basis functions of momentum. These are all irreps that have the same parity under time reversal and inversion since the momentum $\mathbf{k}$ is odd under both. This allows $g+$ and $u-$ irreps but forbids $g-$ and $u+$ irreps. It is also useful to obtain characteristic basis functions for those irreps that have them but it should be kept in mind that these are understood as placeholders for arbitrary sets of functions with the same symmetry under operations from $M$. In this paper, we will usually represent basis functions by the lowest-order polynomials.

(3) Construct the general form of the normal-state Hamiltonian $H_N(\mathbf{k})$ by expanding it into the previously constructed basis,
\begin{equation}
H_N(\mathbf{k}) = \sum_n c_n(\mathbf{k})\, h_n .
\label{2.HNseries.3}
\end{equation}
We enumerate all basis matrices by $\nu$ but the subset that occurs in $H_N(\mathbf{k})$ by $n$. The Hamiltonian and every term in the expansion must be invariant under $M$, i.e., it must transform as an irreducible tensor operator belonging to the trivial irrep $A_\mathrm{triv}$ of $M$. This irrep is even under inversion and under time reversal, i.e., it is $A_{g+}$ or $A_{1g+}$ depending on the group. This requires the form factors $c_n(\mathbf{k})$ to transform as basis functions of the same irrep to which $h_n$ belongs. Moreover, for multidimensional irreps, $c_n(\mathbf{k})$ and $h_n$ must transform as the same component of the irrep and all of them must have the same amplitude if the basis functions and tensor operators are properly normalized. If there is no corresponding basis function for the irrep of some $h_\nu$, this matrix does not occur in Eq.~(\ref{2.HNseries.3}). This excludes $h_\nu$ belonging to $g-$ or $u+$ irreps. Note that $h_0 \equiv \openone$ is always an allowed basis matrix since it is a reducible tensor operator of the trivial irrep $A_\mathrm{triv}$ and $c_0(\mathbf{k}) \sim 1$ is always an appropriate basis function.

As noted above, for the standard case that the internal degrees of freedom include the electron spin, time reversal squares to $-1$ and the dimension $N$ of the internal Hilbert space must be even. Then the number of allowed basis matrices $h_n$ appearing in $H_N(\mathbf{k})$ is $N(N-1)/2$, as shown in Appendix~\ref{app.normal}.

The case of $N=2$ corresponds to spin being the only internal degree of freedom. There is only a single allowed basis matrix, namely $h_0=\openone$. This means that the normal-state Hamiltonian is independent of spin, which is required by TRS. For $N=4$, there are $6$ basis matrices $h_0=\openone$, $h_1$, \dots, $h_5$. These matrices have the special property that $h_1$, \dots, $h_5$ anticommute pairwise \cite{AWH12,Mes14}, while all matrices commute with $h_0$. Results restricted to this case are discussed in Subsection~\ref{sub.4D}.

(4) Construct the allowed pairing states. Here, we have much greater freedom than in constructing $H_N(\mathbf{k})$ since the superconducting state may break symmetries contained in $M$. We write the pairing matrix as
\begin{equation}
D(\mathbf{k}) = \sum_j \delta_j D_j(\mathbf{k}) ,
\label{2.DdeltaDj.2}
\end{equation}
where the $D_j(\mathbf{k})$ are linearly independent matrix-valued functions transforming like (being irreducible tensor operators belonging to) components of a specific irrep $\Gamma_s$, $s=\pm$, of the magnetic point group $M$ and the $\delta_j$ are complex pairing amplitudes. Recall that $D(\mathbf{k})$ transforms like a matrix. $D_j(\mathbf{k})$ can be chosen to be either Hermitian or anti-Hermitian. This can be seen as follows: Note first that Eq.\ (\ref{2.Dfanti.3}) has to be satisfied by each component separately since the $D_j(\mathbf{k})$ are independent functions,
\begin{equation}
U_T\, D_j^T(-\mathbf{k})\, U_T^\dagger = -s_T\, D_j(\mathbf{k}) .
\label{2.Dfanti.4}
\end{equation}
Depending on the irrep $\Gamma_s$, $D_j(\mathbf{k})$ is either even ($s=+$) or odd ($s=-$) under time reversal. Hence, Eq.\ (\ref{2.DUTtrans.3}) implies that
\begin{equation}
U_T\, D_j^*(-\mathbf{k})\, U_T^\dagger = s\, D_j(\mathbf{k}) .
\end{equation}
It follows that $D_j^T(\mathbf{k}) = -s_T\, s\, D_j^*(\mathbf{k})$ and thus
\begin{equation}
D_j^\dagger(\mathbf{k}) = -s_T\, s\, D_j(\mathbf{k}) .
\end{equation}
This equation states that the matrix functions $D_j(\mathbf{k})$ are Hermitian or anti-Hermitian depending on the sign $s_T$ of $\mathcal{T}^2$ and on the pairing state being even or odd under time reversal. Since we consider pure-irrep pairing all appearing $D_j(\mathbf{k})$ have the same sign $-s_T\,s$ under Hermitian conjugation.

For the standard case of $s_T = -1$, $D_j(\mathbf{k})$ is Hermitian (anti-Hermitian) for time-reversal-even (time-re\-ver\-sal-odd) irreps. For a time-reversal-odd irrep $\Gamma_-$, we can pull a common factor of $i$ out of all $D_j(\mathbf{k})$ and absorb it into the order parameters $\delta_j$ in Eq.\ (\ref{2.DdeltaDj.2}). This makes $D_j(\mathbf{k})$ Hermitian and changes the irrep from $\Gamma_-$ to $\Gamma_+$ since the behavior under spatial transformations is unaffected. Hence, for $s_T = -1$ it is sufficient to consider only the time-reversal-even $g+$ and $u+$ irreps for pairing states. This has the desirable consequence that the breaking of TRS is only encoded in the complex order parameters $\delta_j$. If and only if all $\delta_j$ can be made real by a global phase rotation the system respects TRS. This is equivalent to the nonvanishing $\delta_j$ having phase differences of $0$ or~$\pi$.

Conversely, for the nonstandard sign $s_T = +1$, $D_j(\mathbf{k})$ is anti-Hermitian (Hermitian) for time-reversal-even (time-re\-ver\-sal-odd) irreps. Pulling out a factor of $i$, we can make sure that $D_j(\mathbf{k})$ is Hermitian and the irrep is odd under time reversal ($g-$ or $u-$). Again, the breaking of TRS is only encoded in the order parameters $\delta_j$. If and only if all $\delta_j$ can be made purely imaginary by a global phase rotation the system respects TRS. This is again equivalent to the nonvanishing $\delta_j$ having phase differences of $0$ or~$\pi$.

A given model does not necessarily permit all time-re\-ver\-sal-even (or odd) irreps, though. To see this, note that the Hermitian-matrix-valued functions $D_j(\mathbf{k})$ can be expanded into the Hermitian basis matrices $h_\nu$ as
\begin{equation}
D_j(\mathbf{k}) = \sum_\nu d_{j\nu}(\mathbf{k})\, h_\nu ,
\label{2.Ddhexpand.2}
\end{equation}
with real functions $d_{j\nu}(\mathbf{k})$. Consider all products of momentum basis functions $g_l(\mathbf{k})$ belonging to irreps $\Gamma_l$ and of matrices $h_\nu$ belonging to irreps $\Gamma_\nu$. Any such product transforms according to the product representation $\Gamma_l \otimes \Gamma_\nu$, which is generally reducible. A reduction into irreps by standard methods reveals which symmetries of pairing states can occur. Recall that only $g+$ and $u-$ irreps possess momentum basis functions. The possible irreps of basis matrices $h_\nu$ have been obtained in step (1). By reducing all possible products and keeping only those irreps that are even (odd) under time reversal for $s_T=-1$ ($s_T = +1$), we obtain the possible irreps. The coefficients $d_{j\nu}(\mathbf{k})$ in Eq.\ (\ref{2.Ddhexpand.2}) can be constructed out of the functions $g_l(\mathbf{k})$ by standard methods. The full pairing matrix then has the form
\begin{equation}
D(\mathbf{k}) = \sum_\nu \sum_{j=1}^d \delta_j d_{j\nu}(\mathbf{k})\, 
h_\nu \equiv \sum_\nu f_\nu(\mathbf{k})\, h_\nu .
\label{2.Dfhexpand.3}
\end{equation}

The sum in Eq.\ (\ref{2.DdeltaDj.2}) generally contains many terms: While the number of possible basis matrices $h_\nu$ is finite, there are infinitely many smooth functions of momentum that transform according to the same irrep. This implies that TRS can be broken spontaneously for pairing belonging to any irrep, by having amplitudes $\delta_j$ with nontrivial phase differences \cite{endnote.TRSBexample}. Consideration of energetics \cite{VoG85,SiU91,BWW16,ABT17,BAM18} is useful to find plausible modes of TRS breaking. Spontaneous breaking of TRS occurs most naturally for multidimensional irreps, in the form of nontrivial phase factors of contributions belonging to different components of the irrep.

\begin{table}
\caption{\label{tab.pairing.states}Possible combinations of the signs under inversion (parity, $g$ for even, $u$ for odd) and under time reversal ($+$ for even, $-$ for odd) of irreps of pairing states for time reversal squaring to $s_T=\pm 1$. Recall that $s_T=-1$ is the standard case for electrons.}
\begin{ruledtabular}
\begin{tabular}{@{\hspace{1em}}cccc}
$s_T$ & pairing state & coefficients $d_{j\nu}(\mathbf{k})$ & basis matrices $h_\nu$ \\ \hline
$+1$ & $g-$ & $g+$ & $g-$ \\
 & & $u-$ & $u+$ \\
 & $u-$ & $g+$ & $u-$ \\
 & & $u-$ & $g+$ \\
$-1$ & $g+$ & $g+$ & $g+$ \\
 & & $u-$ & $u-$ \\
 & $u+$ & $g+$ & $u+$ \\
 & & $u-$ & $g-$ \\
\end{tabular}
\end{ruledtabular}
\end{table}

Since the normal state is inversion symmetric the superconducting pairing is either even ($g$ irreps) or odd ($u$ irreps) under inversion (parity). Table \ref{tab.pairing.states} shows the possible combinations of signs under inversion and time reversal. We note that for $s_T=-1$ and even-parity pairing, only basis matrices belonging to $g+$ and $u-$ irreps occur. These are the same matrices $h_n$ that appear in the normal-state Hamiltonian $H_N(\mathbf{k})$ in Eq.\ (\ref{2.HNseries.3}). On the other hand, for $s_T=-1$ and odd-parity pairing, only those basis matrices $h_\nu$ that do not occur in $H_N(\mathbf{k})$ can appear in $D(\mathbf{k})$. The numbers of allowed basis matrices are given in Appendix~\ref{app.normal}.

(5) Analyze the nodal structure for each pairing symmetry (irrep) assuming infinitesimal pairing strength. Since we are interested in the fate of BFSs we concentrate on the case $s_T=-1$ and even-parity pairing, while the other cases are briefly discussed in Appendix \ref{app.IPnodes}. In the limit of infinitesimal pairing amplitudes $\delta_j$, superconductivity can be described for each band separately. This is because the superconducting gap is of first order in $\delta_j$, whereas interband effects are of second order. Moreover, there are at most point or line nodes but no BFSs since interband pairing is responsible for the latter~\cite{ABT17,BAM18}.

The normal-state bands are twofold degenerate because of inversion symmetry and TRS. Hence, in an effective description of a single band, the dimension of the internal Hilbert space is $N=2$ and the internal degree of freedom can be described by a pseudospin of length $1/2$ \cite{BAM18,TBA21}. The superconducting pairing must be in the pseudospin-singlet channel since it is of even parity. The pairing matrix is then $f_0(\mathbf{k})\, \sigma_0$. Since $\sigma_0$ belongs to a $g+$ irrep, namely the trivial one, $f_0(\mathbf{k})$ must be a basis function of a $g+$ irrep; see Table \ref{tab.pairing.states}. The symmetry of the pairing state under the magnetic point group can then only be encoded in the function $f_0(\mathbf{k})$. This function thus generically transforms like the pairing matrix $D(\mathbf{k})$ of the full model. Moreover, the symmetry-imposed zeros of $f_0(\mathbf{k})$ in momentum space correspond to gap nodes for infinitesimal pairing (IP nodes). One of the main messages of Refs.\ \cite{BWW16,ABT17,BAM18} was that a momentum-independent but internally anisotropic pairing matrix can lead to a momentum-dependent function $f_0(\mathbf{k})$ and to gap nodes.

A simple but powerful criterion for IP nodes can be obtained as follows: As noted above, for $s_T=-1$ and even-parity pairing, the same basis matrices $h_n$ appear in $H_N(\mathbf{k})$ and $D(\mathbf{k})$. From Eq.\ (\ref{2.HNseries.3}), we know that the normal-state coefficients $c_n(\mathbf{k})$ transform like the basis matrices $h_n$ under the magnetic point group. Hence, the scalar function
\begin{equation}
F(\mathbf{k}) \equiv \sum_n c_n(\mathbf{k})\, f_n(\mathbf{k})
\label{1.Fk.3}
\end{equation}
transforms like the pairing matrix $D(\mathbf{k})$ and thus also like the form factor $f_0(\mathbf{k})$ in the single-band picture and, in particular, has the same symmetry-induced nodes. Therefore, we can use $F(\mathbf{k})$ as a proxy for the IP nodal structure. In fact, instead of the normal-state coefficients $c_n(\mathbf{k})$, we could use any set of basis functions belonging to the same irreps. We will see that an analogous measure emerges naturally for the case of $N=4$. 

(6) Check whether the nodes thereby obtained are inflated if the pairing amplitudes are not infinitesimal. The main tool is the Pfaffian $\Pf \tilde{\mathcal{H}}(\mathbf{k})$ of an antisymmetric Hamiltonian $\tilde{\mathcal{H}}(\mathbf{k})$ that is unitarily equivalent to the BdG Hamiltonian $\mathcal{H}(\mathbf{k})$. Such a unitary transformation is guaranteed to exist if the point group contains the inversion, as shown in \cite{ABT17}. A simpler version of the proof is presented in Appendix~\ref{app.Pfaffian}.

The square of the Pfaffian equals the determinant of the BdG Hamiltonian and thus the product of the quasiparticle energies. Hence, nodes of any kind correspond to $\Pf \tilde{\mathcal{H}}(\mathbf{k}) = 0$. As shown in Appendix \ref{app.Pfaffian}, the Pfaffian is real for even $N$ and imaginary for odd $N$. We define
\begin{equation}
P(\mathbf{k}) \equiv \left\{ \begin{array}{ll}
    \Pf \tilde{\mathcal{H}}(\mathbf{k}) & \mbox{for $N$ even,} \\[0.5ex]
    i\Pf \tilde{\mathcal{H}}(\mathbf{k}) & \mbox{for $N$ odd}
  \end{array} \right.
\end{equation}
to obtain a real quantity. We will simply call $P(\mathbf{k})$ the Pfaffian in the following. The sign of $P(\mathbf{k})$ turns out to depend on the choice of unitary transformation which leads to the antisymmetric matrix $\tilde{\mathcal{H}}(\mathbf{k})$. We choose this transformation in such a way that the Pfaffian is positive at some point far from the normal-state Fermi surface. Since the Pfaffian is a smooth function of momentum this fixes the sign for all $\mathbf{k}$.

For $s_T = -1$ and preserved TRS, $P(\mathbf{k})$ is nonnegative for all $\mathbf{k}$, the topological $\mathbb{Z}_2$ invariant is thus trivial, and there are no topologically protected BFSs, as shown in \cite{ABT17,BAM18}. Conversely, such BFSs are expected for broken TRS. The argument is reviewed in Appendix \ref{app.Pfaffian}. In addition, we there show that the Pfaffian can change sign and BFSs are expected also for $s_T = +1$, regardless of symmetry.

\subsection{Four-dimensional internal Hilbert space}
\label{sub.4D}

Even-parity superconductors with a four-dimensional internal Hilbert space (and time reversal squaring to $-1$) constitute the simplest case beyond the single-band paradigm. According to Appendix \ref{app.normal}, the normal-state Hamiltonian $H_N(\mathbf{k}) = \sum_n c_n(\mathbf{k})\, h_n$ is a superposition of six basis matrices $h_0$, \dots, $h_5$. As noted above, the same six basis matrices appear in the pairing matrix $D(\mathbf{k})$. These matrices realize a nice algebraic structure of $4\times 4$ gamma matrices: One can always choose the $h_n$ in such a way that $h_1$, \dots, $h_5$ anticommute pairwise, while $h_0=\openone$ commutes with any matrix \cite{AWH12,Mes14}; see Appendix \ref{app.algebra}. This implies that for any such model the eigenenergies in the normal state are
\begin{equation}
E_{N\pm}(\mathbf{k}) = c_0(\mathbf{k}) \pm \sqrt{ c_1^2(\mathbf{k}) + \ldots + c_5^2(\mathbf{k}) } ,
\end{equation}
both twofold degenerate. The algebraic structure also allows to derive analytical results for the quasiparticle energies in the superconducting state and for the Pfaffian $P(\mathbf{k})$. We obtain universal results when expressing these quantities in terms of coefficients of basis matrices.

Since we have found negative values of the Pfaffian and thus BFSs there, the same should be true for any model with $N=4$ in this class. For this conclusion to hold, it is important that the prefactors of the matrices are not constrained by symmetries so that all values of the Pfaffian can actually occur.

\begin{table*}
\caption{\label{tab.commute}Commutation relations of the matrices defined in Eqs.\ (\ref{2.BASIS.3a})--(\ref{2.BASIS.3c}). $+$ ($-$) denotes commutation (anticommutation).}
\begin{ruledtabular}
\begin{tabular}{ccccccccccccccccccc}
  & $H_0$ & $H_1$ & $H_2$ & $H_3$ & $H_4$ & $H_5$ & $\Gamma_0$ & $\Gamma_1$ & $\Gamma_2$ & $\Gamma_3$ &
  $\Gamma_4$ & $\Gamma_5$ & $\Phi_0$ & $\Phi_1$ & $\Phi_2$ & $\Phi_3$ & $\Phi_4$ & $\Phi_5$ \\ \hline
$H_0$    & $+$ & $+$ & $+$ & $+$ & $+$ & $+$ & $-$ & $-$ & $-$ & $-$ & $-$ & $-$ & $-$ & $-$ & $-$ & $-$ & $-$ & $-$ \\
$H_1$    & $+$ & $+$ & $-$ & $-$ & $-$ & $-$ & $-$ & $-$ & $+$ & $+$ & $+$ & $+$ & $-$ & $-$ & $+$ & $+$ & $+$ & $+$ \\
$H_2$    & $+$ & $-$ & $+$ & $-$ & $-$ & $-$ & $-$ & $+$ & $-$ & $+$ & $+$ & $+$ & $-$ & $+$ & $-$ & $+$ & $+$ & $+$ \\
$H_3$    & $+$ & $-$ & $-$ & $+$ & $-$ & $-$ & $-$ & $+$ & $+$ & $-$ & $+$ & $+$ & $-$ & $+$ & $+$ & $-$ & $+$ & $+$ \\
$H_4$    & $+$ & $-$ & $-$ & $-$ & $+$ & $-$ & $-$ & $+$ & $+$ & $+$ & $-$ & $+$ & $-$ & $+$ & $+$ & $+$ & $-$ & $+$ \\
$H_5$    & $+$ & $-$ & $-$ & $-$ & $-$ & $+$ & $-$ & $+$ & $+$ & $+$ & $+$ & $-$ & $-$ & $+$ & $+$ & $+$ & $+$ & $-$ \\
$\Gamma_0$ & $-$ & $-$ & $-$ & $-$ & $-$ & $-$ & $+$ & $+$ & $+$ & $+$ & $+$ & $+$ & $-$ & $-$ & $-$ & $-$ & $-$ & $-$ \\
$\Gamma_1$ & $-$ & $-$ & $+$ & $+$ & $+$ & $+$ & $+$ & $+$ & $-$ & $-$ & $-$ & $-$ & $-$ & $-$ & $+$ & $+$ & $+$ & $+$ \\
$\Gamma_2$ & $-$ & $+$ & $-$ & $+$ & $+$ & $+$ & $+$ & $-$ & $+$ & $-$ & $-$ & $-$ & $-$ & $+$ & $-$ & $+$ & $+$ & $+$ \\
$\Gamma_3$ & $-$ & $+$ & $+$ & $-$ & $+$ & $+$ & $+$ & $-$ & $-$ & $+$ & $-$ & $-$ & $-$ & $+$ & $+$ & $-$ & $+$ & $+$ \\
$\Gamma_4$ & $-$ & $+$ & $+$ & $+$ & $-$ & $+$ & $+$ & $-$ & $-$ & $-$ & $+$ & $-$ & $-$ & $+$ & $+$ & $+$ & $-$ & $+$ \\
$\Gamma_5$ & $-$ & $+$ & $+$ & $+$ & $+$ & $-$ & $+$ & $-$ & $-$ & $-$ & $-$ & $+$ & $-$ & $+$ & $+$ & $+$ & $+$ & $-$ \\
$\Phi_0$ & $-$ & $-$ & $-$ & $-$ & $-$ & $-$ & $-$ & $-$ & $-$ & $-$ & $-$ & $-$ & $+$ & $+$ & $+$ & $+$ & $+$ & $+$ \\
$\Phi_1$ & $-$ & $-$ & $+$ & $+$ & $+$ & $+$ & $-$ & $-$ & $+$ & $+$ & $+$ & $+$ & $+$ & $+$ & $-$ & $-$ & $-$ & $-$ \\
$\Phi_2$ & $-$ & $+$ & $-$ & $+$ & $+$ & $+$ & $-$ & $+$ & $-$ & $+$ & $+$ & $+$ & $+$ & $-$ & $+$ & $-$ & $-$ & $-$ \\
$\Phi_3$ & $-$ & $+$ & $+$ & $-$ & $+$ & $+$ & $-$ & $+$ & $+$ & $-$ & $+$ & $+$ & $+$ & $-$ & $-$ & $+$ & $-$ & $-$ \\
$\Phi_4$ & $-$ & $+$ & $+$ & $+$ & $-$ & $+$ & $-$ & $+$ & $+$ & $+$ & $-$ & $+$ & $+$ & $-$ & $-$ & $-$ & $+$ & $-$ \\
$\Phi_5$ & $-$ & $+$ & $+$ & $+$ & $+$ & $-$ & $-$ & $+$ & $+$ & $+$ & $+$ & $-$ & $+$ & $-$ & $-$ & $-$ & $-$ & $+$ \\
\end{tabular}
\end{ruledtabular}
\end{table*}

The BdG Hamiltonian in Eq.\ (\ref{1.HBdG.2}) can be written as a linear combination of 18 Hermitian $8 \times 8$ matrices,
\begin{equation}
\mathcal{H}(\mathbf{k}) = \sum_{n=0}^5 c_n(\mathbf{k})\, H_n
  + \sum_{n=0}^5 f^1_n(\mathbf{k})\, \Gamma_n
  + \sum_{n=0}^5 f^2_n(\mathbf{k})\, \Phi_n .
\label{2.HBdG.4}
\end{equation}
The coefficients are all real functions of momentum. The 18 matrices are
\begin{align}
H_n &= \left(\begin{array}{cc}
    h_n & 0 \\
    0 & - U_P^* h_n^T U_P^T
  \end{array}\right) ,
\label{2.BASIS.3a} \\
\Gamma_n &= \left(\begin{array}{cc}
    0 & h_n U_T \\
    U_T^\dagger h_n & 0
  \end{array}\right) ,
\label{2.BASIS.3b} \\
\Phi_n &= \left(\begin{array}{cc}
    0 & ih_n U_T \\
    -i U_T^\dagger h_n & 0
  \end{array}\right) ,
\label{2.BASIS.3c}
\end{align}
where $n=0,\ldots,5$. The definition of $H_n$ requires some discussion. The matrices $h_n$ are irreducible tensor operators of $g$ or $u$ irreps and
\begin{equation}
H_n = \left\{ \begin{array}{ll}
  \displaystyle \left(\begin{array}{cc}
      h_n & 0 \\
      0 & - h_n^T
    \end{array}\right) & \mbox{for $g$ irreps,} \\[2.5ex]
  \displaystyle \left(\begin{array}{cc}
      h_n & 0 \\
      0 & h_n^T
    \end{array}\right) & \mbox{for $u$ irreps.}
  \end{array}\right.
\end{equation}
For $g$ irreps, the minus sign of $-\mathbf{k}$ in the lower right block of Eq.\ (\ref{1.HBdG.2}) drops out and the first term in Eq.\ (\ref{2.HBdG.4}) is obvious. For $u$ irreps, the form factor $c_n(\mathbf{k})$ is an odd function and the lower right block obtains an additional sign change. Since in Eq.\ (\ref{2.HBdG.4}) the coefficient of $H_n$ is $c_n(\mathbf{k})$ this sign must be incorporated into~$H_n$. The matrices $H_n$, $\Gamma_n$, and $\Phi_n$ are all Hermitian, traceless, square to $\openone$, and either commute or anticommute according to Table \ref{tab.commute}. Moreover, if $A_\alpha$, $\alpha=1,\ldots,18$ denote all 18 matrices, we have
\begin{equation}
\Tr \{A_\alpha, A_\beta\} \equiv \Tr (A_\alpha A_\beta + A_\beta A_\alpha) = 16\, \delta_{\alpha\beta} .
\label{2.TrAA.2}
\end{equation}

The Pfaffian can be expressed in closed form, as discussed in more detail in Appendix \ref{app.analytical.Pfaffian}. We suppress momentum arguments for the rest of this section. It is useful to define the real five-vectors
\begin{align}
\vec{c} &\equiv (c_1,c_2,c_3,c_4,c_5) , \\
\vec{f}^{\,1} &\equiv (f^1_1,f^1_2,f^1_3,f^1_4,f^1_5) , \\
\vec{f}^{\,2} &\equiv (f^2_1,f^2_2,f^2_3,f^2_4,f^2_5)
\end{align}
and the Minkowski-type scalar product
\begin{equation}
\langle A,B\rangle \equiv A_0 B_0 - \vec{A}\cdot\vec{B} .
\label{2.Minkowski.1}
\end{equation}
The Pfaffian can then be written as
\begin{align}
P(\mathbf{k}) &= \langle c,c\rangle^2 + \langle f^1,f^1\rangle^2
  + \langle f^2,f^2\rangle^2 \nonumber \\
&\quad{}+ 4\, \big( \langle c,f^1\rangle^2 + \langle f^1,f^2\rangle^2
  + \langle f^2,c\rangle^2 \big) \nonumber \\
&\quad{}- 2\, \big( \langle c,c\rangle\, \langle f^1,f^1\rangle
  + \langle f^1,f^1\rangle\, \langle f^2,f^2\rangle \nonumber \\
&\quad{}+ \langle f^2,f^2\rangle\, \langle c,c\rangle \big) .
\label{2.Pfr.5}
\end{align}
Another form that will prove useful is
\begin{align}
P(\mathbf{k}) &= \big( \langle c,c\rangle - \langle f^1,f^1\rangle - \langle f^2,f^2\rangle \big)^2
  \nonumber \\
&\quad{}+ 4\, \big( \langle c,f^1\rangle^2 + \langle c,f^2\rangle^2 + \langle f^1,f^2\rangle^2
  \nonumber \\
&\quad{}- \langle f^1,f^1\rangle \langle f^2,f^2\rangle \big) .
\label{2.Pfr.5a}
\end{align}
Nodes are signaled by $P(\mathbf{k})=0$. If $P(\mathbf{k})$ becomes negative for some momenta $\mathbf{k}$ we obtain two-di\-men\-sio\-nal BFSs \cite{ABT17,BAM18}. The algebraic structure and the expressions for the Pfaffian are the same for all models with even-parity superconductors, time reversal squaring to $-1$, and $N=4$. Hence, the conclusion of Refs.\ \cite{ABT17,BAM18} that nodes are inflated into BFSs for TRS-breaking superconducting states applies to all such models.

For infinitesimal pairing, we can neglect terms of fourth order in the amplitudes $f^\alpha_n$ compared to terms of second order in Eq.\ (\ref{2.Pfr.5a}). The result
\begin{align}
P(\mathbf{k}) &\cong \big( \langle c,c\rangle - \langle f^1,f^1\rangle - \langle f^2,f^2\rangle \big)^2
  \nonumber \\
&\quad{}+ 4\, \big( \langle c,f^1\rangle^2 + \langle c,f^2\rangle^2 \big)
\label{2.Pfra.7}
\end{align}
is nonnegative. Hence, the momentum-space volume of the BFSs shrinks to zero for infinitesimal pairing, leaving only point and line nodes. Since the expression in Eq.\ (\ref{2.Pfra.7}) is a sum of squares IP nodes occur when three conditions hold simultaneously. The first reads as
\begin{equation}
\langle c,c\rangle - \langle f^1,f^1\rangle - \langle f^2,f^2\rangle = 0 .
\label{2.IPnodec.3}
\end{equation}
Since $\langle c,c\rangle = E_{N+} E_{N-} = 0$ is a criterion for the normal-state Fermi surface we can say that Eq.\ (\ref{2.IPnodec.3}) describes a \emph{renormalized Fermi surface}. It will be close to the normal-state Fermi surface in the typical case that the pairing energy is small compared to the chemical potential. The second and third condition read as $\langle c,f^1\rangle = \langle c,f^2\rangle = 0$, which are equivalent to
\begin{equation}
\langle c,f\rangle = 0 ,
\end{equation}
where $f\equiv f^1+if^2$. Explicitly, this condition reads as
\begin{equation}
c_0\, (f^1_0 + i f^2_0) - \sum_{n=1}^5 c_n\, (f^1_n + i f^2_n)
  \equiv c_0 f_0 - \sum_{n=1}^5 c_n f_n = 0 .
\label{2.cfN4.4}
\end{equation}
Except for the signs, which do not matter, this agrees with the function $F(\mathbf{k})$ that we found above to encode the IP nodal structure; see Eq.~(\ref{1.Fk.3}).

\section{Applications}
\label{sec.appl}

In the following, we illustrate the general procedure for specific examples. We will mainly consider a familiar setting: the dimension of the internal Hilbert space is $N=4$, resulting from spin and either orbital or basis site, and the model is described by the cubic point group $O_h$. This point group has ten irreps, $A_{1g}$, $A_{2g}$, $E_g$, $T_{1g}$, $T_{2g}$, $A_{1u}$, $A_{2u}$, $E_u$, $T_{1u}$, and $T_{2u}$. For the corresponding gray magnetic point group, the number of irreducible real corepresentations is doubled to $A_{1g+}$, $A_{1g-}$, $A_{2g+}$, etc.

\subsection{Two \textit{s}-orbitals}
\label{sub.same}

We first consider a lattice without basis and with two orbitals per site that are invariant under all point-group transformations. This means that they transform according to the trivial irrep $A_{1g}$ or, in other words, like \textit{s}-orbitals. The interesting point here is that even such a simple model supports nontrivial multiband superconductivity with BFSs.

For the internal Hilbert space, we use the basis $\{ |1{\uparrow}\rangle, |1{\downarrow}\rangle, |2{\uparrow}\rangle, |2{\downarrow}\rangle \}$, where $1$, $2$ refers to the orbital and $\uparrow$, $\downarrow$ to the spin. In this section, the first factor in Kronecker products refers to the orbital and the second to the spin. The matrix representation of the inversion or parity operator $P$ has the trivial form
\begin{equation}
U_P = \openone = \sigma_0 \otimes \sigma_0 .
\end{equation}
The unitary part of the time-reversal operator is
\begin{equation}
U_T = \sigma_0 \otimes i\sigma_2
\end{equation}
since the orbitals are invariant under time reversal, while in the spin sector we have the standard form~$i\sigma_2$.

\begin{table}
\caption{\label{tab.2s.basis}Basis matrices on the internal Hilbert space for the case of two \textit{s}-orbitals and point group $O_h$. The basis matrices are irreducible tensor operators of the irreps listed in the second column. For multidimensional irreps, the states transforming into each other under point-group operations are distinguished by the index in the third column.}
\begin{ruledtabular}
\begin{tabular}{@{\hspace{3em}}ccc@{\hspace{3em}}}
$h_\nu$ & Irrep & Component \\ \hline
$\sigma_0 \otimes \sigma_0$ & $A_{1g+}$ & \\
$\sigma_0 \otimes \sigma_1$ & $T_{1g-}$ & 1 \\
$\sigma_0 \otimes \sigma_2$ & & 2 \\
$\sigma_0 \otimes \sigma_3$ & & 3 \\
$\sigma_1 \otimes \sigma_0$ & $A_{1g+}$ & \\
$\sigma_1 \otimes \sigma_1$ & $T_{1g-}$ & 1 \\
$\sigma_1 \otimes \sigma_2$ & & 2 \\
$\sigma_1 \otimes \sigma_3$ & & 3 \\
$\sigma_2 \otimes \sigma_0$ & $A_{1g-}$ & \\
$\sigma_2 \otimes \sigma_1$ & $T_{1g+}$ & 1 \\
$\sigma_2 \otimes \sigma_2$ & & 2 \\
$\sigma_2 \otimes \sigma_3$ & & 3 \\
$\sigma_3 \otimes \sigma_0$ & $A_{1g+}$ & \\
$\sigma_3 \otimes \sigma_1$ & $T_{1g-}$ & 1 \\
$\sigma_3 \otimes \sigma_2$ & & 2 \\
$\sigma_3 \otimes \sigma_3$ & & 3 \\
\end{tabular}
\end{ruledtabular}
\end{table}

The 16 basis matrices $h_\nu$ of the space of Hermitian $4\times 4$ matrices obtained as Kronecker products are listed in Table \ref{tab.2s.basis}, together with the corresponding irreps. To understand the table, first consider the structural point group. Since the orbitals transform trivially under all point-group elements the spin alone determines the irrep. Then $\sigma_0$ obviously transforms trivially, i.e., according to $A_{1g}$, while $\bsigma=(\sigma_1,\sigma_2,\sigma_3)$ is a pseudovector, which transforms according to $T_{1g}$. Regarding time reversal, $\sigma_0$ in the spin sector is of course even, whereas $\bsigma$ is odd. However, the time-reversal operator $\mathcal{T}$ is antilinear so that the imaginary Pauli matrix $\sigma_2$ in the orbital sector gives another sign change. Note that although the orbital degree of freedom appears to be a trivial spectator, it does lead to the appearance of the additional irreps $A_{1g-}$ and $T_{1g+}$.

Next, we consider momentum basis functions. As noted above, they only exist for $g+$ and $u-$ irreps. Low-or\-der polynomial basis functions can be found in tables \cite{DDJ08,Kat}. It is important to note that for our purposes the constant function and the second-order function $k_x^2+k_y^2+k_z^2$ are allowed basis functions of $A_{1g+}$ but are not listed in some tables. The tables usually do not show a basis function for $A_{1u-}$ since the simplest one is of order $l=9$, specifically $k_xk_yk_z\, [ k_x^4 (k_y^2-k_z^2) + k_y^4 (k_z^2-k_x^2) + k_z^4 (k_x^2-k_y^2) ]$ \cite{KOW77,DDJ08}. The possible irreps of pairing states are now obtained by reducing all products of the allowed irreps of the momentum-dependent form factor and of pairing matrices and excluding the ones that are odd under time reversal and thus violate fermionic antisymmetry. The reduction of the remaining combinations is shown in Table \ref{tab.2s.reduce}. The normal-state Hamiltonian $H_N(\mathbf{k})$ can, and generically does, contain all combinations that transform according to $A_{1g+}$, set in bold face. Only the first row of the table (form-factor irrep $A_{1g+}$) is compatible with purely local pairing, which can thus have $A_{1g+}$ or $T_{1g+}$ symmetry. Note that the latter  is impossible for a single-orbital system.

\begin{table*}
\caption{\label{tab.2s.reduce}Reduction of product representations of the allowed irreps of $\mathbf{k}$-dependent form factors (rows) and basis matrices $h_\nu$ (columns) for two \textit{s}-orbitals. For the form factors, the minimum order of polynomial basis functions is given in the second column. ``\bad'' indicates products that are forbidden since they violate fermionic antisymmetry.}
\begin{ruledtabular}
\begin{tabular}{lc@{\hspace{6em}}llll}
\multicolumn{2}{c}{Form factor:\hspace*{6em}} & \multicolumn{4}{c}{Pairing matrix: Irrep} \\
Irrep & Minimum order $l$ & $A_{1g+}$ & $T_{1g+}$ & $A_{1g-}$ & $T_{1g-}$ 
\\ \hline
$A_{1g+}$ & 0 & {\boldmath$A_{1g+}$} & $T_{1g+}$ & \bad & \bad \\
$A_{2g+}$ & 6 & $A_{2g+}$ & $T_{2g+}$ & \bad & \bad \\
$E_{g+}$ & 2 & $E_{g+}$ & $T_{1g+} \oplus T_{2g+}$ & \bad & \bad \\
$T_{1g+}$ & 4 & $T_{1g+}$ & $\mbox{\boldmath$A_{1g+}$} \oplus E_{g+} \oplus T_{1g+} \oplus T_{2g+}$ &
  \bad & \bad \\
$T_{2g+}$ & 2 & $T_{2g+}$ & $A_{2g+} \oplus E_{g+} \oplus T_{1g+} \oplus T_{2g+}$ &
  \bad & \bad \\
$A_{1u-}$ & 9 & \bad & \bad & $A_{1u+}$ & $T_{1u+}$ \\
$A_{2u-}$ & 3 & \bad & \bad & $A_{2u+}$ & $T_{2u+}$ \\
$E_{u-}$ & 5 & \bad & \bad & $E_{u+}$ & $T_{1u+} \oplus T_{2u+}$ \\
$T_{1u-}$ & 1 & \bad & \bad &
  $T_{1u+}$ & $A_{1u+} \oplus E_{u+} \oplus T_{1u+} \oplus T_{2u+}$ \\
$T_{2u-}$ & 3 & \bad & \bad &
  $T_{2u+}$ & $A_{2u+} \oplus E_{u+} \oplus T_{1u+} \oplus T_{2u+}$ \\
\end{tabular}
\end{ruledtabular}
\end{table*}

The normal-state Hamiltonian contains two types of terms, generated by $A_{1g+}\otimes A_{1g+}$ and by $T_{1g+}\otimes T_{1g+}$, respectively. For the first, there are three basis matrices belonging to $A_{1g+}$ according to Table \ref{tab.2s.basis}, hence we get
\begin{equation}
H_{N1}(\mathbf{k}) = c_{00}(\mathbf{k})\, \sigma_0 \otimes \sigma_0
  + c_{10}(\mathbf{k})\, \sigma_1 \otimes \sigma_0
  + c_{30}(\mathbf{k})\, \sigma_3 \otimes \sigma_0 ,
\end{equation}
where $c_{00}$, $c_{10}$, and $c_{30}$ are generally distinct basis functions of $A_{1g+}$. In other words, they are invariant under all elements of the magnetic point group. The leading polynomial terms read a~\cite{DDJ08,Kat}
\begin{align}
c_{m0}(\mathbf{k}) &= c_{m0}^{(0)} + c_{m0}^{(2)}\, (k_x^2+k_y^2+k_z^2)
  + c_{m0}^{(4)}\, (k_x^4+k_y^4+k_z^4) \nonumber \\
&\quad{} + c_{m0}^{(6)}\, k_x^2k_y^2k_z^2 + \ldots
\label{2.cm0.3}
\end{align}
for $m=0,1,3$. For the second type, we observe that there is a single triplet of matrices belonging to $T_{1g+}$, namely $\sigma_2 \otimes \bsigma$. To obtain an invariant Hamiltonian, they must each be multiplied by the corresponding momentum basis function, which gives
\begin{align}
H_{N2}(\mathbf{k}) &= c_{21}(\mathbf{k})\, \sigma_2 \otimes \sigma_1
  + c_{22}(\mathbf{k})\, \sigma_2 \otimes \sigma_2 \nonumber \\
&\quad{} + c_{23}(\mathbf{k})\, \sigma_2 \otimes \sigma_3 .
\end{align}
The functions $c_{2n}$ are not independent but must transform into each other under the magnetic point group. The leading terms are~\cite{DDJ08,Kat}
\begin{align}
c_{21}(\mathbf{k}) &= c_{2}^{(4)}\, k_yk_z (k_y^2-k_z^2)
  + c_{2}^{(6)}\, k_yk_z (k_y^4-k_z^4) + \ldots ,
\label{2.c21.3} \\
c_{22}(\mathbf{k}) &= c_{2}^{(4)}\, k_zk_x (k_z^2-k_x^2)
  + c_{2}^{(6)}\, k_zk_x (k_z^4-k_x^4) + \ldots , \\
c_{23}(\mathbf{k}) &= c_{2}^{(4)}\, k_xk_y (k_x^2-k_y^2)
  + c_{2}^{(6)}\, k_xk_y (k_x^4-k_y^4) + \ldots
\label{2.c23.3}
\end{align}
The full normal-state Hamiltonian
\begin{equation}
H_N(\mathbf{k}) = H_{N1}(\mathbf{k}) + H_{N2}(\mathbf{k})
  = \sum_{n=0}^5 c_n(\mathbf{k})\, h_n
\end{equation}
is a linear combination of the six basis matrices
\begin{align}
h_0 &\equiv \sigma_0 \otimes \sigma_0 & A_{1g+}, \label{3.basis.2.0} \\
h_1 &\equiv \sigma_1 \otimes \sigma_0 & A_{1g+}, \\
h_2 &\equiv \sigma_3 \otimes \sigma_0 & A_{1g+}, \\
h_3 &\equiv \sigma_2 \otimes \sigma_1 & T_{1g+}, \label{3.basis.2.3} \\
h_4 &\equiv \sigma_2 \otimes \sigma_2 & T_{1g+}, \\
h_5 &\equiv \sigma_2 \otimes \sigma_3 & T_{1g+}, \label{3.basis.2.5}
\end{align}
where the irreps are also given. $h_1$, \dots, $h_5$ anticommute pairwise; see Appendices \ref{app.normal} and~\ref{app.algebra}.

Table \ref{tab.2s.reduce} also provides useful information on superconductivity: (a) All ten irreps that are even under time reversal appear as symmetries of possible pairing states. In fact, every $g+$ irrep $\Gamma_{g+}$ is possible for any model since such a symmetry can be realized by combining an even momentum-space basis function belonging to $\Gamma_{g+}$ with the $A_{1g+}$ basis matrix $h_0=\openone$. On the other hand, $u+$ pairing state require $g-$ basis matrices, here belonging to $A_{1g-}$ and $T_{1g-}$ \cite{endnote.onegminus}. (b) For the even-parity pairing states, only the five $g+$ irreps are relevant. From Table \ref{tab.2s.reduce}, we see that then only the basis matrices transforming according to $A_{1g+}$ or $T_{1g+}$ occur. Further inspection shows that both types of basis matrices contribute to all $g+$ pairing states (all five occur in both columns). We conclude that for pairing states belonging to any single $g+$ irrep, all $A_{1g+}$ and $T_{1g+}$ basis matrices can appear in the pairing matrix
\begin{equation}
D(\mathbf{k}) = \sum_{n=0}^5 f_n(\mathbf{k})\, h_n .
\label{2.Dfh.3}
\end{equation}
What changes between different pairing states are the momentum-dependent form factors $f_n(\mathbf{k})$. In the following, we analyze the nodal structure for several exemplary pairing symmetries.

\subsubsection{$A_{1g+}$ pairing}
\label{subsub.2s.A1g}

We start with the simplest pairing symmetry, $A_{1g+}$. The construction of allowed terms in $D(\mathbf{k})$ is analogous to the construction of $H_N(\mathbf{k})$. From Table \ref{tab.2s.reduce}, we find two contributions to $A_{1g+}$ pairing: (a) form factors that transform according to $A_{1g+}$ combined with the three $A_{1g+}$ basis matrices and (b) a triplet of $T_{1g+}$ form factors combined with the triplet of $T_{1g+}$ basis matrices. We discuss these two contributions in turn. It will prove useful to separate momentum-independent pairing amplitudes denoted by $\delta_{\cdots}$ from suitably normalized momentum basis functions denoted by $d_{\cdots}(\mathbf{k})$, as done in Eq.~(\ref{2.Dfhexpand.3}).

(a) This contribution to the pairing matrix reads as
\begin{align}
D_1(\mathbf{k}) &= \delta_{00} d_{00}(\mathbf{k})\, \sigma_0 \otimes \sigma_0
  + \delta_{10} d_{10}(\mathbf{k})\, \sigma_1 \otimes \sigma_0 \nonumber \\
&\quad{}+ \delta_{30} d_{30}(\mathbf{k})\, \sigma_3 \otimes \sigma_0 ,
\label{3.A1g.D1.2}
\end{align}
where $d_{00}(\mathbf{k})$, $d_{10}(\mathbf{k})$, and $d_{30}(\mathbf{k})$ are basis functions of $A_{1g+}$, and $\delta_{00}$, $\delta_{10}$, and $\delta_{30}$ denote the corresponding pairing amplitudes. The leading polynomial forms of the basis functions are
\begin{equation}
d_{m0}(\mathbf{k}) = d_{m0}^{(0)} + d_{m0}^{(2)}\, (k_x^2+k_y^2+k_z^2) + \ldots ,
\end{equation}
where we can set the three constants $d_{m0}^{(0)}$ to unity as a normalization. The higher-order coefficients are then generally distinct for different $m$. These contributions can be interpreted as \textit{s}-wave pairing since the minimum order of the basis functions is $l=0$. We use the terms \textit{s}-wave, \textit{p}-wave, etc.\ to describe only the momentum dependence, not the symmetry of the full pairing state, for which we always use the irreps. The contribution is evidently spin-singlet pairing because the matrix acting in spin space is $\sigma_0$.

(b) The reducible representation $T_{1g+}\otimes T_{1g+}$ has nine matrix-valued basis functions $d_m(\mathbf{k})\, \sigma_2 \otimes \sigma_n$, $m,n=1,2,3$, where $d_m(\mathbf{k})$ are momentum basis functions of $T_{1g+}$. The reduction $T_{1g+}\otimes T_{1g+} = A_{1g+} \oplus E_{g+} \oplus T_{1g+} \oplus T_{2g+}$ tells us that a basis change to matrix basis functions of the four indicated irreps exists. We here need to find the linear combination of the $d_m(\mathbf{k})\, \sigma_2 \otimes \sigma_n$ that transforms according to $A_{1g+}$. This is simply the sum over products of corresponding components with identical coefficients, i.e., 
\begin{align}
D_2(\mathbf{k}) &= \delta_t\, \big[ d_{21}(\mathbf{k})\, \sigma_2 \otimes \sigma_1
  + d_{22}(\mathbf{k})\, \sigma_2 \otimes \sigma_2 \nonumber \\
&\quad{} + d_{23}(\mathbf{k})\, \sigma_2 \otimes \sigma_3 \big] ,
\label{3.A1g.D2.2}
\end{align}
where $d_{21}(\mathbf{k})$, $d_{22}(\mathbf{k})$, and $d_{23}(\mathbf{k})$ form a triplet of $T_{1g+}$ basis functions and $\delta_t$ is their common amplitude. The leading polynomials are
\begin{align}
d_3(\mathbf{k}) &\equiv d_{21}(\mathbf{k}) = d_t^{(4)}\, k_y k_z (k_y^2-k_z^2) + \ldots ,
\label{3.A1g.D2.3a} \\
d_4(\mathbf{k}) &\equiv d_{22}(\mathbf{k}) = d_t^{(4)}\, k_z k_x (k_z^2-k_x^2) + \ldots , \\
d_5(\mathbf{k}) &\equiv d_{23}(\mathbf{k}) = d_t^{(4)}\, k_x k_y (k_x^2-k_y^2) + \ldots ,
\label{3.A1g.D2.3c}
\end{align}
where we can choose $d_t^{(4)}=1$ as a normalization. This is \textit{g}-wave spin-triplet (hence the subscript ``\textit{t}'') pairing since the minimum order is $l = 4$ and the Pauli matrices $\sigma_1$, $\sigma_2$, $\sigma_3$ act on the spin Hilbert space. This combination is made possible by the nontrivial orbital content.

The full $A_{1g+}$ pairing matrix has the usual form
\begin{equation}
D(\mathbf{k}) \equiv D_1(\mathbf{k}) + D_2(\mathbf{k}) = \sum_{n=0}^5 
f_n(\mathbf{k})\, h_n ,
\end{equation}
where the symmetry properties of the form factors $f_n(\mathbf{k})$ have been obtained above.

We first consider pairing that respects TRS. Then, all $f_n(\mathbf{k})$ can be chosen real. Equation (\ref{2.cfN4.4}) gives the condition for IP nodes. Each pairing form factor $f_n(\mathbf{k})$ is multiplied by the corresponding normal-state form factor $c_n(\mathbf{k})$. The contribution (a) give, to leading order,
\begin{align}
c_0&(\mathbf{k}) f_0(\mathbf{k}) - c_1(\mathbf{k}) f_1(\mathbf{k}) - c_2(\mathbf{k}) f_2(\mathbf{k})
  \nonumber \\
&= c_{00}^{(0)} \delta_{00} - c_{10}^{(0)} \delta_{10} - c_{30}^{(0)} \delta_{30} + \ldots ,
\end{align}
which is generically nonzero and nodeless. The expression can of course have accidental nodes from higher-order terms, which we disregard here.

For the contribution (b), $(c_3,c_4,c_5)$ and $(f_3,f_4,f_5)$ are corresponding basis functions of $T_{1g+}$, and we find
\begin{align}
- c_3&(\mathbf{k}) f_3(\mathbf{k}) - c_4(\mathbf{k}) f_4(\mathbf{k}) - c_5(\mathbf{k}) f_5(\mathbf{k})
  \nonumber \\
&= - c_2^{(4)} \delta_t
 \big[ k_y^2 k_z^2 (k_y^2-k_z^2)^2 + k_z^2 k_x^2 (k_z^2-k_x^2)^2 \nonumber \\
&\quad{} + k_x^2 k_y^2 (k_x^2-k_y^2)^2 \big] + \ldots
\end{align}
This expression vanishes if the conditions
\begin{align}
k_y^2k_z^2 (k_y-k_z)^2 (k_y+k_z)^2 &= 0 , \\
k_z^2k_x^2 (k_z-k_x)^2 (k_z+k_x)^2 &= 0 , \\
k_x^2k_y^2 (k_x-k_y)^2 (k_x+k_y)^2 &= 0
\end{align}
hold simultaneously. This is the case for the $6+8+12=26$ high-symmetry directions in the $O_h$ Brillouin zone. Hence, there are 26 point nodes on a spheroidal normal-state Fermi surface around the $\Gamma$ point. The higher-order terms in the basis functions do not change this picture since the nodes are imposed by $T_{1g+}$ symmetry. Since the conditions only contain squares, they are second-order (``double Weyl'') point nodes~\cite{SiU91,BAM18,RGF10}.

Together with the generically nodeless contribution (a), $\langle c,f\rangle$ can contain first-order line nodes provided that the amplitude $c_2^{(4)} \delta_t$ is sufficiently large and not all terms have the same sign. The location of these line nodes is not fixed by symmetries. In this respect, the situation is similar to the case of mixed singlet-triplet pairing in noncentrosymmetric superconductors \cite{ScB15}. However, there is nothing that prevents a full gap, which is typically energetically favorable.

Breaking of TRS is possible for one-dimensional irreps, see Sec.\ \ref{sec.theory}. However, since the $A_{1g+}$ time-reversal-symmetric state is generically nodeless, the breaking of TRS is not expected to lead to a reduction of the internal energy \cite{SiU91}. If TRS does break, then the condition (\ref{2.cfN4.4}) for IP nodes splits into two independent conditions for the real and imaginary parts of $\langle c,f\rangle$. Hence, TRS-breaking $A_{1g+}$ pairing states are even less likely to have nodes than time-reversal-symmetric ones.

\subsubsection{$A_{2g+}$ pairing}
\label{subsub.2s.A2g}

$A_{2g+}$ pairing is potentially interesting since it is governed by a nontrivial one-dimensional irrep. It appears in two places in Table \ref{tab.2s.reduce}: (a) $A_{2g+}\otimes A_{1g+}$ and (b) $T_{2g+}\otimes T_{1g+}$. Hence, it is incompatible with purely local pairing, for which the first factor must be $A_{1g+}$. We discuss the two cases in turn.

(a) Each of the three $A_{1g+}$ basis matrices is combined with a $A_{2g+}$ form factor, giving
\begin{align}
D_1(\mathbf{k}) &= \delta_{00} d_{00}(\mathbf{k})\, \sigma_0 \otimes \sigma_0
  + \delta_{10} d_{10}(\mathbf{k})\, \sigma_1 \otimes \sigma_0 \nonumber \\
&\quad{}+ \delta_{30} d_{30}(\mathbf{k})\, \sigma_3 \otimes \sigma_0 .
\end{align}
The leading-order polynomial form is
\begin{align}
d_{m0}(\mathbf{k}) &= d_{m0}^{(6)}\, \big[ k_x^4 (k_y^2-k_z^2) + k_y^4 (k_z^2-k_x^2) \nonumber \\
&\quad{} + k_z^4 (k_x^2-k_y^2) \big] + \ldots ,
\end{align}
where we set $d_{m0}^{(6)}=1$ as a normalization. These are \textit{i}-wave ($l=6$) spin-singlet contributions. Based on the rule of thumb that terms of lower order in $\mathbf{k}$ are energetically favored since they have fewer nodes or nodes of lower order and thus lead to higher condensation energy, we expect this contribution to be weak compared to the following one.

(b) The reducible representation $T_{2g+}\otimes T_{1g+}$ has nine matrix-valued basis functions $d_m(\mathbf{k})\, \sigma_2 \otimes \sigma_n$, $m,n=1,2,3$, where $d_m(\mathbf{k})$ are basis functions of $T_{2g+}$. The construction parallels the one for $A_{1g+}$ pairing. The leading polynomial form factors read as
\begin{align}
d_1(\mathbf{k}) &\equiv d_{21}(\mathbf{k}) = d_2^{(2)}\, k_y k_z + \ldots , \\
d_2(\mathbf{k}) &\equiv d_{22}(\mathbf{k}) = d_2^{(2)}\, k_z k_x + \ldots , \\
d_3(\mathbf{k}) &\equiv d_{23}(\mathbf{k}) = d_2^{(2)}\, k_x k_y + \ldots
\end{align}
We choose $d_2^{(2)}=1$ as normalization. The $A_{2g+}$ part of $T_{2g+}\otimes T_{1g+}$ has the matrix-valued basis function
\begin{align}
D_{A_{2g+}}(\mathbf{k}) &= d_1(\mathbf{k})\, h_3 + d_2(\mathbf{k})\, h_4 + d_3(\mathbf{k})\, h_5
  \nonumber \\
&\cong k_y k_z\, \sigma_2 \otimes \sigma_1 + k_z k_x\, \sigma_2 \otimes \sigma_2
  + k_x k_y\, \sigma_2 \otimes \sigma_3 ,
\end{align}
which describes \textit{d}-wave ($l=2$) spin-triplet pairing, allowed due to nontrivial orbital content. Thus the second contribution to the pairing matrix is
\begin{equation}
D_2(\mathbf{k}) \cong \delta_t ( k_y k_z\, \sigma_2 \otimes \sigma_1
  + k_z k_x\, \sigma_2 \otimes \sigma_2 + k_x k_y\, \sigma_2 \otimes \sigma_3 ) .
\end{equation}
The leading order form factors $f_n(\mathbf{k})$ can now be read off. They are summarized in Table~\ref{tab.f.2s.A2g}.

\begin{table}
\caption{\label{tab.f.2s.A2g}Leading-order polynomial forms of the form factors $f_n(\mathbf{k})$ describing $A_{2g+}$ pairing for a model with two \textit{s}-orbitals.}
\begin{ruledtabular}
\begin{tabular}{@{\hspace{2em}}ll@{\hspace{2em}}}
$n$ & $f_n$ \\ \hline
$0$ & $\delta_{00}\, [ k_x^4 (k_y^2-k_z^2) + k_y^4 (k_z^2-k_x^2) + k_z^4 (k_x^2-k_y^2) ]$ \\
$1$ & $\delta_{10}\, [ k_x^4 (k_y^2-k_z^2) + k_y^4 (k_z^2-k_x^2) + k_z^4 (k_x^2-k_y^2) ]$ \\
$2$ & $\delta_{30}\, [ k_x^4 (k_y^2-k_z^2) + k_y^4 (k_z^2-k_x^2) + k_z^4 (k_x^2-k_y^2) ]$ \\
$3$ & $\delta_t\, k_y k_z$ \\
$4$ & $\delta_t\, k_z k_x$ \\
$5$ & $\delta_t\, k_x k_y$ \\
\end{tabular}
\end{ruledtabular}
\end{table}

\begin{table}
\caption{\label{tab.cf.2s.A2g}Leading-order polynomial forms of the products $c_n(\mathbf{k}) f_n(\mathbf{k})$ of form factors describing $A_{2g+}$ pairing for a model with two \textit{s}-orbitals. The amplitudes of the leading terms in $c_n(\mathbf{k})$ have been absorbed into new pairing amplitudes marked by a tilde.}
\begin{ruledtabular}
\begin{tabular}{@{\hspace{2em}}ll@{\hspace{2em}}}
$n$ & $c_n f_n$ \\ \hline
$0$ & $\tdelta_{00}\, [ k_x^4 (k_y^2-k_z^2) + k_y^4 (k_z^2-k_x^2) + k_z^4 
(k_x^2-k_y^2) ]$ \\
$1$ & $\tdelta_{10}\, [ k_x^4 (k_y^2-k_z^2) + k_y^4 (k_z^2-k_x^2) + k_z^4 
(k_x^2-k_y^2) ]$ \\
$2$ & $\tdelta_{30}\, [ k_x^4 (k_y^2-k_z^2) + k_y^4 (k_z^2-k_x^2) + k_z^4 
(k_x^2-k_y^2) ]$ \\
$3$ & $\tdelta_t\, k_y^2 k_z^2 (k_y^2-k_z^2)$ \\
$4$ & $\tdelta_t\, k_z^2 k_x^2 (k_z^2-k_x^2)$ \\
$5$ & $\tdelta_t\, k_x^2 k_y^2 (k_x^2-k_y^2)$ \\
\end{tabular}
\end{ruledtabular}
\end{table}

For time-reversal-symmetric pairing, we can choose all $f_n(\mathbf{k})$ real. For the condition for IP nodes, Eq.\ (\ref{2.cfN4.4}), we require the products $c_n(\mathbf{k}) f_n(\mathbf{k})$, which are listed in Table \ref{tab.cf.2s.A2g}. The amplitudes appearing the these products are distinguished by a tilde. We obtain, to leading order,
\begin{align}
c_0&(\mathbf{k})\, f_0(\mathbf{k}) - \vec c(\mathbf{k})\cdot\vec f(\mathbf{k}) \nonumber \\
&\cong (\tdelta_{00} - \tdelta_{10} - \tdelta_{30} - \tdelta_t)\,
  [ k_x^4 (k_y^2-k_z^2) + k_y^4 (k_z^2-k_x^2) \nonumber \\
&\quad{}+ k_z^4 (k_x^2-k_y^2) ] \nonumber \\
&= -(\tdelta_{00} - \tdelta_{10} - \tdelta_{30} - \tdelta_t)\,
  (k_x-k_y)(k_x+k_y) \nonumber \\
&\quad{}\times (k_y-k_z)(k_y+k_z)(k_z-k_x)(k_z+k_x) .
\label{3.A2g.cf.3}
\end{align}
This product clearly vanishes if two of the three components of $\mathbf{k}$ are equal in magnitude. The IP gap thus generically has six line nodes in the $\{110\}$ planes. They are of first order since $\langle c,f\rangle$ changes sign at the nodes.

The most obvious way to break TRS is to have a nontrivial phase difference between at least two of the amplitudes $\tdelta_{00}$, $\tdelta_{10}$, $\tdelta_{30}$, and $\tdelta_t$. Then IP nodes exist where both the real part and the imaginary part of $\langle c,f\rangle$ vanish. Equation (\ref{3.A2g.cf.3}) shows that the real and imaginary parts have the same symmetry-imposed line nodes so that the TRS-breaking state also has these line nodes for infinitesimal pairing.

To check whether these line nodes are inflated beyond infinitesimal pairing, we consider the Pfaffian. It is useful to keep the full momentum dependence of the normal-state form factors, not just the leading terms. The form factors $c_0(\mathbf{k})$, $c_1(\mathbf{k})$, and $c_2(\mathbf{k})$ are independent functions with $A_{1g+}$ symmetry, while the remaining three form factors can be written as
\begin{align}
c_3(\mathbf{k}) &= a_T(\mathbf{k})\, k_y k_z (k_y^2-k_z^2) , \\
c_4(\mathbf{k}) &= a_T(\mathbf{k})\, k_z k_x (k_z^2-k_x^2) , \\
c_5(\mathbf{k}) &= a_T(\mathbf{k})\, k_x k_y (k_x^2-k_y^2) ,
\end{align}
where $a_T(\mathbf{k})$ is another function with $A_{1g+}$ symmetry. Without loss of generality, we consider the plane $k_x=k_y$, which is nodal for infinitesimal pairing. In this plane, the generalized scalar products read as
\begin{align}
\langle c,c\rangle &= c_0^2(\mathbf{k}) - c_1^2(\mathbf{k}) - c_2^2(\mathbf{k}) \nonumber \\
&\quad{}- 2 a_T^2(\mathbf{k})\, k_x^2 k_z^2 (k_x^2-k_z^2)^2 , \\
\langle c,f_1\rangle &= \langle c,f_2\rangle = 0 , \\
\langle f^1,f^1\rangle &= - (\RR \delta_t)^2\, k_x^2 (2 k_z^2 + k_x^2) ,
\label{3.A2g.f1f1} \\
\langle f^2,f^2\rangle &= - (\II \delta_t)^2\, k_x^2 (2 k_z^2 + k_x^2) , \\
\langle f^1,f^2\rangle &= - \RR\delta_t \II\delta_t\, k_x^2 (2 k_z^2 + k_x^2) ,
\label{3.A2g.f1f2}
\end{align}
where $\mathbf{k}=(k_x,k_x,k_z)$. Equation (\ref{2.Pfr.5a}) then gives the Pfaffian
\begin{align}
P(\mathbf{k})
&= \big( \langle c,c\rangle - \langle f^1,f^1\rangle - \langle f^2,f^2\rangle \big)^2 \nonumber \\
&\quad{}+ 4\, \big( \langle f^1,f^2\rangle^2 - \langle f^1,f^1\rangle \langle f^2,f^2\rangle \big) .
\label{3.A2g.Pf.3}
\end{align}
The first term is a complete square and its zeros define the renormalized Fermi surface discussed above. Using Eqs.\ (\ref{3.A2g.f1f1})--(\ref{3.A2g.f1f2}), the second term obviously vanishes, which can be attributed to the fact that in the plane $k_x=k_y$ only a single pairing channel ($T_{2g+}\otimes T_{1g+}$) contributes to the pairing. The phase of the corresponding amplitude $\delta_t$ can always be chosen real so that TRS breaking does not affect the superconducting state. The upshot is that for noninfinitesimal pairing the Pfaffian still has second-order zeros in the $\{110\}$ planes and thus does not change sign. At least within these planes the line nodes are shifted but neither gapped out nor inflated.

The question arises of what happens in the vicinity of these line nodes when we go off the high-symmetry planes. The first term of the general Pfaffian given in Eq.\ (\ref{2.Pfr.5a}) has second-order zeros at the renormalized Fermi surface. We expand the second term about a point on the plane $k_x=k_y$ by setting $\mathbf{k} = (k_x+q/\sqrt{2},k_x-q/\sqrt{2},k_z)$. The leading form in $q$ reads as
\begin{align}
4\, &\big( \langle c,f^1\rangle^2 + \langle c,f^2\rangle^2
  + \langle f^1,f^2\rangle^2 - \langle f^1,f^1\rangle \langle f^2,f^2\rangle \big) \nonumber \\
&\cong 32\, k_x^2 (k_x^2-k_z^2)^4\, \Big(
  \big| c_0 \delta_{00} - c_1 \delta_{10} - c_2 \delta_{30} - a_T \delta_t \big|^2
  \nonumber \\
&\quad{} + k_x^2 (k_x^2 + 2 k_z^2)\, |\delta_t|^2\,
  \big[ |\delta_{00}|^2 \sin^2(\phi_{00} - \phi_t) \nonumber \\
&\qquad{} - |\delta_{10}|^2 \sin^2(\phi_{10} - \phi_t)
    - |\delta_{30}|^2 \sin^2(\phi_{30} - \phi_t) \big] \Big)\, q^2 ,
\label{3.A2g.Pf2.3}
\end{align}
where $\delta_{00} = |\delta_{00}| e^{i\phi_{00}}$ etc. The expression contains contributions of second and fourth order in the pairing amplitudes. At weak coupling, we can neglect the fourth-order contributions. Then, the leading correction to the Pfaffian away from the $(110)$ plane is nonnegative and generically is strictly positive for $k_x \neq k_z$. Hence, in this case, there is no BFS in the vicinity of the shifted line node, in any direction. On the other hand, for strong coupling, the coefficient in Eq.\ (\ref{3.A2g.Pf2.3}) can become negative. In this case, BFSs can exist on both sides of the $\{110\}$ planes and touching each other at these planes.

For the special case $k_x=k_z$, the whole $q^2$ term in Eq.\ (\ref{3.A2g.Pf2.3}) vanishes. Since we already had assumed $k_x=k_y$ this corresponds to the threefold rotation axis $[111]$. Here, for infinitesimal pairing three nodal lines intersect. We consider $\mathbf{k} = (k_x+q/\sqrt{2}, k_x-q/\sqrt{2},k_x)$. The leading form in the expansion of the second term of the Pfaffian here reads as
\begin{align}
12&8\, k_x^6\, \Big(
  \big| c_0 \delta_{00} - c_1 \delta_{10} - c_2 \delta_{30} - a_T \delta_d \big|^2
  \nonumber \\
&{} + 3\, k_x^4\, |\delta_t|^2\, \big[ |\delta_{00}|^2 \sin^2(\phi_{00} - 
\phi_d) \nonumber \\
&\quad{} - |\delta_{10}|^2 \sin^2(\phi_{10} - \phi_d)
  - |\delta_{30}|^2 \sin^2(\phi_{30} - \phi_d) \big] \Big)\, q^6 .
\end{align}
This term is also nonnegative at weak coupling so that there are no BFSs close to the $\langle 111\rangle$ directions.

In conclusion, the line nodes for the TRS-breaking $A_{2g+}$ pairing state are not inflated into BFSs. However, they are shifted away from the normal-state Fermi surface everywhere but remain within the high-symmetry (mirror) planes. The lack of inflation \emph{within} the high-symmetry planes can be understood on the basis that there is only a single relevant pairing amplitude, which can be chosen real.

\subsubsection{$E_{g+}$ pairing}

Pairing conforming to the two-dimensional irrep $E_{g+}$ is of interest since the breaking of TRS occurs naturally for multidimensional irreps. $E_{g+}$ pairing can emerge from the products (a) $E_{g+} \otimes A_{1g+}$, (b) $T_{1g+} \otimes T_{1g+}$, and (c) $T_{2g+} \otimes T_{1g+}$ in Table \ref{tab.2s.reduce}. Hence, it is incompatible with purely local pairing. We discuss the contributions in turn.

(a) The three $A_{1g+}$ matrices are each combined with a doublet of $E_{g+}$ form factors, giving
\begin{align}
D_1(\mathbf{k})
  &= \left[\delta^1_{00} d^1_{00}(\mathbf{k}) + \delta^2_{00} d^2_{00}(\mathbf{k})\right]
    \sigma_0 \otimes \sigma_0 \nonumber \\
&\quad{} + \left[\delta^1_{10} d^1_{10}(\mathbf{k}) + \delta^2_{10} d^2_{10}(\mathbf{k})\right]
    \sigma_1 \otimes \sigma_0 \nonumber \\
&\quad{} + \left[\delta^1_{30} d^1_{30}(\mathbf{k}) + \delta^2_{30} d^2_{30}(\mathbf{k})\right]
    \sigma_3 \otimes \sigma_0 .
\label{3.Eg.D1.3}
\end{align}
The leading polynomial terms are
\begin{align}
d^1_{m0}(\mathbf{k}) &= d^{(2)}_{m0}\, (k_x^2-k_y^2) + d^{(4)}_{m0}\, (k_x^4-k_y^4)
  + \ldots , \\
d^2_{m0}(\mathbf{k}) &= \frac{d^{(2)}_{m0}}{\sqrt{3}}\,
  (2k_z^2-k_x^2-k_y^2) + \frac{d^{(4)}_{m0}}{\sqrt{3}}\, (2k_z^4-k_x^4-k_y^4) \nonumber \\
&\quad{} + \ldots ,
\end{align}
where we can set the coefficients $d^{(2)}_{m0}$ of the leading terms to unity as a normalization. Recall that the higher-order coefficients are then generally distinct for different $m$. These contributions can be described as \textit{d}-wave ($l=2$) spin-singlet pairing.

(b) The reducible representation $T_{1g+}\otimes T_{1g+}$ has nine matrix-valued basis functions $d_m(\mathbf{k})\, \sigma_2 \otimes \sigma_n$, $m,n=1,2,3$, where $d_m(\mathbf{k})$ are momentum basis functions of $T_{1g+}$. The leading polynomial terms read as
\begin{align}
d_1(\mathbf{k}) &= d^{(4)}\, k_y k_z (k_y^2-k_z^2) + \ldots, \\
d_2(\mathbf{k}) &= d^{(4)}\, k_z k_x (k_z^2-k_x^2) + \ldots, \\
d_3(\mathbf{k}) &= d^{(4)}\, k_x k_y (k_x^2-k_y^2) + \ldots,
\end{align}
where $d^{(4)}$ can be chosen to be unity. The reduction $T_{1g+}\otimes T_{1g+} = A_{1g+} \oplus E_{g+} \oplus T_{1g+} \oplus T_{2g+}$ implies that a basis change to matrix basis functions of the four indicated irreps exists. The linear combinations of the functions $d_m(\mathbf{k})\, \sigma_2 \otimes \sigma_n$ that transform according to $E_{g+}$ are
\begin{align}
D_{x^2-y^2}(\mathbf{k}) &= d_1(\mathbf{k})\, \sigma_2 \otimes \sigma_1
  - d_2(\mathbf{k})\, \sigma_2 \otimes \sigma_2 , \\
D_{3z^2-r^2}(\mathbf{k}) &= \frac{2 d_3(\mathbf{k})}{\sqrt{3}}\, \sigma_2 \otimes \sigma_3
  - \frac{d_1(\mathbf{k})}{\sqrt{3}}\, \sigma_2 \otimes \sigma_1 \nonumber \\
&\quad{}- \frac{d_2(\mathbf{k})}{\sqrt{3}}\, \sigma_2 \otimes \sigma_2 .
\end{align}
These matrix basis functions are no longer simply the product of a scalar momentum-dependent form factor and a momentum-independent matrix. Their contribution to the pairing matrix is
\begin{equation}
D_2(\mathbf{k}) = \delta^1_{2t}\, D_{x^2-y^2}(\mathbf{k})
  + \delta^2_{2t}\, D_{3z^2-r^2}(\mathbf{k}) .
\end{equation}
This describes \textit{g}-wave ($l=4$) spin-triplet pairing, made possible by nontrivial orbital content.

(c) The analysis for $T_{2g+} \otimes T_{1g+} = A_{2g+} \oplus E_{g+} \oplus T_{1g+} \oplus T_{2g+}$ is analogous, except that now the momentum basis functions belong to $T_{2g+}$,
\begin{align}
d'_1(\mathbf{k}) &= d^{\prime(2)}\, k_y k_z + \ldots, \\
d'_2(\mathbf{k}) &= d^{\prime(2)}\, k_z k_x + \ldots, \\
d'_3(\mathbf{k}) &= d^{\prime(2)}\, k_x k_y + \ldots,
\end{align}
where we choose $d^{\prime(2)}=1$. We have the matrix basis functions
\begin{align}
D'_{x^2-y^2}(\mathbf{k}) &= \frac{2 d'_3(\mathbf{k})}{\sqrt{3}}\, \sigma_2 \otimes \sigma_3
  - \frac{d'_1(\mathbf{k})}{\sqrt{3}}\, \sigma_2 \otimes \sigma_1 \nonumber \\
&\quad{} - \frac{d'_2(\mathbf{k})}{\sqrt{3}}\, \sigma_2 \otimes \sigma_2 , \\
D'_{3z^2-r^2}(\mathbf{k}) &= -d'_1(\mathbf{k})\, \sigma_2 \otimes \sigma_1
  + d'_2(\mathbf{k})\, \sigma_2 \otimes \sigma_2 .
\end{align}
Note that the forms of the expressions for the two components of $E_g$, i.e., for the $x^2-y^2$ and the $3z^2-r^2$ matrix basis functions, are interchanged compared to case (b). To determine the correct components, their behavior under twofold rotation about the $[110]$ direction has been examined. Furthermore, to find the relative factor, which turns out to be $-1$, the behavior under threefold rotation about $[111]$ has been considered. The contribution to the pairing matrix is
\begin{equation}
D_3(\mathbf{k}) = \delta^1_{2t'}\, D'_{x^2-y^2}(\mathbf{k})
  + \delta^2_{2t'}\, D'_{3z^2-r^2}(\mathbf{k}) .
\end{equation}
This is \textit{d}-wave spin-triplet pairing, again made possible by nontrivial orbital content.

The matrix $D(\mathbf{k}) = D_1(\mathbf{k}) + D_2(\mathbf{k}) + D_3(\mathbf{k})$ is evidently of the form of Eq.\ (\ref{2.Dfh.3}). The form factors $f_n(\mathbf{k})$ are complicated functions of $\mathbf{k}$ with nodes in different places. The leading-order polynomial forms are listed in Table~\ref{tab.f.2s.Eg}.

\begin{table}
\caption{\label{tab.f.2s.Eg}Leading-order polynomial forms of the form factors $f_n(\mathbf{k})$ describing $E_{g+}$ pairing for a model with two \textit{s}-orbitals.}
\begin{ruledtabular}
\begin{tabular}{@{\hspace{1em}}ll@{\hspace{1em}}}
$n$ & $f_n$ \\ \hline
$0$ & $\delta^1_{00}\, (k_x^2 - k_y^2)
  + \frac{\delta^2_{00}}{\sqrt{3}}\, (2k_z^2 - k_x^2 - k_y^2)$ \\
$1$ & $\delta^1_{10}\, (k_x^2 - k_y^2)
  + \frac{\delta^2_{10}}{\sqrt{3}}\, (2k_z^2 - k_x^2 - k_y^2)$ \\
$2$ & $\delta^1_{30}\, (k_x^2 - k_y^2)
  + \frac{\delta^2_{30}}{\sqrt{3}}\, (2k_z^2 - k_x^2 - k_y^2)$ \\
$3$ & $\Big( \delta^1_{2t} - \frac{\delta^2_{2t}}{\sqrt{3}} \Big) \, k_y k_z (k_y^2-k_z^2)
  + \Big({-} \frac{\delta^1_{2t'}}{\sqrt{3}} - \delta^2_{2t'} \Big) \, k_y k_z$ \\
$4$ & $\Big({-} \delta^1_{2t} - \frac{\delta^2_{2t}}{\sqrt{3}} \Big) \, k_z k_x (k_z^2-k_x^2)
  + \Big({-} \frac{\delta^1_{2t'}}{\sqrt{3}} + \delta^2_{2t'} \Big) \, k_z k_x$ \\
$5$ & $\frac{2\delta^2_{2t}}{\sqrt{3}}\, k_x k_y (k_x^2-k_y^2)
  + \frac{2\delta^1_{2t'}}{\sqrt{3}}\, k_x k_y$ \\
\end{tabular}
\end{ruledtabular}
\end{table}

\begin{table}
\caption{\label{tab.cf.2s.Eg}Leading-order polynomial forms of the products $c_n(\mathbf{k}) f_n(\mathbf{k})$ of form factors describing $E_{g+}$ pairing for a model with two \textit{s}-orbitals. The amplitudes of the leading terms in $c_n(\mathbf{k})$ have been absorbed into new pairing amplitudes marked by a tilde.}
\begin{ruledtabular}
\begin{tabular}{@{\hspace{2em}}ll@{\hspace{2em}}}
$n$ & $c_n f_n$ \\ \hline
$0$ & $\tdelta^1_{00}\, (k_x^2 - k_y^2)
  + \frac{\tdelta^2_{00}}{\sqrt{3}}\, (2k_z^2 - k_x^2 - k_y^2)$ \\
$1$ & $\tdelta^1_{10}\, (k_x^2 - k_y^2)
  + \frac{\tdelta^2_{10}}{\sqrt{3}}\, (2k_z^2 - k_x^2 - k_y^2)$ \\
$2$ & $\tdelta^1_{30}\, (k_x^2 - k_y^2)
  + \frac{\tdelta^2_{30}}{\sqrt{3}}\, (2k_z^2 - k_x^2 - k_y^2)$ \\
$3$ & $\Big( \tdelta^1_{2t} - \frac{\tdelta^2_{2t}}{\sqrt{3}} \Big) \, k_y^2 k_z^2 (k_y^2-k_z^2)^2$ \\
  & $\quad{}+ \Big({-} \frac{\tdelta^1_{2t'}}{\sqrt{3}}
    - \tdelta^2_{2t'} \Big) \, k_y^2 k_z^2 (k_y^2-k_z^2)$ \\
$4$ & $\Big({-} \tdelta^1_{2t} - \frac{\tdelta^2_{2t}}{\sqrt{3}} \Big) \, 
k_z^2 k_x^2 (k_z^2-k_x^2)^2$ \\
  & $\quad{}+ \Big({-} \frac{\tdelta^1_{2t'}}{\sqrt{3}}
    + \tdelta^2_{2t'} \Big)\, k_z^2 k_x^2 (k_z^2-k_x^2)$ \\
$5$ & $\frac{2 \tdelta^2_{2t}}{\sqrt{3}}\, k_x^2 k_y^2 (k_x^2-k_y^2)^2
  + \frac{2 \tdelta^1_{2t'}}{\sqrt{3}}\, k_x^2 k_y^2 (k_x^2-k_y^2)$ \\
\end{tabular}
\end{ruledtabular}
\end{table}

In the condition for IP nodes, Eq.\ (\ref{2.cfN4.4}), each pairing form factor $f_n(\mathbf{k})$ is multiplied by the normal-state form factor $c_n(\mathbf{k})$. We list the leading-order polynomial form of these products in Table \ref{tab.cf.2s.Eg}. The contributions for $n=0,1,2$ have the same form and can be grouped together with new amplitudes $\tdelta^1_0$ and $\tdelta^2_0$. With this, we obtain, to leading order,
\begin{align}
&c_0(\mathbf{k})\, f_0(\mathbf{k}) - \vec c(\mathbf{k}) \cdot \vec f(\mathbf{k})
  \cong \tdelta^1_0\, (k_x^2 - k_y^2) \nonumber \\
&{} + \tdelta^1_{2t}\, k_z^2 (k_x^2 - k_y^2)
    (k_x^4 + k_y^4 + k_z^4 + k_x^2 k_y^2 - 2 k_x^2 k_z^2 - 2 k_y^2 k_z^2) 
\nonumber \\
&{} + \frac{\tdelta^1_{2t'}}{\sqrt{3}}\, (k_x^2 - k_y^2)
  (k_z^4 - 2 k_x^2 k_y^2 - k_x^2 k_z^2 - k_y^2 k_z^2) \nonumber \\
&{} + \frac{\tdelta^2_0}{\sqrt{3}}\, (2k_z^2 - k_x^2 - k_y^2) \nonumber \\
&{} + \frac{\tdelta^2_{2t}}{\sqrt{3}}\, \big({-}2 k_x^6 k_y^2 + 4 k_x^4 k_y^4 - 2 k_x^2 k_y^6 + k_x^6 k_z^2
  + k_y^6 k_z^2 \nonumber \\
&\quad{} - 2 k_x^4 k_z^4 - 2 k_y^4 k_z^4 + k_x^2 k_z^6 + k_y^2 k_z^6\big) 
\nonumber \\
&{} + \tdelta^2_{2t'}\, k_z^2 (k_x^4 + k_y^4 - k_x^2 k_z^2 - k_y^2 k_z^2) .
\end{align}
We first discuss time-reversal-symmetric pairing, for which all amplitudes can be chosen real. Based on the order alone, we expect that typically $\tdelta^1_0$ and $\tdelta^2_0$ dominate. Unless both amplitudes vanish these contributions give two first-order line nodes---the function changes sign at these nodes. Note that all contributions belonging to the first component of $E_{g+}$ have nodal planes at $k_y=\pm k_x$. This is because these nodes are symmetry induced and must lie in the diagonal mirror planes. On the other hand, the contributions belonging to the second component also have two nodal surfaces (two line nodes on the normal-state Fermi surface) but these are not pinned to high-symmetry planes. Hence, higher-order terms can shift them around. Their only symmetry-enforced property is to pass through the $\langle 111\rangle$ directions, where they intersect with the nodes of the first component. If both components have nonzero amplitudes there are still generically two line nodes, which have to pass through the $\langle 111\rangle$ directions~\cite{endnote.rotated.Eg}.

If TRS is broken and only $\tdelta^1_0$ and $\tdelta^2_0$ are nonzero there must be a nontrivial phase difference between these two amplitudes and we find different line nodes in the real and imaginary parts, resulting in point nodes at their crossings. Since the real and imaginary parts are zero in the $\langle 111\rangle$ directions, we obtain at least eight point nodes in these directions. All higher-order terms are zero there so that they cannot shift or gap out these point nodes.

We next turn to the possibility of BFSs. Without loss of generality, we consider the IP point node in the $[111]$ direction. For $\mathbf{k} = k\,(1,1,1)/\sqrt{3} \equiv k\, \hat{\mathbf{n}}_{111}$, the normal-state form factors $c_0(\mathbf{k}) \equiv c_{00}(\mathbf{k})$, $c_1(\mathbf{k}) \equiv c_{10}(\mathbf{k})$, and $c_2(\mathbf{k}) \equiv c_{30}(\mathbf{k})$ are independent even functions of $k$; see Eq.\ (\ref{2.cm0.3}). On the other hand, $c_3(\mathbf{k}) \equiv c_{21}(\mathbf{k})$, $c_4(\mathbf{k}) \equiv c_{22}(\mathbf{k})$, and $c_5(\mathbf{k}) \equiv c_{23}(\mathbf{k})$ vanish in this direction; see Eqs.\ (\ref{2.c21.3})--(\ref{2.c23.3}). Furthermore, Table \ref{tab.f.2s.Eg} shows that
\begin{align}
f_0(\mathbf{k}) &= f_1(\mathbf{k}) = f_2(\mathbf{k}) = 0 ,
  \label{3.Eg.111.f0} \\
f_3(\mathbf{k}) &= - \frac{\delta^1_{2t'}}{3\sqrt{3}}\, k^2 - \frac{\delta^2_{2t'}}{3}\, k^2 , \\
f_4(\mathbf{k}) &= - \frac{\delta^1_{2t'}}{3\sqrt{3}}\, k^2 + \frac{\delta^2_{2t'}}{3}\, k^2 , \\
f_5(\mathbf{k}) &= \frac{2\delta^1_{2t'}}{3\sqrt{3}}\, k^2 .
  \label{3.Eg.111.f5}
\end{align}
This implies that
\begin{align}
\langle c,c\rangle &= c_0^2(k) - c_1^2(k) - c_2^2(k) , \\
\langle c,f^1\rangle &= \langle c,f^2\rangle = 0 , \\
\langle f^1,f^1\rangle
  &= - \frac{2k^2}{9} \big[ (\RR \delta^1_{2t'})^2 + (\RR \delta^2_{2t'})^2 \big] , \\
\langle f^2,f^2\rangle
  &= - \frac{2k^2}{9} \big[ (\II \delta^1_{2t'})^2 + (\II \delta^2_{2t'})^2 \big] , \\
\langle f^1,f^2\rangle
  &= -\frac{2k^2}{9} \big[ \RR \delta^1_{2t'} \II \delta^1_{2t'}
   + \RR \delta^2_{2t'} \II \delta^2_{2t'} \big] .
\end{align}
Equation (\ref{2.Pfr.5a}) then gives
\begin{align}
P&(k\, \hat{\mathbf{n}}_{111}) \nonumber \\
&= \bigg[ c_0^2(k) - c_1^2(k) - c_2^2(k) + \frac{2}{9}\, k^2\, |\delta^1_{2t'}|^2
  + \frac{2}{9}\, k^2\, |\delta^2_{2t'}|^2 \bigg]^{\!2} \nonumber \\
&\quad{}- \frac{16}{81}\, k^4\, |\delta^1_{2t'}|^2 |\delta^2_{2t'}|^2 \sin^2(\phi_1 - \phi_2) ,
\label{3.Eg.P111.3}
\end{align}
where the two relevant pairing amplitudes are written as $\delta^{1,2}_{2t'} = |\delta^{1,2}_{2t'}|\, e^{i\phi_{1,2}}$. The first term has a second-order zero at the renormalized normal-state Fermi surface. The second term is negative whenever the phase difference between $\delta^1_{2t'}$ and $\delta^2_{2t'}$ is not an integer multiple of $\pi$. This is generically the case for broken TRS. This means that in the vicinity of the renormalized normal-state Fermi surface we find a region with $P(k\, \hat{\mathbf{n}}_{111})<0$ and thus the point node is inflated into a BFS pierced by the $[111]$ axis~\cite{endnote.Psmooth}.

If the superconducting energy scale becomes comparable to normal-state energies the BFSs are no longer spheroidal pockets close to the IP point nodes. The BFSs might then merge and could move either to the $\Gamma$ point or to the edge of the Brillouin zone and annihilate there \cite{BAM18}. We now check whether this can happen. On the unrenormalized normal-state Fermi surface in the $[111]$ direction, $c_0^2 - c_1^2 - c_2^2$ vanishes. The Pfaffian can then be written as
\begin{align}
P(k_F\, \hat{\mathbf{n}}_{111}) &= \frac{4}{81}\, k_F^4 \big( |\delta^1_{2t'}|^2
  - |\delta^2_{2t'}|^2 \big)^2 \nonumber \\
&\quad{} + \frac{16}{81}\, k_F^4\, |\delta^1_{2t'}|^2 |\delta^2_{2t'}|^2
  \cos^2(\phi_1 - \phi_2) .
\end{align}
This means that for the special TRS-breaking state with $|\delta^1_{2t'}| = |\delta^2_{2t'}|$ and phase difference $\pm \pi$ or equivalent, the Pfaffian vanishes at $k=k_F$ so that the BFS must touch the normal-state Fermi surface. In this case, the BFSs cannot annihilate for strong pairing. The special conditions of equal amplitudes and phase difference of $\pm \pi$ are quite natural from the point of view of energetics~\cite{VoG85,SiU91,BWW16,ABT17,BAM18}.

In conclusion, at infinitesimal pairing, the gap generically has point nodes in the $\langle 111\rangle$ directions if TRS is broken. These nodes are expected to be inflated into BFSs if the amplitudes $\delta^1_{2t'}$ and $\delta^2_{2t'}$ are both nonzero and have a nontrivial phase difference. All other amplitudes do not contribute to the inflation of nodes along the $\langle 111\rangle$ directions since the corresponding form factors $f_n(\mathbf{k})$ vanish there. For the energetically favored $(1,i)$ state, the BFSs stick to the normal-state Fermi surface at the former point nodes.

\subsubsection{$T_{1g+}$ pairing}
\label{subsub.2s.T1g}

The analysis for the three-dimensional irrep $T_{1g+}$ is analogous and we will be brief. All functions of momentum are represented by the lowest-order polynomials of correct symmetry. $T_{1g+}$ pairing appears in the following products in Table \ref{tab.2s.reduce}: (a) $A_{1g+}\otimes T_{1g+}$, (b) $E_{g+}\otimes T_{1g+}$, (c) $T_{1g+}\otimes A_{1g+}$, (d) $T_{1g+}\otimes T_{1g+}$, (e) $T_{2g+}\otimes T_{1g+}$. $T_{1g+}$ symmetry is possible even for purely local pairing due to the mo\-men\-tum-in\-de\-pen\-dent contribution~(a).

(a) For the contribution $A_{1g+}\otimes T_{1g+}$, we find the matrix basis functions
\begin{align}
D_{x,A_{1g+}}(\mathbf{k}) &\cong h_3 = \sigma_2 \otimes \sigma_1 , \\
D_{y,A_{1g+}}(\mathbf{k}) &\cong h_4 = \sigma_2 \otimes \sigma_2 , \\
D_{z,A_{1g+}}(\mathbf{k}) &\cong h_5 = \sigma_2 \otimes \sigma_3 .
\end{align}
These describe \textit{s}-wave spin-triplet pairing, made possible by the nontrivial orbital structure.

(b) For $E_{g+}\otimes T_{1g+}$, the basis functions read as
\begin{align}
D_{x,E_{g+}}(\mathbf{k}) &\cong \frac{1}{3}\, (2k_x^2-k_y^2-k_z^2)\, h_3 \nonumber \\
&= \frac{1}{3}\, (2k_x^2-k_y^2-k_z^2)\, \sigma_2 \otimes \sigma_1 , \\
D_{y,E_{g+}}(\mathbf{k}) &\cong \frac{1}{3}\, (2k_y^2-k_z^2-k_x^2)\, h_4 \nonumber \\
&= \frac{1}{3}\, (2k_y^2-k_z^2-k_x^2)\, \sigma_2 \otimes \sigma_2 , \\
D_{z,E_{g+}}(\mathbf{k}) &\cong \frac{1}{3}\, (2k_z^2-k_x^2-k_y^2)\, h_5 \nonumber \\
&= \frac{1}{3}\, (2k_z^2-k_x^2-k_y^2)\, \sigma_2 \otimes \sigma_3 .
\end{align}
This is \textit{d}-wave spin-triplet pairing.

(c) $T_{1g+}\otimes A_{1g+}$ involves three triplets of basis functions,
\begin{align}
D_{x,m0}(\mathbf{k}) &\cong k_yk_z (k_y^2-k_z^2)\, \sigma_m \otimes \sigma_0 , \\
D_{y,m0}(\mathbf{k}) &\cong k_zk_x (k_z^2-k_x^2)\, \sigma_m \otimes \sigma_0 , \\
D_{z,m0}(\mathbf{k}) &\cong k_xk_y (k_x^2-k_y^2)\, \sigma_m \otimes \sigma_0 ,
\end{align}
for $m=0,1,3$. This is \textit{g}-wave spin-singlet pairing.

(d) For $T_{1g+}\otimes T_{1g+}$, we get the basis functions
\begin{align}
&D_{x,T_{1g+}}(\mathbf{k}) \cong k_z k_x (k_z^2-k_x^2)\, h_5 - k_x k_y (k_x^2-k_y^2)\, h_4 \nonumber \\
&\quad= k_z k_x (k_z^2-k_x^2)\, \sigma_2 \otimes \sigma_3
  - k_x k_y (k_x^2-k_y^2)\, \sigma_2 \otimes \sigma_2 , \\
&D_{y,T_{1g+}}(\mathbf{k}) \cong k_x k_y (k_x^2-k_y^2)\, h_3 - k_y k_z (k_y^2-k_z^2)\, h_5 \nonumber \\
&\quad= k_x k_y (k_x^2-k_y^2)\, \sigma_2 \otimes \sigma_1
  - k_y k_z (k_y^2-k_z^2)\, \sigma_2 \otimes \sigma_3 , \\
&D_{z,T_{1g+}}(\mathbf{k}) \cong k_y k_z (k_y^2-k_z^2)\, h_4 - k_z k_x (k_z^2-k_x^2)\, h_3 \nonumber \\
&\quad= k_y k_z (k_y^2-k_z^2)\, \sigma_2 \otimes \sigma_2
  - k_z k_x (k_z^2-k_x^2)\, \sigma_2 \otimes \sigma_1 .
\end{align}
This is \textit{g}-wave spin-triplet pairing.

(e) For $T_{2g+}\otimes T_{1g+}$, we get the basis functions
\begin{align}
D_{x,T_{2g+}}(\mathbf{k}) &\cong k_z k_x\, h_5 + k_x k_y\, h_4 \nonumber \\
&= k_z k_x\, \sigma_2 \otimes \sigma_3 + k_x k_y\, \sigma_2 \otimes \sigma_2 , \\
D_{y,T_{2g+}}(\mathbf{k}) &\cong k_x k_y\, h_3 + k_y k_z\, h_5 \nonumber \\
&= k_x k_y\, \sigma_2 \otimes \sigma_1 + k_y k_z\, \sigma_2 \otimes \sigma_3 , \\
D_{z,T_{2g+}}(\mathbf{k}) &\cong k_y k_z\, h_4 + k_z k_x\, h_3 \nonumber \\
&= k_y k_z\, \sigma_2 \otimes \sigma_2 + k_z k_x\, \sigma_2 \otimes \sigma_1 .
\end{align}
This is \textit{d}-wave spin-triplet pairing.

\begin{table}
\caption{\label{tab.f.2s.T1g}Leading-order polynomial forms of the form factors $f_n(\mathbf{k})$ describing $T_{1g+}$ pairing for a model with two \textit{s}-orbitals.}
\begin{ruledtabular}
\begin{tabular}{@{\hspace{1em}}ll@{\hspace{1em}}}
$n$ & $f_n$ \\ \hline
$0$ & $\delta_{x,00}\, k_y k_z (k_y^2-k_z^2)
  + \delta_{y,00}\, k_z k_x (k_z^2-k_x^2)$ \\
  & $\quad{}+ \delta_{z,00}\, k_x k_y (k_x^2-k_y^2)$ \\
$1$ & $\delta_{x,10}\, k_y k_z (k_y^2-k_z^2)
  + \delta_{y,10}\, k_z k_x (k_z^2-k_x^2)$ \\
  & $\quad{}+ \delta_{z,10}\, k_x k_y (k_x^2-k_y^2)$ \\
$2$ & $\delta_{x,30}\, k_y k_z (k_y^2-k_z^2)
  + \delta_{y,30}\, k_z k_x (k_z^2-k_x^2)$ \\
  & $\quad{}+ \delta_{z,30}\, k_x k_y (k_x^2-k_y^2)$ \\
$3$ & $\delta_{x,A_{1g+}} + \frac{\delta_{x,E_{g+}}}{3}\, (2k_x^2-k_y^2-k_z^2)$ \\
  & $\quad{}+ \delta_{y,T_{1g+}}\, k_x k_y (k_x^2-k_y^2)
  - \delta_{z,T_{1g+}}\, k_z k_x (k_z^2-k_x^2)$ \\
  & $\quad{}+ \delta_{y,T_{2g+}}\, k_x k_y + \delta_{z,T_{2g+}}\, k_z k_x$ \\
$4$ & $\delta_{y,A_{1g+}} + \frac{\delta_{y,E_{g+}}}{3}\, (2k_y^2-k_z^2-k_x^2)$ \\
  & $\quad{}+ \delta_{z,T_{1g+}}\, k_y k_z (k_y^2-k_z^2)
  - \delta_{x,T_{1g+}}\, k_x k_y (k_x^2-k_y^2)$ \\
  & $\quad{}+ \delta_{z,T_{2g+}}\, k_y k_z + \delta_{x,T_{2g+}}\, k_x k_y$ \\
$5$ & $\delta_{z,A_{1g+}} + \frac{\delta_{z,E_{g+}}}{3}\, (2k_z^2-k_x^2-k_y^2)$ \\
  & $\quad{}+ \delta_{x,T_{1g+}}\, k_z k_x (k_z^2-k_x^2)
  - \delta_{y,T_{1g+}}\, k_y k_z (k_y^2-k_z^2)$ \\
  & $\quad{}+ \delta_{x,T_{2g+}}\, k_z k_x + \delta_{y,T_{2g+}}\, k_y k_z$ \\
\end{tabular}
\end{ruledtabular}
\end{table}

\begin{table}
\caption{\label{tab.cf.2s.T1g}Leading-order polynomial forms of the products $c_n(\mathbf{k}) f_n(\mathbf{k})$ of form factors describing $T_{1g+}$ pairing for a model with two \textit{s}-orbitals. The amplitudes of the 
leading terms in $c_n(\mathbf{k})$ have been absorbed into new pairing amplitudes marked by a tilde.}
\begin{ruledtabular}
\begin{tabular}{ll}
$n$ & $c_n f_n$ \\ \hline
$0$ & $\tdelta_{x,00}\, k_y k_z (k_y^2-k_z^2)
  + \tdelta_{y,00}\, k_z k_x (k_z^2-k_x^2)$ \\
  & $\quad{}+ \tdelta_{z,00}\, k_x k_y (k_x^2-k_y^2)$ \\
$1$ & $\tdelta_{x,10}\, k_y k_z (k_y^2-k_z^2)
  + \tdelta_{y,10}\, k_z k_x (k_z^2-k_x^2)$ \\
  & $\quad{}+ \tdelta_{z,10}\, k_x k_y (k_x^2-k_y^2)$ \\
$2$ & $\tdelta_{x,30}\, k_y k_z (k_y^2-k_z^2)
  + \tdelta_{y,30}\, k_z k_x (k_z^2-k_x^2)$ \\
  & $\quad{}+ \tdelta_{z,30}\, k_x k_y (k_x^2-k_y^2)$ \\
$3$ & $\tdelta_{x,A_{1g+}}\, k_y k_z (k_y^2-k_z^2)$ \\
  & $\quad{}+ \frac{\tdelta_{x,E_{g+}}}{3}\, k_y k_z (2k_x^2-k_y^2-k_z^2)(k_y^2-k_z^2)$ \\
  & $\quad{}+ \tdelta_{y,T_{1g+}}\, k_x k_y^2 k_z (k_x^2-k_y^2)(k_y^2-k_z^2)$ \\
  & $\quad{}- \tdelta_{z,T_{1g+}}\, k_x k_y k_z^2 (k_z^2-k_x^2)(k_y^2-k_z^2)$ \\
  & $\quad{}+ \tdelta_{y,T_{2g+}}\, k_x k_y^2 k_z (k_y^2-k_z^2)
  + \tdelta_{z,T_{2g+}}\, k_x k_y k_z^2 (k_y^2-k_z^2)$ \\
$4$ & $\tdelta_{y,A_{1g+}}\, k_z k_x (k_z^2-k_x^2)$ \\
  & $\quad{}+ \frac{\tdelta_{y,E_{g+}}}{3}\, k_z k_x (2k_y^2-k_z^2-k_x^2)(k_z^2-k_x^2)$ \\
  & $\quad{}+ \tdelta_{z,T_{1g+}} k_x k_y k_z^2 (k_y^2-k_z^2)(k_z^2-k_x^2)$ \\
  & $\quad{}- \tdelta_{x,T_{1g+}} k_x^2 k_y k_z (k_x^2-k_y^2)(k_z^2-k_x^2)$ \\
  & $\quad{}+ \tdelta_{z,T_{2g+}} k_x k_y k_z^2 (k_z^2-k_x^2)
  + \tdelta_{x,T_{2g+}}\, k_x^2 k_y k_z (k_z^2-k_x^2)$ \\
$5$ & $\tdelta_{z,A_{1g+}}\, k_x k_y (k_x^2-k_y^2)$ \\
  & $\quad{}+ \frac{\tdelta_{z,E_{g+}}}{3}\, k_x k_y (2k_z^2-k_x^2-k_y^2)(k_x^2-k_y^2)$ \\
  & $\quad{}+ \tdelta_{x,T_{1g+}}\, k_x^2 k_y k_z (k_z^2-k_x^2)(k_x^2-k_y^2)$ \\
  & $\quad{}- \tdelta_{y,T_{1g+}}\, k_x k_y^2 k_z (k_y^2-k_z^2)(k_x^2-k_y^2)$ \\
  & $\quad{}+ \tdelta_{x,T_{2g+}}\, k_x^2 k_y k_z (k_x^2-k_y^2)
  + \tdelta_{y,T_{2g+}}\, k_x k_y^2 k_z (k_x^2-k_y^2)$ \\
\end{tabular}
\end{ruledtabular}
\end{table}

The resulting form factors are summarized in Table \ref{tab.f.2s.T1g}. To determine the IP nodes, we require the products $c_n(\mathbf{k})\, f_n(\mathbf{k})$, which are listed in Table \ref{tab.cf.2s.T1g}. Defining $\tdelta_{\nu,0} \equiv \tdelta_{\nu,00} - \tdelta_{\nu,10} - \tdelta_{\nu,30} - \tdelta_{\nu,A_{1g+}}$ for $\nu=x,y,z$, we obtain
\begin{align}
c_0&(\mathbf{k})\, f_0(\mathbf{k}) - \vec c(\mathbf{k}) \cdot \vec f(\mathbf{k}) \nonumber \\
&= \bigg[ \tdelta_{x,0} - \frac{\tdelta_{x,E_{g+}}}{3}\, (2k_x^2-k_y^2-k_z^2)
  + \tdelta_{x,T_{2g+}}\, k_x^2 \bigg] \nonumber \\
&\quad{}\times k_y k_z (k_y^2-k_z^2) + \ldots ,
\label{3.T1g.cf.3}
\end{align}
where two terms with cyclically permuted indices $x$, $y$, and $z$ have been suppressed.  Note that contribution (d) has dropped out. This is an artifact of having used the same leading-order basis functions for $c_n(\mathbf{k})$ and $f_n(\mathbf{k})$. Using different ones, we see that the terms do not cancel. They do not change the following discussion, though.

If only the $\tdelta_x$ amplitudes are different from zero, i.e., for pairing of $(1,0,0)$ type \cite{SiU91,BWW16,BAM18}, we expect four first-order line nodes in the planes $k_y=0$, $k_z=0$, $k_y=k_z$, and $k_y=-k_z$. This is a new example of a state that is necessarily nodal even for purely local pairing, in which case $\tdelta_{x,0}$ is the only nonvanishing amplitude. Time-reversal-symmetric superpositions of $(1,0,0)$, $(0,1,0)$, and $(0,0,1)$ pairing generically also have four line nodes.

If TRS is broken, the states with $(1,i,0)$ and $(1,\omega,\omega^2)$ where $\omega = e^{2\pi i/3}$ are plausible \cite{VoG85,SiU91,BWW16}. The $(1,i,0)$ state has IP nodes where both the real and the imaginary part of $c_0 (\mathbf{k})\, f_0(\mathbf{k}) - \vec c(\mathbf{k}) \cdot \vec f(\mathbf{k})$ vanish. This leads to 18 point nodes in the $\langle 001\rangle$, $\langle 101\rangle$, $\langle 111\rangle$ directions outside of the 
$k_z=0$ plane, and one line node in the $k_z=0$ plane. Compare the $T_{2g+}$, $(1,i,0)$ pairing state for the $j=3/2$ example \cite{ABT17,BAM18}, where we found two point nodes and one line node in the $k_z=0$ plane.

\begin{figure}
\includegraphics[width=0.8\columnwidth]{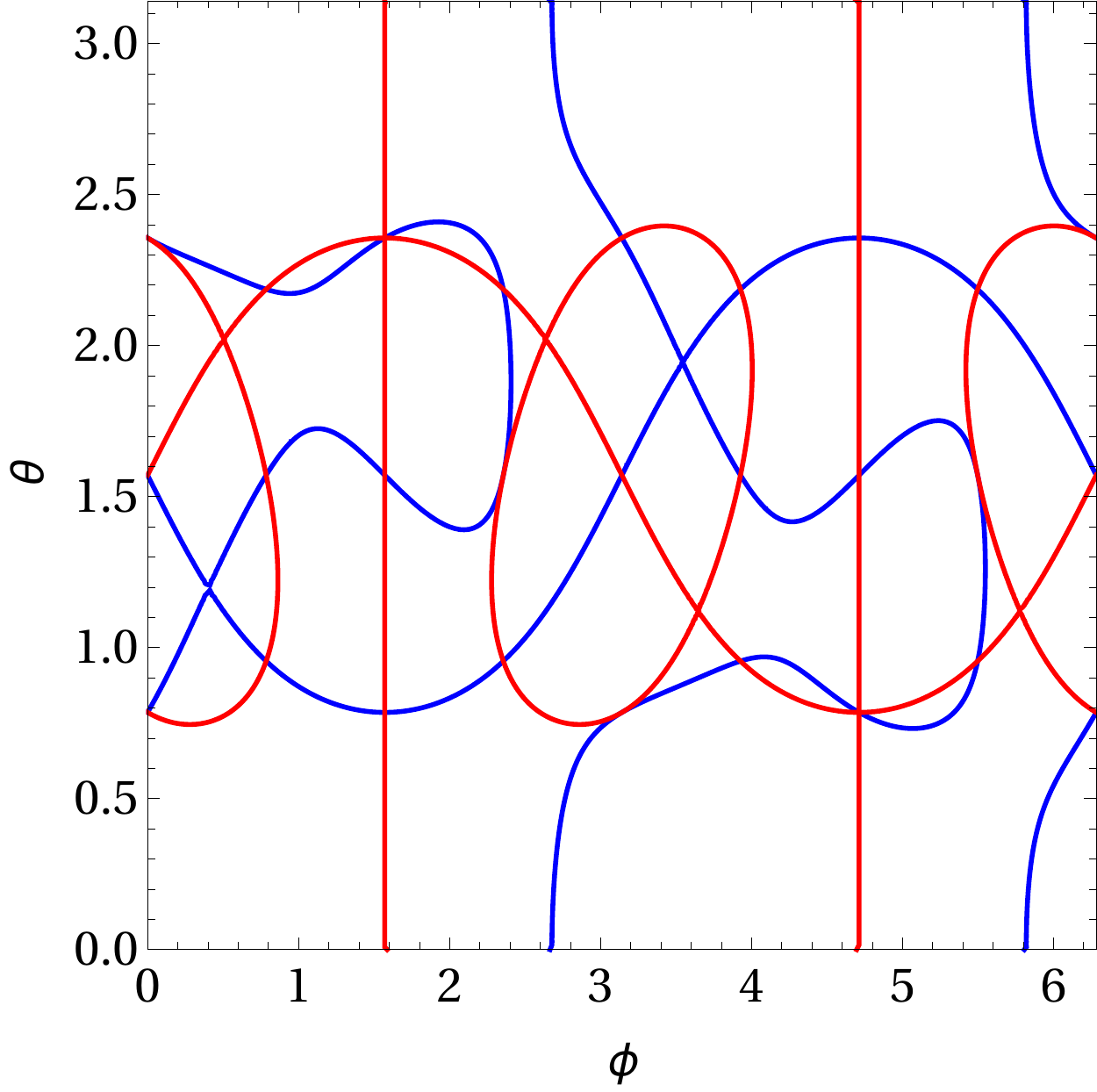}
\caption{\label{fig.1www.nodes}Zeros of real (blue) and imaginary (red) parts of $\langle c,f\rangle$ for the $T_{1g+}$ pairing state with order parameter $1,\omega,\omega^2$, with $\omega=e^{2\pi i/3}$, as functions of the spherical polar angles $\theta$, $\phi$ of $\mathbf{k}$. Lowest-order polynomial basis functions and the parameter values $|\mathbf{k}|=1$, $\tdelta_{i,E_{g+}}/\tdelta_{i,0}=0.5$, and $\tdelta_{i,T_{2g+}}/\tdelta_{i,0}=0.3$ for $i=x,y,z$ have been used.}
\end{figure}

For the $(1,\omega,\omega^2)$ state, there are point nodes where both the real and the imaginary part vanish. Figure \ref{fig.1www.nodes} illustrates the zeros of the real and imaginary parts for a typical parameter set. For generic parameters, there are point nodes in the 26 cubic high-symmetry directions. Six of these are special in that either the zero contours of the real and imaginary part are cotangent or in that the imaginary zero contour has a self crossing. In these cases, the quasiparticle dispersion close to the point node is linear in all directions except along a single axis, where it is quadratic to leading order. The other 20 point nodes show linear dispersion.

For noninfinitesimal pairing, we expect the nodes to be inflated. We here only consider the $(1,i,0)$ state. We write the pairing amplitudes as $\delta_{x,00} = \delta_{00}$, $\delta_{y,00} = i\delta_{00}$, and $\delta_{z,00} = 0$, etc. The superconducting form factors then read as
\begin{align}
f_0(\mathbf{k}) &= \delta_{00}\, \big[ k_y k_z (k_y^2-k_z^2)
  + i\, k_z k_x (k_z^2-k_x^2) \big] , \\
f_1(\mathbf{k}) &= \delta_{10}\, \big[ k_y k_z (k_y^2-k_z^2)
  + i\, k_z k_x (k_z^2-k_x^2) \big] , \\
f_2(\mathbf{k}) &= \delta_{30}\, \big[ k_y k_z (k_y^2-k_z^2)
  + i\, k_z k_x (k_z^2-k_x^2) \big] , \\
f_3(\mathbf{k}) &= \delta_{A_{1g+}} + \frac{\delta_{E_{g+}}}{3}\, (2k_x^2-k_y^2-k_z^2) \nonumber \\
&\quad{}+ i\, \delta_{T_{1g+}}\, k_x k_y (k_x^2-k_y^2)
  + i\, \delta_{T_{2g+}}\, k_x k_y , \\
f_4(\mathbf{k}) &= i\delta_{A_{1g+}} + \frac{i\,\delta_{E_{g+}}}{3}\, (2k_y^2-k_z^2-k_x^2) \nonumber \\
&\quad{}- \delta_{T_{1g+}}\, k_x k_y (k_x^2-k_y^2) + \delta_{T_{2g+}}\, k_x k_y , \\
f_5(\mathbf{k}) &= \delta_{T_{1g+}}\, \big[ k_z k_x (k_z^2-k_x^2) - i\, 
k_y k_z (k_y^2-k_z^2) \big]
  \nonumber \\
&\quad{}+ \delta_{T_{2g+}}\, \big[ k_z k_x + i\, k_y k_z \big] .
\end{align}
Equation (\ref{3.T1g.cf.3}) shows that $\langle c,f^1\rangle = \langle c,f^2\rangle = 0$ remains valid in the radial direction through all point nodes. To go on, we have to distinguish between the inequivalent point nodes. In the $[001]$ direction, we have
\begin{align}
f_0(\mathbf{k}) &= f_1(\mathbf{k}) = f_2(\mathbf{k}) = f_5(\mathbf{k}) = 0 , \\
f_3(\mathbf{k}) &= \delta_{A_{1g+}} - \frac{\delta_{E_{g+}}}{3}\, k^2 , \\
f_4(\mathbf{k}) &= i\delta_{A_{1g+}} - \frac{i\delta_{E_{g+}}}{3}\, k^2 .
\end{align}
We find $\langle f^1,f^2\rangle = 0$ and $\langle f^1,f^1\rangle = \langle f^2,f^2\rangle = -(\delta_{A_{1g+}}-\delta_{E_{g+}}k^2/3)^2$ and thus for the Pfaffian 
\begin{equation}
P(\mathbf{k}) = \langle c,c\rangle \left[ \langle c,c\rangle
    + 4 \left( \delta_{A_{1g+}} - \frac{\delta_{E_{g+}}}{3}\, k^2 \right)^{\!2} \right] .
\end{equation}
The first factor changes sign at the normal-state Fermi surface. The second is generically nonzero there and thus the Pfaffian also changes sign at the normal-state Fermi surface. Hence, the point nodes in the $\langle 001\rangle$ directions are inflated and touch the normal-state Fermi surface at the IP point nodes.

In the $[101]$ direction, we have $\mathbf{k}=k\,(1,0,1)/\sqrt{2}$ and
\begin{align}
f_0(\mathbf{k}) &= f_1(\mathbf{k}) = f_2(\mathbf{k}) = 0 , \\
f_3(\mathbf{k}) &= \delta_{A_{1g+}} + \frac{\delta_{E_{g+}}}{6}\, k^2 , \\
f_4(\mathbf{k}) &= i\delta_{A_{1g+}} - \frac{i\delta_{E_{g+}}}{3}\, k^2 , \\
f_5(\mathbf{k}) &= \frac{\delta_{T_{2g+}}}{2}\, k^2 .
\end{align}
This implies that $\langle f^1,f^2\rangle = 0$ but $\langle f^1,f^1\rangle$ and $\langle f^2,f^2\rangle$ are generally unequal. The Pfaffian thus reads as
\begin{equation}
P(\mathbf{k})
  = \left[ \langle c,c\rangle - \langle f^1,f^1\rangle - \langle f^2,f^2\rangle \right]^2
  - 4\, \langle f^1,f^1\rangle \langle f^2,f^2\rangle .
\end{equation}
The first term goes to zero on a renormalized Fermi surface. The second term
\begin{align}
- 4\, &\langle f^1,f^1\rangle \langle f^2,f^2\rangle
  = -4 \left( \delta_{A_{1g+}} - \frac{\delta_{E_{g+}}}{3}\, k^2 \right)^{\!2} \nonumber \\
&{}\times \left[ \left( \delta_{A_{1g+}} + \frac{\delta_{E_{g+}}}{6}\, k^2 \right)^{\!2}
    + \frac{\delta_{T_{2g+}}^2}{4}\, k^4 \right]
\end{align}
generically is strictly negative. Then, the Pfaffian becomes negative and we find a BFS for a range of $k$ values in the vicinity of but not usually touching the normal-state Fermi surface. The Pfaffian is identical for the directions $[\bar{1}01]$, $[011]$, $[0\bar{1}1]$, and their negatives.

Finally, along $[111]$, we have $\mathbf{k}=k\,(1,1,1)/\sqrt{3}$ and
\begin{align}
f_0(\mathbf{k}) &= f_1(\mathbf{k}) = f_2(\mathbf{k}) = 0 , \\
f_3(\mathbf{k}) &= \delta_{A_{1g+}} + \frac{i\delta_{T_{2g+}}}{3}\, k^2 , \\
f_4(\mathbf{k}) &= i\delta_{A_{1g+}} + \frac{\delta_{T_{2g+}}}{3}\, k^2 , \\
f_5(\mathbf{k}) &= \frac{1+i}{3}\, \delta_{T_{2g+}}\, k^2 .
\end{align}
We thus obtain
\begin{equation}
\langle f^1,f^2\rangle
  = -\frac{1}{3}\, \delta_{T_{2g+}} k^2 \left( 2\, \delta_{A_{1g+}}
    + \frac{1}{3}\, \delta_{T_{2g+}} k^2 \right)
\end{equation}
and
\begin{equation}
\langle f^1,f^1\rangle = \langle f^2,f^2\rangle
  = - \left(\delta_{A_{1g+}}^2 + \frac{2}{9}\, \delta_{T_{2g+}}^2 k^4 \right) .
\end{equation}
The Pfaffian is
\begin{align}
P(\mathbf{k}) &= \left[ \langle c,c\rangle - \langle f^1,f^1\rangle - \langle f^2,f^2\rangle \right]^2 \nonumber \\
&\quad{} + 4 \left[ \langle f^1,f^2\rangle^2 - \langle f^1,f^1\rangle \langle f^2,f^2\rangle \right] ,
\label{3.T1g.line.inf.3}
\end{align}
wherein the second term evaluates to
\begin{align}
4 &\left[ \langle f^1,f^2\rangle^2 - \langle f^1,f^1\rangle \langle f^2,f^2\rangle \right] \nonumber \\
&= -\frac{4}{27} \left( 3\, \delta_{A_{1g+}} - \delta_{T_{2g+}}\, k^2 \right)^2 \nonumber \\
&\quad{}\times \left[ 2\, \delta_{A_{1g+}}^2 + ( \delta_{A_{1g+}} + \delta_{T_{2g+}}\, k^2 )^2 \right] .
\end{align}
Since this is generally negative we also expect BFSs that do not touch the normal-state Fermi surface in the $\langle 111\rangle$ directions.

For the equatorial line node, we take $\mathbf{k}=(k_x,k_y,0)$. The superconducting form factors are
\begin{align}
f_0(\mathbf{k}) &= f_1(\mathbf{k}) = f_2(\mathbf{k}) = f_5(\mathbf{k}) = 0 , \\
f_3(\mathbf{k}) &= \delta_{A_{1g+}} + \frac{\delta_{E_{g+}}}{3}\, (2k_x^2-k_y^2) \nonumber \\
&\quad{}+ i\, \delta_{T_{1g+}}\, k_x k_y (k_x^2-k_y^2)
  + i\, \delta_{T_{2g+}}\, k_x k_y , \\
f_4(\mathbf{k}) &= i\delta_{A_{1g+}} + \frac{i\,\delta_{E_{g+}}}{3}\, (2k_y^2-k_x^2) \nonumber \\
&\quad{}- \delta_{T_{1g+}}\, k_x k_y (k_x^2-k_y^2) + \delta_{T_{2g+}}\, k_x k_y .
\end{align}
This gives
\begin{align}
\langle f^1,&f^2\rangle
  = -\frac{1}{3}\, k_x k_y \big[ 6\, \delta_{A_{1g+}} \delta_{T_{2g+}} \nonumber \\
&{}+ 3\, \delta_{E_{g+}} \delta_{T_{1g+}} (k_x^2-k_y^2)^2
  + \delta_{E_{g+}} \delta_{T_{2g+}} (k_x^2+k_y^2) \big] , \\
\langle f^1,&f^1\rangle = -\frac{1}{9} \left[ 3 \delta_{A_{1g+}}
    + \delta_{E_{g+}} (2k_x^2 - k_y^2) \right]^2 \nonumber \\
&{}- \left[ \delta_{T_{1g+}} (k_x^2-k_y^2) - \delta_{T_{2g+}} \right]^2 k_x^2 k_y^2 , \\
\langle f^2,&f^2\rangle = -\frac{1}{9} \left[ 3 \delta_{A_{1g+}}
    + \delta_{E_{g+}} (2k_y^2 - k_x^2) \right]^2 \nonumber \\
&{}- \left[ \delta_{T_{1g+}} (k_x^2-k_y^2) + \delta_{T_{2g+}} \right]^2 k_x^2 k_y^2 .
\end{align}
The Pfaffian again has the form of Eq.\ (\ref{3.T1g.line.inf.3}), where the second term
\begin{align}
4 &\left[ \langle f^1,f^2\rangle^2 - \langle f^1,f^1\rangle \langle f^2,f^2\rangle \right] \nonumber \\
&= -\frac{4}{81}\, \Big[ 9\, \delta_{A_{1g+}}^2
  + 2\, \delta_{A_{1g+}} \delta_{E_{g+}} (k_x^2 + k_y^2) \nonumber \\
&\quad{}+ \delta_{E_{g+}}^2 (2k_x^2-k_y^2)(2k_y^2-k_x^2) \nonumber \\
&\quad{}+ 9\, \delta_{T_{1g+}}^2\, k_x^2 k_y^2 (k_x^2 - k_y^2)^2
  - 9\, \delta_{T_{2g+}}^2\, k_x^2 k_y^2 \Big]^2
\end{align}
is non-positive and generically negative so that the line node is inflated by noninfinitesimal pairing. The resulting BFS is toroidal but may be pinched off, i.e., have self crossings, at some momenta, but these self crossings would be accidental. Since the first term in Eq.\ (\ref{3.T1g.line.inf.3}) becomes zero close to but not at the normal-state Fermi surface the BFS generically does not touch the normal-state Fermi surface.

In summary, the point and line nodes of the $T_{1g+}$, $(1,i,0)$ pairing state are all inflated by noninfinitesimal pairing. Only the inflated point nodes on the $k_z$-axis touch the normal-state Fermi surface, the other pockets are shifted away from it and could thus be annihilated at strong coupling.

\subsection{Two orbitals of opposite parity}
\label{sub.opposite}

To have a single even and a single odd orbital per unit cell for the point group $O_h$, the first must transform like a one-dimensional $g$ irrep and the second like a one-dimensional $u$ irrep. The most natural possibilities are $A_{1g}$ (\textit{s}-orbital) and $A_{2u}$ ($f_{xyz}$-orbital). Now the inversion or parity matrix is nontrivial:
\begin{equation}
U_P = \sigma_3 \otimes \sigma_0 .
\end{equation}
Moreover, the $f_{xyz}$-orbital is odd under fourfold rotations but even under threefold rotations. Models of this symmetry have been analyzed in the context of Dirac and Weyl semimetals~\cite{AMV18}.

The unitary part of time reversal is
\begin{equation}
U_T = \sigma_0 \otimes i\sigma_2
\end{equation}
since the orbitals are invariant. The basis matrices can be written as Kronecker products, which transform according to the irreps as summarized in Table \ref{tab.sf.basis}. Compared to the example of two \textit{s}-orbitals, Table \ref{tab.2s.basis}, the Pauli matrices $\sigma_1$ and $\sigma_2$ for the orbital degree of freedom are now odd under inversion. The new element is that $u$ irreps occur, due to the nontrivial parity operator. This provides additional possibilities for the products of irreps of $\mathbf{k}$-dependent form factors and basis matrices. Table \ref{tab.sf.reduce} shows all relevant reductions of product representations.

\begin{table}
\caption{\label{tab.sf.basis}Basis matrices on the internal Hilbert space for the case of one \textit{s}-orbital and one $f_{xyz}$-orbital and point group $O_h$. The basis matrices are irreducible tensor operators of the irreps listed in the second column. For multidimensional irreps, the states transforming into each other under point-group operations are distinguished by the index in the third column.}
\begin{ruledtabular}
\begin{tabular}{@{\hspace{3em}}ccc@{\hspace{3em}}}
$h_\nu$ & Irrep & Component \\ \hline
$\sigma_0 \otimes \sigma_0$ & $A_{1g+}$ & \\
$\sigma_0 \otimes \sigma_1$ & $T_{1g-}$ & 1 \\
$\sigma_0 \otimes \sigma_2$ & & 2 \\
$\sigma_0 \otimes \sigma_3$ & & 3 \\
$\sigma_1 \otimes \sigma_0$ & $A_{2u+}$ & \\
$\sigma_1 \otimes \sigma_1$ & $T_{2u-}$ & 1 \\
$\sigma_1 \otimes \sigma_2$ & & 2 \\
$\sigma_1 \otimes \sigma_3$ & & 3 \\
$\sigma_2 \otimes \sigma_0$ & $A_{2u-}$ & \\
$\sigma_2 \otimes \sigma_1$ & $T_{2u+}$ & 1 \\
$\sigma_2 \otimes \sigma_2$ & & 2 \\
$\sigma_2 \otimes \sigma_3$ & & 3 \\
$\sigma_3 \otimes \sigma_0$ & $A_{1g+}$ & \\
$\sigma_3 \otimes \sigma_1$ & $T_{1g-}$ & 1 \\
$\sigma_3 \otimes \sigma_2$ & & 2 \\
$\sigma_3 \otimes \sigma_3$ & & 3 \\
\end{tabular}
\end{ruledtabular}
\end{table}

\begin{table*}
\caption{\label{tab.sf.reduce}Reduction of product representations of the allowed irreps of $\mathbf{k}$-dependent form factors (rows) and basis matrices $h_\nu$ (columns) for one \textit{s}-orbital and one $f_{xyz}$-orbital. For the form factors, the minimum order of polynomial basis functions is given in the second column. ``\bad'' indicates products that are forbidden since they violate fermionic antisymmetry.}
\begin{ruledtabular}
\begin{tabular}{lc@{\hspace{0em}}llllll}
\multicolumn{2}{c}{Form factor:\hspace*{0em}} & \multicolumn{6}{c}{Pairing matrix: Irrep} \\
Irrep & Min.\ $l$ & $A_{1g+}$ & $T_{1g-}$ & $A_{2u+}$ & $T_{2u+}$ & $A_{2u-}$ & $T_{2u-}$ \\ \hline
$A_{1g+}$ & 0 & {\boldmath$A_{1g+}$} & \bad & $A_{2u+}$ & $T_{2u+}$ & \bad & \bad \\
$A_{2g+}$ & 6 & $A_{2g+}$ & \bad & $A_{1u+}$ & $T_{1u+}$ & \bad & \bad \\
$E_{g+}$ & 2 & $E_{g+}$ & \bad & $E_{u+}$ & $T_{1u+} \oplus T_{2u+}$ & \bad & \bad \\
$T_{1g+}$ & 4 & $T_{1g+}$ & \bad & $T_{2u+}$ & $A_{2u+} \oplus E_{u+} \oplus T_{1u+} \oplus T_{2u+}$ & \bad & \bad \\
$T_{2g+}$ & 2 & $T_{2g+}$ & \bad & $T_{1u+}$ & $A_{1u+} \oplus E_{u+} \oplus T_{1u+} \oplus T_{2u+}$ & \bad & \bad \\
$A_{1u-}$ & 9 & \bad & $T_{1u+}$ & \bad & \bad & $A_{2g+}$ & $T_{2g+}$ \\
$A_{2u-}$ & 3 & \bad & $T_{2u+}$ & \bad & \bad & {\boldmath$A_{1g+}$} & $T_{1g+}$ \\
$E_{u-}$ & 5 & \bad & $T_{1u+} \oplus T_{2u+}$ & \bad & \bad & $E_{g+}$ & 
$T_{1g+} \oplus T_{2g+}$ \\
$T_{1u-}$ & 1 & \bad & $A_{1u+} \oplus E_{u+} \oplus T_{1u+} \oplus T_{2u+}$ & \bad & \bad & $T_{2g+}$ & $A_{2g+} \oplus E_{g+} \oplus T_{1g+} \oplus T_{2g+}$ \\
$T_{2u-}$ & 3 & \bad & $A_{2u+} \oplus E_{u+} \oplus T_{1u+} \oplus T_{2u+}$ & \bad & \bad & $T_{1g+}$ & $\mbox{\boldmath$A_{1g+}$} \oplus E_{g+} \oplus T_{1g+} \oplus T_{2g+}$ \\
\end{tabular}
\end{ruledtabular}
\end{table*}

The normal-state Hamiltonian is a linear combination of all basis matrices that allow to form products with full $A_{1g+}$ symmetry, marked in bold face in Table \ref{tab.sf.reduce}. These basis matrices, together with their irreps, are
\begin{align}
h_0 &\equiv \sigma_0 \otimes \sigma_0 & A_{1g+}, \\
h_1 &\equiv \sigma_3 \otimes \sigma_0 & A_{1g+}, \\
h_2 &\equiv \sigma_2 \otimes \sigma_0 & A_{2u-}, \\
h_3 &\equiv \sigma_1 \otimes \sigma_1 & T_{2u-}, \\
h_4 &\equiv \sigma_1 \otimes \sigma_2 & T_{2u-}, \\
h_5 &\equiv \sigma_1 \otimes \sigma_3 & T_{2u-} .
\end{align}
There are thus six matrices that satisfy the same algebra as for the case of two \textit{s}-orbitals; see Appendices \ref{app.normal} and \ref{app.algebra}. The normal-state Hamiltonian reads as
\begin{align}
&H_N(\mathbf{k}) \nonumber \\
&\:= c_{00}(\mathbf{k})\, \sigma_0 \otimes \sigma_0
  + c_{30}(\mathbf{k})\, \sigma_3 \otimes \sigma_0
  + c_{20}(\mathbf{k})\, \sigma_2 \otimes \sigma_0 \nonumber \\
&\:\quad{} + c_{11}(\mathbf{k})\, \sigma_1 \otimes \sigma_1
  + c_{12}(\mathbf{k})\, \sigma_1 \otimes \sigma_2
  + c_{13}(\mathbf{k})\, \sigma_1 \otimes \sigma_3 ,
\end{align}
where the leading polynomial forms are
\begin{align}
c_{00}(\mathbf{k}) &= c_{00}^{(0)} + c_{00}^{(2)}\, (k_x^2 + k_y^2 + k_z^2) + \ldots , \\
c_{30}(\mathbf{k}) &= c_{30}^{(0)} + c_{30}^{(2)}\, (k_x^2 + k_y^2 + k_z^2) + \ldots , \\
c_{20}(\mathbf{k}) &= c_{20}^{(3)}\, k_x k_y k_z + \ldots , \\
c_{11}(\mathbf{k}) &= c_1^{(3)}\, k_x (k_y^2-k_z^2) + \ldots , \\
c_{12}(\mathbf{k}) &= c_1^{(3)}\, k_y (k_z^2-k_x^2) + \ldots , \\
c_{13}(\mathbf{k}) &= c_1^{(3)}\, k_z (k_x^2-k_y^2) + \ldots
\end{align}
Four of the form factors are odd in momentum. They do not break inversion symmetry since they multiply orbital matrices that are also odd under inversion.

Turning to superconducting pairing, it is interesting that local pairing is now either trivial ($A_{1g+}$) or has odd parity ($A_{2u+}$, $T_{2u+}$), as seen from the first row of Table \ref{tab.sf.reduce}. Furthermore, for even-parity pairing ($g+$ irreps), only the basis matrices belonging to $A_{1g+}$, $T_{2u-}$, and $A_{2u-}$ can occur, i.e., the same matrices $h_0$, \dots, $h_5$ as in $H_N(\mathbf{k})$. Since $\mathcal{C}P$ squares to $+\openone$ the Hamiltonian can be unitarily transformed into antisymmetric form, guaranteeing the existence of a Pfaffian \cite{ABT17}. In the present example, where $U_P = \sigma_3 \otimes \sigma_0$, the matrix $\Omega$ mediating this transformation reads as
\begin{equation}
\Omega = \frac{1}{\sqrt{2}} \left(\begin{array}{cc} 1 & 1 \\ i & -i \end{array}\right)
  \otimes \exp\!\left(-i\,\frac{\pi}{2}\, \frac{\sigma_3}{2}\right) \otimes \sigma_0 .
\end{equation}
This specific form does not affect the eigenvalues, though. Since the algebra of the basis matrices is unchanged, the expressions for the eigenvalues, the Pfaffian, and the condition for IP nodes remain unchanged. In the following, we briefly discuss the $A_{2g+}$ and $E_{g+}$ pairing states and compare them to the case of two \textit{s}-orbitals.

\subsubsection{$A_{2g+}$ pairing}

$A_{2g+}$ appears in three places in Table \ref{tab.sf.reduce}: (a) $A_{2g+}\otimes A_{1g+}$, (b) $A_{1u-}\otimes A_{2u-}$, and (c) $T_{1u-}\otimes T_{2u-}$. Note that the minimum orders of form factors are (a) $6$, (b) $9$, and (c) $1$ so that one expects that the $T_{1u-}\otimes T_{2u-}$ contribution typically dominates.

(a) For $A_{2g+}\otimes A_{1g+}$:
\begin{equation}
D_{A_{2g+}}(\mathbf{k})= \delta_{00} d_{00}(\mathbf{k})\, \sigma_{0} \otimes \sigma_{0}
  + \delta_{30} d_{30}(\mathbf{k})\, \sigma_{3} \otimes \sigma_{0} .
\end{equation}
To the leading order, $d_{m0}(\mathbf{k})$ takes the form
\begin{equation}
d_{m0}(\mathbf{k}) \cong d_{m0}^{(6)}\, \big[ k_{x}^4 (k_{y}^2-k_{z}^2)
  + k_{y}^4 (k_{z}^2-k_{x}^2) + k_{z}^4 (k_{x}^2-k_{y}^2) \big] .
\end{equation}
$d_{m0}^{(6)}$ is set to unity.

(b) For $A_{1u-}\otimes A_{2u-}$:
\begin{equation}
D_{A_{1u-}}(\mathbf{k}) = \delta_{A_{1u-}} d_{A_{1u-}}(\mathbf{k})\, \sigma_{2} \otimes \sigma_{0} ,
\end{equation}
with $d_{A_{1u-}}(\mathbf{k})$ to the leading order given by
\begin{align}
d_{A_{1u-}}(\mathbf{k}) &\cong d_{A_{1u-}}^{(9)}\, k_{x} k_{y} k_{z}\, \big[
  k_{x}^4 (k_{y}^2-k_{z}^2) + k_{y}^4 (k_{z}^2-k_{x}^2) \nonumber \\
&\quad{}+ k_{z}^4 (k_{x}^2-k_{y}^2) \big] .
\end{align}
We set $d_{A_{1u-}}^{(9)}$ to unity.

(c) For $T_{1u-}\otimes T_{2u-}$:
\begin{equation}
D_{T_{1u-}}(\mathbf{k}) \cong \delta_{T_{1u-}} ( k_{x}\, \sigma_{1}\otimes \sigma_{1}
  + k_{y}\, \sigma_{1}\otimes \sigma_{2} + k_{z}\, \sigma_{1}\otimes \sigma_{3} )
\end{equation}
to leading order. The components are assigned such that the whole term $D_{T_{1u-}}(\mathbf{k})$ changes sign under any four-fold rotation~\cite{endnote.opp.C4z}.

\begin{table}
\caption{\label{tab.f.sf.A2g}Leading-order polynomial forms of the form factors $f_n(\mathbf{k})$ describing $A_{2g+}$ pairing for a model with one \textit{s}-orbital and one $f_{xyz}$-orbital.}
\begin{ruledtabular}
\begin{tabular}{ll}
$n$ & $f_n$ \\ \hline
$0$ & $\delta_{00}\, [k_{x}^4(k_{y}^2-k_{z}^2)+k_{y}^4(k_{z}^2-k_{x}^2)+k_{z}^4(k_{x}^2-k_{y}^2)]$ \\
$1$ & $\delta_{30}\, [k_{x}^4(k_{y}^2-k_{z}^2)+k_{y}^4(k_{z}^2-k_{x}^2)+k_{z}^4(k_{x}^2-k_{y}^2)]$ \\
$2$ & $\delta_{A_{1u-}}\, k_xk_yk_z\,[k_{x}^4(k_{y}^2-k_{z}^2)+k_{y}^4(k_{z}^2-k_{x}^2)+k_{z}^4(k_{x}^2-k_{y}^2)] $ \\
$3$ & $\delta_{T_{1u-}}\, k_x$ \\ 
$4$ & $\delta_{T_{1u-}}\, k_y$ \\  
$5$ & $\delta_{T_{1u-}}\, k_z$ \\
\end{tabular}
\end{ruledtabular}
\end{table}

\begin{table}
\caption{\label{tab.cf.sf.A2g}Leading-order polynomial forms of the products $c_n(\mathbf{k}) f_n(\mathbf{k})$ of form factors describing $A_{2g+}$ pairing for a model with one \textit{s}-orbital and one $f_{xyz}$-orbital. The amplitudes of the leading terms in $c_n(\mathbf{k})$ have been absorbed into new pairing amplitudes marked by a tilde.}
\begin{ruledtabular}
\begin{tabular}{ll}
$n$ & $c_n f_n$ \\ \hline
$0$ & $\tdelta_{00}\, [k_{x}^4(k_{y}^2-k_{z}^2)+k_{y}^4(k_{z}^2-k_{x}^2)+k_{z}^4(k_{x}^2-k_{y}^2)] $ \\
$1$ & $\tdelta_{30}\, [k_{x}^4(k_{y}^2-k_{z}^2)+k_{y}^4(k_{z}^2-k_{x}^2)+k_{z}^4(k_{x}^2-k_{y}^2)] $\\
$2$ & $\tdelta_{A_{1u-}}\, k_x^2k_y^2k_z^2\,[k_{x}^4(k_{y}^2-k_{z}^2)+k_{y}^4(k_{z}^2-k_{x}^2)+k_{z}^4(k_{x}^2-k_{y}^2)] $ \\
$3$ & $\tdelta_{T_{1u-}}\, k_x^2(k_{y}^2-k_{z}^2)$ \\
$4$ & $\tdelta_{T_{1u-}}\, k_y^2(k_{z}^2-k_{x}^2)$ \\
$5$ & $\tdelta_{T_{1u-}}\, k_z^2(k_{x}^2-k_{y}^2)$ \\
\end{tabular}
\end{ruledtabular}
\end{table}

The resulting superconducting form factors $f_{n}$ and the products $c_{n} f_{n}$, which are required to determine the IP nodes, are listed in Tables \ref{tab.f.sf.A2g} and \ref{tab.cf.sf.A2g}, respectively, to the leading order. The condition for IP nodes reads as
\begin{align}
&c_0(\mathbf{k})\, f_0(\mathbf{k}) - \vec c(\mathbf{k}) \cdot \vec f(\mathbf{k})
  = \big( \tdelta_{00} - \tdelta_{30} - \tdelta_{A_{1u-}} k_{x}^2 k_{y}^2 k_{z}^2 \big) \nonumber \\
&\qquad{}\times \big[ k_{x}^4 (k_{y}^2-k_{z}^2) + k_{y}^4 (k_{z}^2-k_{x}^2)
  + k_{z}^4 (k_{x}^2-k_{y}^2) \big] = 0 .
\label{3.A2g.IPnodes.3}
\end{align}
Contribution (c) has dropped out. This is again an artifact of using the same basis functions for $c_n$ and $f_n$. Going beyond leading order, a contribution remains but does not affect the conclusions. For example, the $T_{1u-}$ basis functions $k_x^3$, $k_y^3$, $k_z^3$ generate another term proportional to $k_{x}^4 (k_{y}^2-k_{z}^2) + k_{y}^4 (k_{z}^2-k_{x}^2) + k_{z}^4 (k_{x}^2-k_{y}^2)$. Equation (\ref{3.A2g.IPnodes.3}) is satisfied whenever any two of the components of $\mathbf{k}$ are equal. Thus there are line nodes in the $\{110\}$ planes for infinitesimal pairing.

Form factors in the $(110)$ plane read as
\begin{align}
f_0(\mathbf{k}) &= f_1(\mathbf{k}) = f_2(\mathbf{k}) = 0 , \\
f_3(\mathbf{k}) &= f_4(\mathbf{k}) = \delta_{T_{1u-}}k_{x} , \\
f_5(\mathbf{k}) &= \delta_{T_{1u-}}k_{z} ,
\end{align}
which gives
\begin{align}
\langle f^1 ,f^1 \rangle &= -(\RR \delta_{T_{1u-}})^2 (2 k_{x}^2+k_{z}^2) , \\
\langle f^2 ,f^2 \rangle &= -(\II \delta_{T_{1u-}})^2 (2 k_{x}^2+k_{z}^2) , \\
\langle f^1 ,f^2 \rangle &= -\RR \delta_{T_{1u-}}\, \II \delta_{T_{1u-}} (2 k_{x}^2+k_{z}^2)
\end{align}
and also $\langle c ,f^1 \rangle=\langle c ,f^2 \rangle=0$. In this case, the Pfaffian simplifies to the form of Eq.\ (\ref{3.A2g.Pf.3}). The second term $4\,[\langle f^1 ,f^2 \rangle^2-\langle f^1 ,f^1 \rangle \langle f^2 ,f^2 \rangle]$ vanishes, which implies that there is no inflation of the line nodes in the mirror plane. The vanishing can be attributed to the fact that in the $\{110\}$ planes, only a single amplitude $\delta_{T_{1u-}}$ leads to a superconducting gap, the phase of which can always be chosen real so that the TRS breaking is irrelevant. On the other hand, $\langle f^1 ,f^1 \rangle +\langle f^2 ,f^2 \rangle$ is nonzero so that the nodes are shifted. This is analogous to the case of two \textit{s}-orbitals.

\subsubsection{$E_{g+}$ pairing}

In Table \ref{tab.sf.reduce}, pairing with $E_{g+}$ symmetry occurs in (a) $E_{g+}\otimes A_{1g+}$, (b) $E_{u-}\otimes A_{2u-}$, (c) $T_{1u-}\otimes T_{2u-}$, and (d) $T_{2u-}\otimes T_{2u-}$. The matrix-valued basis functions are given in the following to leading order only.

(a) For $E_{g+}\otimes A_{1g+}$:
\begin{align}
D_{x^2-y^2,00}(\mathbf{k}) &\cong (k_x^2-k_y^2)\, \sigma_0 \otimes \sigma_0 , \\
D_{3z^2-r^2,00}(\mathbf{k}) &\cong \frac{1}{\sqrt{3}}\, (2k_z^2-k_x^2-k_y^2)\, \sigma_0 \otimes \sigma_0 , \\
D_{x^2-y^2,30}(\mathbf{k}) &\cong (k_x^2-k_y^2)\, \sigma_3 \otimes \sigma_0 , \\
D_{3z^2-r^2,30}(\mathbf{k}) &\cong \frac{1}{\sqrt{3}}\, (2k_z^2-k_x^2-k_y^2)\, \sigma_3 \otimes \sigma_0 .
\end{align}

(b) For $E_{u-}\otimes A_{2u-}$:
\begin{align}
D_{x^2-y^2,20}(\mathbf{k}) &\cong k_xk_yk_z (k_x^2-k_y^2)\, \sigma_2 \otimes \sigma_0 , \\
D_{3z^2-r^2,20}(\mathbf{k}) &\cong \frac{1}{\sqrt{3}}\, k_xk_yk_z (2k_z^2-k_x^2-k_y^2)\,
  \sigma_2 \otimes \sigma_0 .
\end{align}
Note that $k_xk_yk_z\, \sigma_2\times \sigma_0$ is invariant under $O_h$.

(c) For $T_{1u-}\otimes T_{2u-}$:
\begin{align}
D_{x^2-y^2,T_{1u-}}(\mathbf{k}) &\cong \frac{1}{\sqrt{3}}\, (-k_x\, \sigma_1 \otimes \sigma_1
  - k_y\, \sigma_1 \otimes \sigma_2 \nonumber \\
&\quad{}+ 2\, k_z\, \sigma_1 \otimes \sigma_3 ) , \\
D_{3z^2-r^2,T_{1u-}}(\mathbf{k}) &\cong - (k_x\, \sigma_1 \otimes \sigma_1
  - k_y\, \sigma_1 \otimes \sigma_2) .
\end{align}
For this and the following contribution, the transformation properties under three- and four-fold rotations have been used to determine the two components.

(d) For $T_{2u-}\otimes T_{2u-}$:
\begin{align}
&D_{x^2-y^2,T_{2u-}}(\mathbf{k}) \cong \big[ k_x (k_y^2-k_z^2)\, \sigma_1 
\otimes \sigma_1 \nonumber \\
&\quad{}- k_y (k_z^2-k_x^2)\, \sigma_1 \otimes \sigma_2 \big] , \\
&D_{3z^2-r^2,T_{2u-}}(\mathbf{k}) \cong \frac{1}{\sqrt{3}}\, \big[ {-}k_x 
(k_y^2-k_z^2)\,
  \sigma_1 \otimes \sigma_1 \nonumber \\
&\quad{}- k_y (k_z^2-k_x^2)\, \sigma_1 \otimes \sigma_2
  + 2\, k_z (k_x^2-k_y^2)\, \sigma_1 \otimes \sigma_3 \big] .
\end{align}

The resulting superconducting form factors $f_n(\mathbf{k})$ are given in Table \ref{tab.f.sf.Eg} and the products $c_n(\mathbf{k}) f_n(\mathbf{k})$ appearing in the condition (\ref{2.cfN4.4}) for IP nodes are shown in Table \ref{tab.cf.sf.Eg}. The total contribution to $\langle c,f\rangle$ from $x^2-y^2$ basis functions (with amplitudes $\tdelta^1_{\cdots}$) has two symmetry-imposed first-order line nodes at $k_y=\pm k_x$. In a time-reversal-symmetric state, the inclusion of $3z^2-r^2$ basis functions (with amplitudes $\tdelta^2_{\cdots}$) generically leads to two line nodes elsewhere on the normal-state Fermi surface. The nodes must intersect with the $\langle 111\rangle$ axes, though, since there the full expression vanishes. TRS-breaking states generically lead to point nodes in the $\langle 111\rangle$ directions. This is for example the case for the generalized $(1,i)$-type state. These point nodes are solely determined by symmetry and therefore agree with the case of two \textit{s}-orbitals.

\begin{table}
\caption{\label{tab.f.sf.Eg}Leading-order polynomial forms of the form factors $f_n(\mathbf{k})$ describing $E_{g+}$ pairing for a model with one \textit{s}-orbital and one $f_{xyz}$-orbital.}
\begin{ruledtabular}
\begin{tabular}{ll}
$n$ & $f_n$ \\ \hline
$0$ & $\delta^1_{00}\, (k_x^2-k_y^2)
  + \frac{\delta^2_{00}}{\sqrt{3}}\, (2k_z^2 - k_x^2 - k_y^2)$ \\
$1$ & $\delta^1_{30}\, (k_x^2-k_y^2)
  + \frac{\delta^2_{30}}{\sqrt{3}}\, (2k_z^2 - k_x^2 - k_y^2)$ \\
$2$ & $\delta^1_{20}\, k_xk_yk_z (k_x^2 - k_y^2)
  + \frac{\delta^2_{20}}{\sqrt{3}}\, k_xk_yk_z (2k_z^2 - k_x^2 - k_y^2)$ \\
$3$ & $\Big({-} \frac{\delta^1_{T_{1u-}}}{\sqrt{3}} - \delta^2_{T_{1u-}} \Big)\, k_x
  + \Big( \delta^1_{T_{2u-}} - \frac{\delta^2_{T_{2u-}}}{\sqrt{3}} \Big)\, k_x (k_y^2-k_z^2)$ \\ 
$4$ & $\Big({-} \frac{\delta^1_{T_{1u-}}}{\sqrt{3}} + \delta^2_{T_{1u-}} \Big)\, k_y
  + \Big({-} \delta^1_{T_{2u-}} - \frac{\delta^2_{T_{2u-}}}{\sqrt{3}} \Big)\, k_y (k_z^2-k_x^2)$ \\
$5$ & $\frac{2\, \delta^1_{T_{1u-}}}{\sqrt{3}}\, k_z
  + \frac{2\, \delta^2_{T_{2u-}}}{\sqrt{3}}\, k_z (k_x^2-k_y^2)$ \\
\end{tabular}
\end{ruledtabular}
\end{table}

\begin{table}
\caption{\label{tab.cf.sf.Eg}Leading-order polynomial forms of the products $c_n(\mathbf{k}) f_n(\mathbf{k})$ of form factors describing $E_{g+}$ pairing for a model with one \textit{s}-orbital and one $f_{xyz}$-orbital. The amplitudes of the leading terms in $c_n(\mathbf{k})$ have been absorbed into new pairing amplitudes marked by a tilde.}
\begin{ruledtabular}
\begin{tabular}{@{\hspace{1em}}ll@{\hspace{1em}}}
$n$ & $c_n f_n$ \\ \hline
$0$ & $\tdelta^1_{00}\, (k_x^2-k_y^2)
  + \frac{\tdelta^2_{00}}{\sqrt{3}}\, (2k_z^2 - k_x^2 - k_y^2)$ \\
$1$ & $\tdelta^1_{30}\, (k_x^2-k_y^2)
  + \frac{\tdelta^2_{30}}{\sqrt{3}}\, (2k_z^2 - k_x^2 - k_y^2)$ \\
$2$ & $\tdelta^1_{20}\, k_x^2k_y^2k_z^2 (k_x^2 - k_y^2)
  + \frac{\tdelta^2_{20}}{\sqrt{3}}\, k_x^2k_y^2k_z^2 (2k_z^2 - k_x^2 - k_y^2)$ \\
$3$ & $\Big({-} \frac{\tdelta^1_{T_{1u-}}}{\sqrt{3}} - \tdelta^2_{T_{1u-}} \Big)\, k_x^2 (k_y^2-k_z^2)$ \\
  & $\quad{}+ \Big( \tdelta^1_{T_{2u-}} - \frac{\tdelta^2_{T_{2u-}}}{\sqrt{3}} \Big)\,
    k_x^2 (k_y^2-k_z^2)^2$ \\
$4$ & $\Big({-} \frac{\tdelta^1_{T_{1u-}}}{\sqrt{3}} + \tdelta^2_{T_{1u-}} \Big)\, k_y^2 (k_z^2-k_x^2)$ \\
  & $\quad{}+ \Big({-} \tdelta^1_{T_{2u-}} - \frac{\tdelta^2_{T_{2u-}}}{\sqrt{3}} \Big)\,
    k_y^2 (k_z^2-k_x^2)^2$ \\
$5$ & $\frac{2\, \tdelta^1_{T_{1u-}}}{\sqrt{3}}\, k_z^2 (k_x^2-k_y^2)
  + \frac{2\, \tdelta^2_{T_{2u-}}}{\sqrt{3}}\, k_z^2 (k_x^2-k_y^2)^2$ \\
\end{tabular}
\end{ruledtabular}
\end{table}

For noninfinitesimal pairing that breaks TRS, the point nodes are inflated. This is seen by considering the Pfaffian on the high-symmetry axis $\mathbf{k}=k\,(1,1,1)/\sqrt{3}$ through a IP point node. On this axis, we 
have, to leading order,
\begin{align}
f_0(\mathbf{k}) &= f_1(\mathbf{k}) = f_2(\mathbf{k}) = 0 , \\
f_3(\mathbf{k}) &= -\frac{\delta^1_{T_{1u-}}}{3}\, k
  - \frac{\delta^2_{T_{1u-}}}{\sqrt{3}}\, k , \\
f_4(\mathbf{k}) &= -\frac{\delta^1_{T_{1u-}}}{3}\, k
  + \frac{\delta^2_{T_{1u-}}}{\sqrt{3}}\, k , \\
f_5(\mathbf{k}) &= \frac{2\, \delta^1_{T_{1u-}}}{3}\, k .
\end{align}
For the $(1,i)$ pairing state with $\delta^2_{T_{1u-}} = i\delta^1_{T_{1u-}}$, $\delta^1_{T_{1u-}}\in \mathbb{R}$, we find
\begin{align}
\langle f^1,f^1\rangle &= \langle f^2,f^2\rangle = -\frac{2k^2}{3}\, \big(\delta^1_{T_{1u-}}\big)^2 , \\
\langle f^1,f^2\rangle &= 0 .
\end{align}
On the other hand, the normal-state form factors are, to leading order,
\begin{align}
c_0(\mathbf{k}) &= c_{00}^{(0)} , \\
c_1(\mathbf{k}) &= c_{30}^{(0)} , \\
c_2(\mathbf{k}) &= \frac{1}{3\sqrt{3}}\, c_{20}^{(3)}\, k^3 , \\
c_3(\mathbf{k}) &= c_4(\mathbf{k}) = c_5(\mathbf{k}) = 0 .
\end{align}
We thus find $\langle c,f^1\rangle = \langle c,f^2\rangle = 0$ and the analysis is analogous to the one for the $(1,i)$ state for two \textit{s}-orbitals. Hence, we expect BFSs that touch the normal-state Fermi surface at the IP nodes.

\subsection{Two-side basis: Diamond structure}
\label{sub.diamond}

Another origin of internal degrees of freedom is a nontrivial basis of the crystal. This is a good place to consider an example: the diamond structure with one \textit{s}-orbital per basis site. The space group is 227, belonging to the point group $O_h$. We write matrices as Kronecker products of a matrix acting on site space and a matrix on spin space.

\begin{table}
\caption{\label{tab.diamond.basis}Basis matrices on the internal Hilbert space for the case of \textit{s}-orbitals forming a diamond structure. The basis matrices are irreducible tensor operators of the irreps listed in the second column. For multidimensional irreps, the states transforming into each other under point-group operations are distinguished by the index in the third column.}
\begin{ruledtabular}
\begin{tabular}{@{\hspace{3em}}ccc@{\hspace{3em}}}
$h_\nu$ & Irrep & Component \\ \hline
$\sigma_0 \otimes \sigma_0$ & $A_{1g+}$ & \\
$\sigma_0 \otimes \sigma_1$ & $T_{1g-}$ & 1 \\
$\sigma_0 \otimes \sigma_2$ & & 2 \\
$\sigma_0 \otimes \sigma_3$ & & 3 \\
$\sigma_1 \otimes \sigma_0$ & $A_{1g+}$ & \\
$\sigma_1 \otimes \sigma_1$ & $T_{1g-}$ & 1 \\
$\sigma_1 \otimes \sigma_2$ & & 2 \\
$\sigma_1 \otimes \sigma_3$ & & 3 \\
$\sigma_2 \otimes \sigma_0$ & $A_{2u-}$ & \\
$\sigma_2 \otimes \sigma_1$ & $T_{2u+}$ & 1 \\
$\sigma_2 \otimes \sigma_2$ & & 2 \\
$\sigma_2 \otimes \sigma_3$ & & 3 \\
$\sigma_3 \otimes \sigma_0$ & $A_{2u+}$ & \\
$\sigma_3 \otimes \sigma_1$ & $T_{2u-}$ & 1 \\
$\sigma_3 \otimes \sigma_2$ & & 2 \\
$\sigma_3 \otimes \sigma_3$ & & 3 \\
\end{tabular}
\end{ruledtabular}
\end{table}

The parity matrix $U_P = \sigma_1 \otimes \sigma_0$ is nontrivial since inversion interchanges the basis sites. This case has also been analyzed in the context of semimetals~\cite{AMV18}. Moreover, the fourfold axes also interchange the basis sites, whereas the threefold axes do not. Time reversal is unchanged, $U_T = \sigma_0 \otimes i\sigma_2$. The basis matrices are listed in Table \ref{tab.diamond.basis}. This is the same scheme as for two orbitals of opposite parity, see Table \ref{tab.sf.basis}, except that the Pauli matrices $\sigma_1$ and $\sigma_3$ in the first (orbital/site) factor are interchanged. Thus the results for the pairing can be mapped over from Sec.\ \ref{sub.opposite} without effort.

\subsection{Effective spin 3/2}
\label{sub.j32}

Here, we consider electrons with effective angular momentum $j=3/2$. It is of interest to check whether the results obtained for local pairing in such a model \cite{ABT17,BAM18} are robust under nonlocal pairing and which additional pairing states are allowed for nonlocal pairing. The Hilbert space for $j=3/2$ is four dimensional. In this case, it is useful to express all matrices as polynomials of the standard angular-momentum-$3/2$ matrices
\begin{align}
J_x &= \left(\begin{array}{cccc}
  0 & \sqrt{3}/2 & 0 & 0 \\
  \sqrt{3}/2 & 0 & 1 & 0 \\
  0 & 1 & 0 & \sqrt{3}/2 \\
  0 & 0 & \sqrt{3}/2 & 0
  \end{array}\right) , \\
J_y &= \left(\begin{array}{cccc}
  0 & -i\, \sqrt{3}/2 & 0 & 0 \\
  i\, \sqrt{3}/2 & 0 & -i & 0 \\
  0 & i & 0 & -i\,\sqrt{3}/2 \\
  0 & 0 & i\,\sqrt{3}/2 & 0
  \end{array}\right) , \\
J_z &= \left(\begin{array}{cccc}
  3/2 & 0 & 0 & 0 \\
  0 & 1/2 & 0 & 0 \\
  0 & 0 & -1/2 & 0 \\
  0 & 0 & 0 & -3/2
  \end{array}\right) ,
\end{align}
and the $4\times 4$ identity matrix $J_0 \equiv \openone$.

\begin{table}
\caption{\label{tab.j32.basis}Basis matrices on the internal Hilbert space for the case of electrons with angular momentum $j=3/2$. The basis matrices are irreducible tensor operators of the irreps listed in the second column. For multidimensional irreps, the states transforming into each other under point-group operations are distinguished by the index in the third column.}
\begin{ruledtabular}
\begin{tabular}{@{\hspace{1em}}lcc@{\hspace{1em}}}
$h_\nu$ & Irrep & Component \\ \hline
$J_0$ & $A_{1g+}$ & \\
$\frac{2}{\sqrt{5}}\, J_x$ & $T_{1g-}$ & 1 \\
$\frac{2}{\sqrt{5}}\, J_y$ & & 2 \\
$\frac{2}{\sqrt{5}}\, J_z$ & & 3 \\
$\frac{1}{\sqrt{3}}\, (J_yJ_z + J_zJ_y)$ & $T_{2g+}$ & 1 \\
$\frac{1}{\sqrt{3}}\, (J_zJ_x + J_xJ_z)$ & & 2 \\
$\frac{1}{\sqrt{3}}\, (J_xJ_y + J_yJ_x)$ & & 3 \\
$\frac{1}{\sqrt{3}}\, (J_x^2 - J_y^2)$ & $E_{g+}$ & 1 \\
$\frac{1}{3}\, (2J_z^2 - J_x^2 - J_y^2)$ & & 2 \\
$\frac{2}{\sqrt{3}}\, (J_xJ_yJ_z + J_zJ_yJ_x)$ & $A_{2g-}$ & \\
$\frac{8}{\sqrt{365}}\, J_x^3$ & $T_{1g-}$ & 1 \\
$\frac{8}{\sqrt{365}}\, J_y^3$ & & 2 \\
$\frac{8}{\sqrt{365}}\, J_z^3$ & & 3 \\
$\frac{1}{\sqrt{3}}\, [ J_x(J_y^2-J_z^2) + (J_y^2-J_z^2)J_x ]$ & $T_{2g-}$ & 1 \\
$\frac{1}{\sqrt{3}}\, [ J_y(J_z^2-J_x^2) + (J_z^2-J_x^2)J_y ]$ & & 2 \\
$\frac{1}{\sqrt{3}}\, [ J_z(J_x^2-J_y^2) + (J_x^2-J_y^2)J_z ]$ & & 3 \\
\end{tabular}
\end{ruledtabular}
\end{table}

The parity matrix is trivial, $U_P = \openone = J_0$, since the angular momentum is invariant under inversion. The unitary part of the time-reversal operator now reads as
\begin{equation}
U_T = e^{i J_y \pi}
  = \begin{pmatrix}
    0 & 0 & 0 & 1 \\
    0 & 0 & -1 & 0 \\
    0 & 1 & 0 & 0 \\
    -1 & 0 & 0 & 0
  \end{pmatrix} .
\end{equation}
The 16 basis matrices $h_\nu$ of the space of Hermitian $4\times 4$ matrices are listed in Table \ref{tab.j32.basis}, together with the corresponding irreps. We normalize the basis matrices in such a way that $\Tr h_\nu^2 = 4$. Apart from this, the entries in the table follow from known basis functions \cite{Kat}, taking into account that $\mathbf{J}=(J_x,J_y,J_z)$ is even under inversion and odd under time reversal, and symmetrizing products of angular-momentum matrices so as to generate Hermitian matrices~\cite{SRV17}. A new feature is the presence of basis matrices belonging to the two-dimensional irrep $E_{g+}$.

\begin{table*}
\caption{\label{tab.j32.reduce}Reduction of product representations of the allowed irreps of $\mathbf{k}$-dependent form factors (rows) and basis matrices $h_\nu$ (columns) for electrons with angular momentum $j=3/2$. For the form factors, the minimum order of polynomial basis functions is given in the second column. ``\bad'' indicates products that are forbidden since they violate fermionic antisymmetry. For brevity, the symbols $\mathcal{A} \equiv A_{1u+} \oplus E_{u+} \oplus T_{1u+} \oplus T_{2u+}$ and $\mathcal{B} \equiv A_{2u+} \oplus E_{u+} \oplus T_{1u+} \oplus T_{2u+}$ are used.}
\begin{ruledtabular}
\begin{tabular}{lc@{\hspace{0em}}llllll}
\multicolumn{2}{c}{Form factor:\hspace*{0em}} & \multicolumn{6}{c}{Pairing matrix: Irrep} \\
Irrep & Min.\ $l$ & $A_{1g+}$ & $T_{2g+}$ & $E_{g+}$ & $T_{1g-}$ & $A_{2g-}$ & $T_{2g-}$ \\ \hline
$A_{1g+}$ & 0 & {\boldmath$A_{1g+}$} & $T_{2g+}$ & $E_{g+}$ & \bad & \bad 
& \bad \\
$A_{2g+}$ & 6 & $A_{2g+}$ & $T_{1g+}$ & $E_{g+}$ & \bad & \bad & \bad \\
$E_{g+}$ & 2 & $E_{g+}$ & $T_{1g+} \oplus T_{2g+}$ & $\mbox{\boldmath$A_{1g+}$} \oplus A_{2g+} \oplus E_{g+}$
  & \bad & \bad & \bad \\
$T_{1g+}$ & 4 & $T_{1g+}$ & $A_{2g+} \oplus E_{g+} \oplus T_{1g+} \oplus T_{2g+}$ & $T_{1g+} \oplus T_{2g+}$ 
  & \bad & \bad & \bad \\
$T_{2g+}$ & 2 & $T_{2g+}$ & $\mbox{\boldmath$A_{1g+}$} \oplus E_{g+} \oplus T_{1g+} \oplus T_{2g+}$ &
  $T_{1g+} \oplus T_{2g+}$ & \bad & \bad & \bad \\
$A_{1u-}$ & 9 & \bad & \bad & \bad & $T_{1u+}$ & $A_{2u+}$ & $T_{2u+}$ \\
$A_{2u-}$ & 3 & \bad & \bad & \bad & $T_{2u+}$ & $A_{1u+}$ & $T_{1u+}$ \\
$E_{u-}$ & 5 & \bad & \bad & \bad & $T_{1u+} \oplus T_{2u+}$ & $E_{u+}$ & 
$T_{1u+} \oplus T_{2u+}$ \\
$T_{1u-}$ & 1 & \bad & \bad & \bad & $\mathcal{A}$ & $T_{2u+}$ & $\mathcal{B}$ \\
$T_{2u-}$ & 3 & \bad & \bad & \bad & $\mathcal{B}$ & $T_{1u+}$ & $\mathcal{A}$ \\
\end{tabular}
\end{ruledtabular}
\end{table*}

Table \ref{tab.j32.reduce} shows all relevant reductions of product representations. The normal-state Hamiltonian can only contain the highlighted $A_{1g+}$ combinations and local pairing is only compatible with the first row of the table---this reproduces the known three irreps $A_{1g+}$, $T_{2g+}$, and $E_{g+}$ \cite{ABT17,BAM18}. Again, all ten time-reversal-even pairing symmetries can occur and we restrict ourselves to $g+$ irreps (even parity). All of these occur for any of the three irreps $A_{1g+}$, $T_{2g+}$, and $E_{g+}$ of basis matrices.

The normal-state Hamiltonian $H_N(\mathbf{k})$ is a linear combination of the basis matrices
\begin{align}
h_0 &\equiv J_0 & A_{1g+}, \\
h_1 &\equiv \frac{1}{\sqrt{3}}\, (J_yJ_z + J_zJ_y) & T_{2g+}, \label{j32.h1} \\
h_2 &\equiv \frac{1}{\sqrt{3}}\, (J_zJ_x + J_xJ_z) & T_{2g+}, \\
h_3 &\equiv \frac{1}{\sqrt{3}}\, (J_xJ_y + J_yJ_x) & T_{2g+}, \label{j32.h3} \\
h_4 &\equiv \frac{1}{\sqrt{3}}\, (J_x^2 - J_y^2) & E_{g+}, \\
h_5 &\equiv \frac{1}{3}\, (2J_z^2 - J_x^2 - J_y^2) & E_{g+} ,
\end{align}
which again satisfy the universal algebra; see Appendices \ref{app.normal} and \ref{app.algebra}. The normal-state Hamiltonian contains these matrices with form factors $c_0(\mathbf{k}),\ldots,c_5(\mathbf{k})$, which must transform in the same way as $h_0,\ldots,h_5$.

\subsubsection{$A_{2g+}$ pairing}
\label{subsub.j32.A2g}

$A_{2g+}$ pairing appears in three places in Table \ref{tab.j32.reduce}: (a) $A_{2g+}\otimes A_{1g+}$, (b) $E_{g+}\otimes E_{g+}$, and (c) $T_{1g+}\otimes T_{2g+}$. This is a potentially interesting pairing state since it is impossible for purely local pairing and it is an example of a nontrivial one-dimensional irrep. In the following, we give the basis functions to the leading order only.

(a) For $A_{2g+}\otimes A_{1g+}$:
\begin{align}
D_{A_{2g+}}(\mathbf{k}) &\cong \big[ k_{x}^4 (k_{y}^2-k_{z}^2) + k_{y}^4 (k_{z}^2-k_{x}^2) \nonumber \\
&\quad{} + k_{z}^4 (k_{x}^2-k_{y}^2) \big]\, h_0 . 
\end{align}

(b) For $E_{g+}\otimes E_{g+}$:
\begin{align}
D_{E_{g+}}(\mathbf{k}) &\cong (k_x^2 - k_y^2)\, h_5 - \frac{1}{\sqrt{3}}\, (2k_z^2 - k_x^2 - k_y^2)\, h_4 .
\end{align}

(c) For $T_{1g+}\otimes T_{2g+}$:
\begin{align}
D_{T_{1g+}}(\mathbf{k}) &\cong  k_y k_z (k_y^2 - k_z^2)\, h_1 + k_z k_x (k_z^2 - k_x^2)\, h_2 \nonumber \\
&\quad{} + k_x k_y (k_x^2 - k_y^2)\, h_3 .
\end{align}
The assignment of the components is done in such a way that the form factor changes sign under any fourfold rotation.

\begin{table}
\caption{\label{tab.f.j32.A2g}Leading-order polynomial forms of the form factors $f_n(\mathbf{k})$ describing $A_{2g+}$ pairing for electrons with 
angular momentum $j=3/2$.}
\begin{ruledtabular}
\begin{tabular}{@{\hspace{1em}}ll@{\hspace{1em}}}
$n$ & $f_n$ \\ \hline
$0$ & $\delta_{A_{2g+}}\,[k_{x}^4(k_{y}^2-k_{z}^2)+k_{y}^4(k_{z}^2-k_{x}^2)+k_{z}^4(k_{x}^2-k_{y}^2)] $ \\
$1$ & $\delta_{T_{1g+}}\, k_y k_z (k_y^2-k_z^2)$ \\
$2$ & $\delta_{T_{1g+}}\, k_z k_x (k_z^2-k_x^2) $ \\
$3$ & $\delta_{T_{1g+}}\, k_x k_y (k_x^2-k_y^2)$ \\ 
$4$ & $- \frac{\delta_{E_{g+}}}{\sqrt{3}}\, (2 k_z^2-k_x^2-k_y^2)$ \\
$5$ & $\delta_{E_{g+}} (k_x^2-k_y^2)$ \\
\end{tabular}
\end{ruledtabular}
\end{table}

\begin{table}
\caption{\label{tab.cf.j32.A2g}Leading-order polynomial forms of the products $c_n(\mathbf{k}) f_n(\mathbf{k})$ of form factors describing $A_{2g+}$ pairing for electrons with angular momentum $j=3/2$. The amplitudes of the leading terms in $c_n(\mathbf{k})$ have been absorbed into new pairing amplitudes marked by a tilde.}
\begin{ruledtabular}
\begin{tabular}{@{\hspace{1em}}ll@{\hspace{1em}}}
$n$ & $c_n f_n$ \\ \hline
$0$ & $\tdelta_{A_{2g+}}\, [k_{x}^4(k_{y}^2-k_{z}^2)+k_{y}^4(k_{z}^2-k_{x}^2)+k_{z}^4(k_{x}^2-k_{y}^2)] $ \\
$1$ & $\tdelta_{T_{1g+}}\, k_y^2 k_z^2(k_y^2-k_z^2)$\\
$2$ & $\tdelta_{T_{1g+}}\, k_z^2 k_x^2 (k_z^2-k_x^2)$\\
$3$ & $\tdelta_{T_{1g+}}\, k_x k_y (k_x^2-k_y^2)$ \\
$4$ & $- \frac{\tdelta_{E_{g+}}}{\sqrt{3}}\, (k_x^2-k_y^2)(2 k_z^2-k_x^2-k_y^2)$ \\
$5$ & $\frac{\tdelta_{E_{g+}}}{\sqrt{3}}\, (k_x^2-k_y^2)(2 k_z^2-k_x^2-k_y^2)$ \\
\end{tabular}
\end{ruledtabular}
\end{table}

The resulting superconducting form factors $f_{n}$ and the products $c_{n} f_{n}$, which are required to determine the IP nodes, are listed in Tables \ref{tab.f.j32.A2g} and \ref{tab.cf.j32.A2g}, respectively, to the leading order. The condition for IP nodes reads as
\begin{align}
c_0&(\mathbf{k})\, f_0(\mathbf{k}) - \vec c(\mathbf{k}) \cdot \vec f(\mathbf{k}) \nonumber \\
&= \tdelta_{A_{2g+}}\, \big[ k_{x}^4 (k_{y}^2-k_{z}^2) + k_{y}^4 (k_{z}^2-k_{x}^2)
  + k_{z}^4 (k_{x}^2-k_{y}^2) \big] \nonumber \\
&\quad{}- \tdelta_{T_{1g+}}\, \big[ k_y^2k_z^2 (k_y^2-k_z^2)+ k_z^2k_x^2 (k_z^2-k_x^2) \nonumber \\
&\quad{}+ k_x^2k_y^2 (k_x^2-k_y^2)\big] = 0 ,
\end{align}
where the contributions of type (b) cancel. This is again an artifact of having used the same basis functions for $c_n$ and $f_n$. For general basis functions, the terms do not cancel, but they do not change the conclusions. The above expression vanishes whenever any two of the three components of $\mathbf{k}$ are equal. There are line nodes in the $\{110\}$ planes for infinitesimal pairing.

Next, we consider the TRS-breaking state where the amplitude from $T_{1g+}\otimes T_{2g+}$ has a phase shift of $\pi/2$ relative to the amplitude from $E_{g+}\otimes E_{g+}$. The real and imaginary parts of the condition for IP nodes have the same momentum dependence. Hence, the line nodes of the real and imaginary part coincide and the TRS-breaking state retains the six line nodes in the mirror planes. The form factors in the $(110)$ plane read as
\begin{align}
f_0(\mathbf{k}) &= f_3(\mathbf{k}) = f_5(\mathbf{k}) = 0 , \\
f_1(\mathbf{k}) &=- i \delta_{T_{1g+}} k_zk_x (k_z^2-k_x^2) , \\
f_2(\mathbf{k}) &= i \delta_{T_{1g+}} k_zk_x (k_z^2-k_x^2) , \\
f_4(\mathbf{k}) &= -\frac{2}{\sqrt{3}}\, \delta_{E_{g+}} (k_z^2-k_x^2) ,
\end{align}
with $\delta_{T_{1g+}}$ and $\delta_{E_{g+}}$ real, which gives
\begin{align}
\langle f^1 ,f^1 \rangle &= -\frac{4}{3}\, \delta_{E_{g+}}^2 (k_z^2-k_x^2)^2 , \\
\langle f^2 ,f^2 \rangle &= -2\, \delta_{T_{1g+}}^2 k_z^2 k_x^2 (k_z^2-k_x^2)^2 , \\
\langle f^1 ,f^2 \rangle &= 0 .
\end{align}
and also $\langle c ,f^1 \rangle=\langle c ,f^2 \rangle=0$. In this case, the Pfaffian simplifies to
\begin{align}
P(\textbf{k}) &= (\langle c ,c \rangle-\langle f^1 ,f^1 \rangle-\langle f^2 ,f^2 \rangle)^2
  - 4\langle f^1 ,f^1 \rangle \langle f^2 ,f^2 \rangle .
\end{align}
The second term
\begin{align}
-4\langle f^1 ,f^1 \rangle \langle f^2 ,f^2 \rangle =-\frac{32}{3}\,
  \delta_{E_{g+}}^2 \delta_{T_{1g+}}^2 k_z^2 k_x^2 (k_z^2-k_x^2)^4
\end{align}
is generically negative. Thus we expect all the line nodes to inflate for strong coupling unlike for the two examples of $A_{2g+}$ pairing discussed above. The inflated line nodes are not attached to the normal-state Fermi surface. However, the inflation vanishes in special high-symmetry directions: On the $[111]$ axis, $\langle f^1 ,f^1 \rangle=\langle f^2 ,f^2 \rangle=0$, thus the nodes are not inflated and stick to the normal-state Fermi surface. For $[001]$ and $[110]$, nodes are also not inflated because there is only a single amplitude from $E_{g+}\otimes E_{g+}$ and its phase can be gauged away. Moreover, along these directions $\langle f^1 ,f^1 \rangle$ and $\langle f^2 ,f^2 \rangle$ are not both zero. Thus here the nodes are neither inflated nor attached to the normal-state Fermi surface. The vanishing inflation in high-symmetry directions implies that the BFSs have self-touching points there. Interestingly, $[111]$ is the direction in which three weak-coupling line nodes intersect whereas two intersect in the $[001]$ direction and there is no intersection in the $[110]$ direction.

\subsubsection{$E_{g+}$ pairing}

We consider $E_{g+}$ pairing as an example for a symmetry that is also possible for purely local pairing. The question is what changes for nonlocal pairing. $E_{g+}$ appears in six places in Table \ref{tab.j32.reduce}: (a) $A_{1g+}\otimes E_{g+}$, (b) $A_{2g+}\otimes E_{g+}$, (c) $E_{g+}\otimes A_{1g+}$, (d) $E_{g+}\otimes E_{g+}$, (e) $T_{1g+}\otimes T_{2g+}$, and (f) $T_{2g+}\otimes T_{2g+}$. The matrix-valued basis functions are given in the following to leading order only.

(a) For $A_{1g+}\otimes E_{g+}$, we find constants to leading order:
\begin{align}
D_{x^2-y^2,A_{1g+}}(\mathbf{k}) &\cong h_4 = J_x^2 - J_y^2 , \\
D_{3z^2-r^2,A_{1g+}}(\mathbf{k}) &\cong h_5 = \frac{1}{\sqrt{3}}\, (2J_z^2 - J_x^2 - J_y^2) .
\end{align}
These are of course the contributions from local pairing~\cite{ABT17,BAM18}.

(b) For $A_{2g+}\otimes E_{g+}$:
\begin{align}
&D_{x^2-y^2,A_{2g+}}(\mathbf{k}) \nonumber \\
&\quad\cong \big[ k_x^4 (k_y^2-k_z^2) + k_y^4 (k_z^2-k_x^2) + k_z^4 (k_x^2-k_y^2) \big]\, h_5 \\
&D_{3z^2-r^2,A_{2g+}}(\mathbf{k}) \nonumber \\
&\quad\cong -\big[ k_x^4 (k_y^2-k_z^2) + k_y^4 (k_z^2-k_x^2) + k_z^4 (k_x^2-k_y^2) \big]\, h_4 .
\end{align}

(c) For $E_{g+}\otimes A_{1g+}$:
\begin{align}
D_{x^2-y^2,0}(\mathbf{k}) &\cong (k_x^2 - k_y^2)\, h_0 , \\
D_{3z^2-r^2,0}(\mathbf{k}) &\cong \frac{1}{\sqrt{3}}\, (2k_z^2 - k_x^2 - k_y^2)\, h_0 .
\end{align}

(d) For $E_{g+}\otimes E_{g+}$:
\begin{align}
D_{x^2-y^2,E_{g+}}(\mathbf{k}) &\cong \frac{1}{\sqrt{3}}\, (2k_z^2 - k_x^2 - k_y^2)\, h_4
  + (k_x^2 - k_y^2)\, h_5 , \\
D_{3z^2-r^2,E_{g+}}(\mathbf{k}) &\cong (k_x^2 - k_y^2)\, h_4
  - \frac{1}{\sqrt{3}}\, (2k_z^2 - k_x^2 - k_y^2)\, h_5 .
\end{align}

(e) For $T_{1g+}\otimes T_{2g+}$:
\begin{align}
&D_{x^2-y^2,T_{1g+}}(\mathbf{k}) \cong \frac{1}{\sqrt{3}}\, \big[ k_y k_z 
(k_y^2-k_z^2)\, h_1 \nonumber \\
&\quad{}+ k_z k_x (k_z^2-k_x^2)\, h_2 - 2\, k_x k_y (k_x^2-k_y^2)\, h_3 \big] , \\
&D_{3z^2-r^2,T_{1g+}}(\mathbf{k}) \cong k_y k_z (k_y^2-k_z^2)\, h_1 - k_z 
k_x (k_z^2-k_x^2)\, h_2 .
\end{align}

(f) For $T_{2g+}\otimes T_{2g+}$:
\begin{align}
D_{x^2-y^2,T_{2g+}}(\mathbf{k}) &\cong - k_y k_z\, h_1 + k_z k_x\, h_2 , 
\label{3.32.Eg.f.1} \\
D_{3z^2-r^2,T_{2g+}}(\mathbf{k}) &\cong \frac{1}{\sqrt{3}}\,
 ( k_y k_z\, h_1 + k_z k_x\, h_2 - 2\, k_x k_y\, h_3 ) .
\label{3.32.Eg.f.2}
\end{align}

The resulting superconducting form factors $f_n(\mathbf{k})$ are given in Table \ref{tab.f.j32.Eg} and the products $c_n(\mathbf{k}) f_n(\mathbf{k})$ appearing in the condition (\ref{2.cfN4.4}) for IP nodes are shown in Table \ref{tab.cf.j32.Eg}. The analysis is analogous to the previous cases of $E_{g+}$ pairing: By inserting $k_y=\pm k_x$, one can see that the contribution to $\langle c,f\rangle$ from $x^2-y^2$ basis functions (with amplitudes $\tdelta^1_{\cdots}$) has two symmetry-imposed first-order line nodes for $k_y=\pm k_x$. In a time-reversal-symmetric state, the inclusion of $3z^2-r^2$ basis functions generically leads to two line nodes elsewhere on the normal-state Fermi surface. The nodes must intersect with the $\langle 111\rangle$ axes. TRS-breaking states generically lead to point nodes in the $\langle 111\rangle$ directions, for example for order parameters proportional to $(1,i)$. These point nodes are solely determined by symmetry. The presence of these eight point nodes was also found for purely local pairing \cite{BAM18}. We thus find that the inclusion of nonlocal pairing does not change the nodal structure.

\begin{table}
\caption{\label{tab.f.j32.Eg}Leading-order polynomial forms of the form factors $f_n(\mathbf{k})$ describing $E_{g+}$ pairing for electrons with angular momentum $j=3/2$.}
\begin{ruledtabular}
\begin{tabular}{ll}
$n$ & $f_n$ \\ \hline
$0$ & $\delta^1_0\, (k_x^2-k_y^2)
  + \frac{\delta^2_0}{\sqrt{3}}\, (2k_z^2 - k_x^2 - k_y^2)$ \\
$1$ & $\Big( \frac{\delta^1_{T_{1g+}}}{\sqrt{3}} + \delta^2_{T_{1g+}} \Big)\, k_y k_z (k_y^2-k_z^2)
  + \Big( {-} \delta^1_{T_{2g+}} + \frac{\delta^2_{T_{2g+}}}{\sqrt{3}} \Big)\,  k_y k_z$ \\
$2$ & $\Big( \frac{\delta^1_{T_{1g+}}}{\sqrt{3}} - \delta^2_{T_{1g+}} \Big)\, k_z k_x (k_z^2-k_x^2)
  + \Big( \delta^1_{T_{2g+}} + \frac{\delta^2_{T_{2g+}}}{\sqrt{3}} \Big)\, k_z k_x$ \\
$3$ & $-\frac{2\delta^1_{T_{1g+}}}{\sqrt{3}}\, k_x k_y (k_x^2-k_y^2)
  - \frac{2\delta^2_{T_{2g+}}}{\sqrt{3}}\, k_x k_y$ \\
$4$ & $\delta^1_{A_{1g+}}
  - \delta^2_{A_{2g+}}\, \big[k_x^4 (k_y^2-k_z^2) + k_y^4 (k_z^2-k_x^2) + 
k_z^4 (k_x^2-k_y^2)\big]$ \\
  & $\quad{}+ \frac{\delta^1_{E_{g+}}}{\sqrt{3}}\, (2k_z^2 - k_x^2 - k_y^2)
  + \delta^2_{E_{g+}} (k_x^2-k_y^2)$ \\
$5$ & $\delta^2_{A_{1g+}}
  + \delta^1_{A_{2g+}}\, \big[k_x^4 (k_y^2-k_z^2) + k_y^4 (k_z^2-k_x^2) + 
k_z^4 (k_x^2-k_y^2)\big]$ \\
  & $\quad{}+ \delta^1_{E_{g+}} (k_x^2-k_y^2)
  - \frac{\delta^2_{E_{g+}}}{\sqrt{3}}\, (2k_z^2 - k_x^2 - k_y^2)$ \\
\end{tabular}
\end{ruledtabular}
\end{table}

\begin{table}
\caption{\label{tab.cf.j32.Eg}Leading-order polynomial forms of the products $c_n(\mathbf{k}) f_n(\mathbf{k})$ of form factors describing $E_{g+}$ pairing for electrons with angular momentum $j=3/2$. The amplitudes of the leading terms in $c_n(\mathbf{k})$ have been absorbed into new pairing amplitudes marked by a tilde.}
\begin{ruledtabular}
\begin{tabular}{ll}
$n$ & $c_n f_n$ \\ \hline
$0$ & $\tdelta^1_0\, (k_x^2-k_y^2)
  + \frac{\tdelta^2_0}{\sqrt{3}}\, (2k_z^2 - k_x^2 - k_y^2)$ \\
$1$ & $\Big( \frac{\tdelta^1_{T_{1g+}}}{\sqrt{3}} + \tdelta^2_{T_{1g+}} \Big)\, k_y^2 k_z^2 (k_y^2-k_z^2)
  + \Big( {-} \tdelta^1_{T_{2g+}} + \frac{\tdelta^2_{T_{2g+}}}{\sqrt{3}} \Big)\,  k_y^2 k_z^2$ \\
$2$ & $\Big( \frac{\tdelta^1_{T_{1g+}}}{\sqrt{3}} - \tdelta^2_{T_{1g+}} \Big)\, k_z^2 k_x^2 (k_z^2-k_x^2)
  + \Big( \tdelta^1_{T_{2g+}} + \frac{\tdelta^2_{T_{2g+}}}{\sqrt{3}} \Big)\, k_z^2 k_x^2$ \\
$3$ & $-\frac{2\tdelta^1_{T_{1g+}}}{\sqrt{3}}\, k_x^2 k_y^2 (k_x^2-k_y^2)
  - \frac{2\tdelta^2_{T_{2g+}}}{\sqrt{3}}\, k_x^2 k_y^2$ \\
$4$ & $\tdelta^1_{A_{1g+}} (k_x^2-k_y^2) - \tdelta^2_{A_{2g+}} (k_x^2-k_y^2)$ \\
  & $\qquad{}\times \big[k_x^4 (k_y^2-k_z^2) + k_y^4 (k_z^2-k_x^2) + k_z^4 (k_x^2-k_y^2)\big]$ \\
  & $\quad{}+ \frac{\tdelta^1_{E_{g+}}}{\sqrt{3}}\, (k_x^2-k_y^2)(2k_z^2 - k_x^2 - k_y^2)
  + \tdelta^2_{E_{g+}} (k_x^2-k_y^2)^2$ \\
$5$ & $\frac{\tdelta^2_{A_{1g+}}}{\sqrt{3}}\, (2k_z^2 - k_x^2 - k_y^2)
  + \frac{\tdelta^1_{A_{2g+}}}{\sqrt{3}}\, (2k_z^2 - k_x^2 - k_y^2)$ \\
  & $\qquad{}\times \big[k_x^4 (k_y^2-k_z^2) + k_y^4 (k_z^2-k_x^2) + k_z^4 (k_x^2-k_y^2)\big]$ \\
  & $\quad{}+ \frac{\tdelta^1_{E_{g+}}}{\sqrt{3}}\, (k_x^2-k_y^2)(2k_z^2 - k_x^2 - k_y^2)$ \\
  & $\quad{}- \frac{\tdelta^2_{E_{g+}}}{3}\, (2k_z^2 - k_x^2 - k_y^2)^2$ \\
\end{tabular}
\end{ruledtabular}
\end{table}

For purely local pairing, the point nodes are inflated into BFSs for noninfinitesimal pairing \cite{BAM18}. We briefly sketch the analysis when nonlocal pairing is included. We consider the Pfaffian on the $[111]$ axis, $\mathbf{k} = k\,(1,1,1)/\sqrt{3}$. Table \ref{tab.f.j32.Eg} then shows that
\begin{align}
f_0(\mathbf{k}) &= 0 , \\
f_1(\mathbf{k}) &= -\frac{\delta^1_{T_{2g+}}}{3}\, k^2 + \frac{\delta^2_{T_{2g+}}}{3\sqrt{3}}\, k^2 , \\
f_2(\mathbf{k}) &= \frac{\delta^1_{T_{2g+}}}{3}\, k^2 + \frac{\delta^2_{T_{2g+}}}{3\sqrt{3}}\, k^2 , \\
f_3(\mathbf{k}) &= -\frac{2\, \delta^2_{T_{2g+}}}{3\sqrt{3}}\, k^2 , \\
f_4(\mathbf{k}) &= \delta^1_{A_{1g+}} , \\
f_5(\mathbf{k}) &= \delta^2_{A_{1g+}} .
\end{align}
For the generalized $(1,i)$ pairing state with $\delta^2_{A_{1g+}} = i\delta^1_{A_{1g+}}$, $\delta^2_{T_{2g+}} = i\delta^1_{T_{2g+}}$, and $\delta^1_{A_{1g+}}, \delta^1_{T_{2g+}} \in \mathbb{R}$, we find
\begin{align}
\langle f^1,f^1\rangle &= \langle f^2,f^2\rangle = -\frac{2k^4}{9}\, \big( \delta^1_{T_{2g+}} \big)^2
  - \big( \delta^1_{A_{1g+}} \big)^2 , \\
\langle f^1,f^2\rangle &= 0 .
\end{align}
On the other hand, the normal-state form factors are, to leading order,
\begin{align}
c_0(\mathbf{k}) &= c_0^{(0)} , \\
c_1(\mathbf{k}) &= c_2(\mathbf{k}) = c_3(\mathbf{k}) = \frac{c_2^{(2)}}{3}\, k^2 , \\
c_4(\mathbf{k}) &= c_5(\mathbf{k}) = 0 .
\end{align}
It follows that
\begin{align}
\langle c,c\rangle &= \big(c_0^{(0)}\big)^2 - \frac{\big(c_2^{(2)}\big)^2}{9}\, k^4 , \\
\langle c,f^1\rangle &= \langle c,f^2\rangle = 0 .
\end{align}
The analysis is thus analogous to the previous two examples with $E_{g+}$ pairing. The point nodes are inflated into BFSs, which touch the normal-state Fermi surface. Hence, the inclusion of nonlocal pairing does not affect the phenomenology for this pairing state. Note that only two of the six contributions (a)--(f) lead to inflation in the $[111]$ direction, namely the local $A_{1g+}\otimes E_{g+}$ contribution and the $T_{2g+}\otimes T_{2g+}$ contribution. For this to happen, there must be a pair of nonzero amplitudes with nontrivial phase difference in at least one of these two channels.

\subsubsection{$T_{1g+}$ pairing}

The irrep $T_{1g+}$ provides an example for a pairing state that cannot occur for local pairing in the $j=3/2$ model but unlike $A_{2g+}$ is multidimensional. $T_{1g+}$ pairing emerges in seven places in Table \ref{tab.j32.reduce}: (a) $A_{2g+}\otimes T_{2g+}$, (b) $E_{g+}\otimes T_{2g+}$, (c) $T_{1g+}\otimes A_{1g+}$, (d) $T_{1g+}\otimes T_{2g+}$, (e) $T_{1g+}\otimes E_{g+}$, (f) $T_{2g+}\otimes T_{2g+}$, and (g) $T_{2g+}\otimes E_{g+}$. We give the the matrix-valued basis functions to the leading order only.

(a) For $A_{2g+}\otimes T_{2g+}$:
\begin{align}
& D_{x,A_{2g+}}(\mathbf{k}) \nonumber\\
&\quad\cong \big[ k_x^4 (k_y^2-k_z^2) + k_y^4 (k_z^2-k_x^2) + k_z^4 (k_x^2-k_y^2) \big]\, h_1 , \\
& D_{y,A_{2g+}}(\mathbf{k}) \nonumber \\
&\quad\cong \big[ k_x^4 (k_y^2-k_z^2) + k_y^4 (k_z^2-k_x^2) + k_z^4 (k_x^2-k_y^2) \big]\, h_2 , \\
& D_{z,A_{2g+}}(\mathbf{k}) \nonumber \\
&\quad\cong \big[ k_x^4 (k_y^2-k_z^2) + k_y^4 (k_z^2-k_x^2) + k_z^4 (k_x^2-k_y^2) \big]\, h_3 .
\end{align}

(b) For $E_{g+}\otimes T_{2g+}$:
\begin{align}
D_{x,E_{g+}}(\mathbf{k}) &\cong (k_y^2-k_z^2)\, h_1 , \\
D_{y,E_{g+}}(\mathbf{k}) &\cong (k_z^2-k_x^2)\, h_2 , \\
D_{z,E_{g+}}(\mathbf{k}) &\cong (k_x^2-k_y^2)\, h_3 .
\end{align}

(c) For $T_{1g+}\otimes A_{1g+}$:
\begin{align}
D_{x,T_{1g+}}(\mathbf{k}) &\cong k_y k_z (k_y^2-k_z^2)\, h_0 , \\
D_{y,T_{1g+}}(\mathbf{k}) &\cong k_z k_x (k_z^2-k_x^2)\, h_0 , \\
D_{z,T_{1g+}}(\mathbf{k}) &\cong k_x k_y (k_x^2-k_y^2)\, h_0 .
\end{align}

(d) For $T_{1g+}\otimes T_{2g+}$:
\begin{align}
D^{\prime}_{x,T_{1g+}}(\mathbf{k}) &\cong k_z k_x (k_z^2-k_x^2)\, h_3 + k_x k_y (k_x^2-k_y^2)\, h_2 ,
  \\
D^{\prime}_{y,T_{1g+}}(\mathbf{k}) &\cong k_x k_y (k_x^2-k_y^2)\, h_1 + k_y k_z (k_y^2-k_z^2)\, h_3 ,
  \\
D^{\prime}_{z,T_{1g+}}(\mathbf{k}) &\cong k_y k_z (k_y^2-k_z^2)\, h_2 + k_z k_x (k_z^2-k_x^2)\, h_1 .
\end{align}

(e) For $T_{1g+}\otimes E_{g+}$:
\begin{align}
D^{\prime\prime}_{x,T_{1g+}}(\mathbf{k}) &\cong k_y k_z (k_y^2-k_z^2)\bigg(\frac{\sqrt{3}}{2}\, h_4
  - \frac{1}{2} \, h_5\bigg) , \\
D^{\prime\prime}_{y,T_{1g+}}(\mathbf{k}) &\cong -k_z k_x (k_z^2-k_x^2)\bigg(\frac{\sqrt{3}}{2}\, h_4
  + \frac{1}{2} \, h_5\bigg) , \\
D^{\prime\prime}_{z,T_{1g+}}(\mathbf{k}) &\cong k_x k_y (k_x^2-k_y^2)\, h_5 .
\end{align}

(f) For $T_{2g+}\otimes T_{2g+}$:
\begin{align}
D_{x,T_{2g+}}(\mathbf{k}) &\cong k_z k_x \, h_3 - k_x k_y \, h_2 , \\
D_{y,T_{2g+}}(\mathbf{k}) &\cong k_x k_y \, h_1 - k_y k_z \, h_3 , \\
D_{z,T_{2g+}}(\mathbf{k}) &\cong k_y k_z \, h_2 - k_z k_x \, h_1 .
\end{align}

(g) For $T_{2g+}\otimes E_{g+}$:
\begin{align}
D^{\prime}_{x,T_{2g+}}(\mathbf{k}) &\cong -k_y k_z
  \bigg(\frac{1}{2} \, h_4 + \frac{\sqrt{3}}{2}\, h_5\bigg) , \\
D^{\prime}_{y,T_{2g+}}(\mathbf{k}) &\cong  k_z k_x
  \bigg({-}\frac{1}{2} \, h_4 + \frac{\sqrt{3}}{2}\, h_5\bigg) , \\
D^{\prime}_{z,T_{2g+}}(\mathbf{k}) &\cong  k_x k_y \, h_4 .
\end{align}

\begin{table}
\caption{\label{tab.f.j32.T1g}Leading-order polynomial forms of the form factors $f_n(\mathbf{k})$ describing $T_{1g+}$ pairing for electrons with angular momentum $j=3/2$.}
\begin{ruledtabular}
\begin{tabular}{ll}
$n$ & $f_n$ \\ \hline
$0$ & $\delta_{x,T_{1g+}} k_y k_z (k_y^2-k_z^2)
  + \delta_{y,T_{1g+}} k_z k_x (k_z^2-k_x^2)$ \\
  & ${}+ \delta_{z,T_{1g+}} k_x k_y (k_x^2-k_y^2)$ \\
$1$ & $\delta_{x,A_{2g+}} \big[k_x^4 (k_y^2-k_z^2) + k_y^4 (k_z^2-k_x^2) + k_z^4 (k_x^2-k_y^2)\big]$ \\
  & ${}+ \delta_{y,T_{1g+}'} k_x k_y (k_x^2-k_y^2)
  + \delta_{z,T_{1g+}'} k_z k_x (k_z^2-k_x^2)$ \\
  & ${}+ \delta_{y,T_{2g+}} k_x k_y - \delta_{z,T_{2g+}} k_z k_x
  + \delta_{x,E_{g+}} (k_y^2-k_z^2)$ \\
$2$ & $\delta_{y,A_{2g+}} \big[k_x^4 (k_y^2-k_z^2) + k_y^4 (k_z^2-k_x^2) + k_z^4 (k_x^2-k_y^2)\big]$ \\
  & ${}+ \delta_{z,T_{1g+}'} k_y k_z (k_y^2-k_z^2)
  + \delta_{x,T_{1g+}'} k_x k_y (k_x^2-k_y^2)$ \\
  & ${}+ \delta_{z,T_{2g+}} k_y k_z - \delta_{x,T_{2g+}} k_x k_y
  + \delta_{y,E_{g+}} (k_z^2-k_x^2)$ \\
$3$ & $\delta_{z,A_{2g+}} \big[k_x^4 (k_y^2-k_z^2) + k_y^4 (k_z^2-k_x^2) + k_z^4 (k_x^2-k_y^2)\big]$ \\
  & ${}+ \delta_{x,T_{1g+}'} k_z k_x (k_z^2-k_x^2)
  + \delta_{y,T_{1g+}'} k_y k_z (k_y^2-k_z^2)$ \\
  & ${}+ \delta_{x,T_{2g+}} k_z k_x - \delta_{y,T_{2g+}} k_y k_z
  + \delta_{z,E_{g+}} (k_x^2-k_y^2)$ \\
$4$ & $\frac{\sqrt{3}}{2} \big[\delta_{x,T_{1g+}''} k_y k_z (k_y^2-k_z^2)
  - \delta_{y,T_{1g+}''} k_z k_x (k_z^2-k_x^2)\big]$ \\
  & ${}- \frac{1}{2}(\delta_{x,T_{2g+}'} k_y k_z
  + \delta_{y,T_{2g+}'} k_z k_x) + \delta_{z,T_{2g+}'} k_x k_y$ \\
$5$ & $-\frac{1}{2} \big[\delta_{x,T_{1g+}''} k_y k_z (k_y^2-k_z^2)
  + \delta_{y,T_{1g+}''} k_z k_x (k_z^2-k_x^2)\big]$ \\
  & ${}+ \delta_{z,T_{1g+}''} k_x k_y(k_x^2-k_y^2)
  - \frac{\sqrt{3}}{2}(\delta_{x,T_{2g+}'} k_y k_z - \delta_{y,T_{2g+}'} k_z k_x)$ \\
\end{tabular}
\end{ruledtabular}
\end{table}

\begin{table}
\caption{\label{tab.cf.j32.T1g}Leading-order polynomial forms of the products $c_n(\mathbf{k}) f_n(\mathbf{k})$ of form factors describing $T_{1g+}$ pairing for electrons with angular momentum $j=3/2$. The amplitudes of the leading terms in $c_n(\mathbf{k})$ have been absorbed into new pairing amplitudes marked by a tilde.}
\begin{ruledtabular}
\begin{tabular}{@{}ll@{}}
$n$ & $c_n f_n$ \\ \hline
$0$  & $\tdelta_{x,T_{1g+}} k_y k_z (k_y^2-k_z^2)+\,
  \tdelta_{y,T_{1g+}} k_z k_x (k_z^2-k_x^2)$ \\
  & ${}+ \tdelta_{z,T_{1g+}} k_x k_y (k_x^2-k_y^2)$ \\
$1$ & $\tdelta_{x,A_{2g+}}\,k_y k_z\, \big[k_x^4 (k_y^2-k_z^2)
  + k_y^4 (k_z^2-k_x^2) + k_z^4 (k_x^2-k_y^2)\big]$ \\
  & ${}+ \tdelta_{y,T_{1g+}'} k_x k_y^2 k_z (k_x^2-k_y^2)
  + \tdelta_{z,T_{1g+}'} k_x k_y k_z^2 (k_z^2-k_x^2)$ \\
  & ${}+ \tdelta_{y,T_{2g+}} k_x k_y^2 k_z
  - \tdelta_{z,T_{2g+}} k_x k_y k_z^2 + \tdelta_{x,E_{g+}}\,k_y k_z (k_y^2-k_z^2)$ \\ 
$2$ & $\tdelta_{y,A_{2g+}} k_z k_x \big[k_x^4 (k_y^2-k_z^2)
  + k_y^4 (k_z^2-k_x^2) + k_z^4 (k_x^2-k_y^2)\big]$ \\
  & ${}+ \tdelta_{z,T_{1g+}'} k_x k_y k_z^2 (k_y^2-k_z^2)
  + \tdelta_{x,T_{1g+}'} k_x^2 k_y k_z (k_x^2-k_y^2)$ \\
  & ${}+ \tdelta_{z,T_{2g+}} k_x k_y k_z^2
  - \tdelta_{x,T_{2g+}} k_x^2 k_y k_z
  + \tdelta_{y,E_{g+}} k_z k_x (k_z^2-k_x^2)$ \\
$3$ & $\tdelta_{z,A_{2g+}} k_x k_y \big[k_x^4 (k_y^2-k_z^2)
  + k_y^4 (k_z^2-k_x^2) + k_z^4 (k_x^2-k_y^2)\big]$ \\
  & ${}+ \tdelta_{x,T_{1g+}'} k_x^2 k_y k_z (k_z^2-k_x^2)
  + \tdelta_{y,T_{1g+}'} k_x k_y^2 k_z (k_y^2-k_z^2)$ \\
  & ${}+ \tdelta_{x,T_{2g+}} k_x^2 k_y k_z - \tdelta_{y,T_{2g+}} k_x k_y^2 k_z
  + \tdelta_{z,E_{g+}} k_x k_y (k_x^2-k_y^2)$ \\
$4$ & $\frac{\sqrt{3}}{2} \big[\tdelta_{x,T_{1g+}''} k_y k_z (k_y^2-k_z^2)
  - \tdelta_{y,T_{1g+}''} k_z k_x (k_z^2-k_x^2)\big] (k_x^2-k_y^2)$ \\
  & ${}- \frac{1}{2}(\tdelta_{x,T_{2g+}'} k_y k_z
  + \tdelta_{y,T_{2g+}'} k_z k_x) (k_x^2-k_y^2)$ \\
  & ${}+ \tdelta_{z,T_{2g+}'} k_x k_y (k_x^2-k_y^2)$ \\
$5$ & $- \frac{1}{2\sqrt{3}} \big[\tdelta_{x,T_{1g+}''} k_y k_z (k_y^2-k_z^2)
  + \tdelta_{y,T_{1g+}''} k_z k_x (k_z^2-k_x^2)\big]$ \\
  & ${}\times (2 k_z^2-k_x^2-k_y^2)
  + \frac{\tdelta_{z,T_{1g+}''}}{\sqrt{3}} k_x k_y (k_x^2-k_y^2)
  (2 k_z^2-k_x^2-k_y^2)$ \\
  & ${}- \frac{1}{2}\, (\tdelta_{x,T_{2g+}'} k_y k_z
  - \tdelta_{y,T_{2g+}'} k_z k_x) (2 k_z^2-k_x^2-k_y^2)$ \\
\end{tabular}
\end{ruledtabular}
\end{table}

The analysis is analogous to the one for a system with two \textit{s}-orbitals; see Sec.\ \ref{subsub.2s.T1g}. The resulting superconducting form factors $f_n(\mathbf{k})$ are given in Table \ref{tab.f.j32.T1g} and the products $c_n(\mathbf{k}) f_n(\mathbf{k})$ appearing in the condition (\ref{2.cfN4.4}) for IP nodes are shown in Table \ref{tab.cf.j32.T1g}. Hence,
\begin{align}
c_0&(\mathbf{k})\, f_0(\mathbf{k}) - \vec c(\mathbf{k}) \cdot \vec f(\mathbf{k}) \nonumber \\
  &= \bigg[ \tdelta_{x,T_{1g+}} - \tdelta_{x,E_{g+}}
  + \tdelta_{x,A_{2g+}} (k_z^2-k_x^2)(k_x^2-k_y^2) \nonumber \\
&{}+ \tdelta_{x,T_{1g+}'}\, k_x^2
  - \frac{\tdelta_{x,T_{1g+}''}}{\sqrt{3}}\, (2 k_x^2-k_y^2-k_z^2) \nonumber \\
&{}- \tdelta_{x,T_{2g+}'} \bigg]\,  k_y k_z (k_y^2-k_z^2) + \ldots ,
\end{align}
where two terms with cyclically permuted indices $x$, $y$, and $z$ have been suppressed. For broken TRS, the pairing states $(1,i,0)$ and $(1,\omega,\omega^2)$ with $\omega=e^{2\pi i/3}$, are plausible \cite{VoG85,SiU91,BWW16}. We here consider the simpler $(1,i,0)$ state, which has 18 point nodes in the $\langle 001\rangle$, $\langle 101\rangle$, and $\langle 111\rangle$ directions and one line node in the $k_z=0$ plane.

For noninfinitesimal pairing, we expect the nodes to be inflated. For the nodes along the $[001]$ direction, the form factors read as
\begin{align}
f_0(\mathbf{k}) &= f_3(\mathbf{k}) = f_4(\mathbf{k}) = f_5(\mathbf{k}) = 0 , \\
f_1(\mathbf{k}) &= -\delta_{E_{g+}} k^2 , \\
f_2(\mathbf{k}) &= i\delta_{E_{g+}} k^2 .
\end{align}
We find $\langle f^1, f^1 \rangle=\langle f^2, f^2 \rangle=-\delta^2_{E_{g+}} k^4$ and $\langle f^1, f^2 \rangle=0$ as well as $\langle c^1, f^1 \rangle=\langle c^2, f^2 \rangle=0$. In this case, the Pfaffian simplifies to
\begin{align}
P(\mathbf{k})=\langle c, c \rangle\, \big(\langle c, c \rangle + 4 \delta^2_{E_{g+}} k^4 \big) .
\end{align}
The first factor changes sign at the normal-state Fermi surface but the second factor does not and thus the Pfaffian changes sign at the normal-state Fermi surface. Hence, the point nodes in the $\langle 001 \rangle$ directions are inflated and remain attached to the normal-state Fermi surfaces for arbitrarily strong coupling.

In the $[101]$ direction, we have $\mathbf{k}=k(1,0,1)/\sqrt{2}$ and the form factors read as
\begin{align}
f_0(\mathbf{k}) &= f_1(\mathbf{k})=f_2(\mathbf{k})= 0 , \\
f_3(\mathbf{k}) &= \frac{\delta_{T_{2g+}}}{2}\, k^2 , \\
f_4(\mathbf{k}) &= - i\, \frac{\delta_{T_{2g+}'}}{4}\, k^2  ,  \\
f_5(\mathbf{k}) &= i\, \frac{\sqrt{3}\,\delta_{T_{2g+}'}}{4}\, k^2 .
\end{align}
This implies $\langle f^1, f^2 \rangle =0$ but $\langle f^1, f^1 \rangle$ and $\langle f^2, f^2 \rangle$ are generally unequal. The Pfaffian is thus $P(\mathbf{k}) = \left[ \langle c,c\rangle - \langle f^1,f^1\rangle - \langle f^2,f^2\rangle \right]^2 - 4\, \langle f^1,f^1\rangle \langle f^2,f^2\rangle$. The first term vanishes on a renormalized Fermi surface. The second term
\begin{equation}
- 4\, \langle f^1,f^1\rangle \langle f^2,f^2\rangle
  = -\frac{1}{4}\, \delta_{T_{2g+}}^2 \delta_{T_{2g+}'}^2 k^8
\end{equation}
is generically negative. Hence, the Pfaffian generically becomes negative in the vicinity of the normal-state Fermi surface but the BFS does not usually touch it. The Pfaffian is identical for all $\langle 101\rangle$ directions that are not in the $k_z=0$ plane.

Along $[111]$, we have $\mathbf{k}=k\,(1,1,1)/\sqrt{3}$ and
\begin{align}
f_0(\mathbf{k}) &= 0 , \\
f_1(\mathbf{k}) &= i\, \frac{\delta_{T_{2g+}}}{3} \,k^2 , \\
f_2(\mathbf{k}) &= - \frac{\delta_{T_{2g+}}}{3} \,k^2 , \\
f_3(\mathbf{k}) &= \frac{\delta_{T_{2g+}}}{3} \,k^2- i\frac{\delta_{T_{2g+}}}{3} \,k^2 , \\
f_4(\mathbf{k}) &= - \frac{\delta_{T_{2g+}'}}{6} \,k^2- i \frac{\delta_{T_{2g+}'}}{6} \,k^2 , \\
f_5(\mathbf{k}) &= - \frac{\delta_{T_{2g+}'}}{2\sqrt{3}} \,k^2
  + i \frac{\delta_{T_{2g+}'}}{2\sqrt{3}} \,k^2 .
\end{align}
We thus obtain
\begin{equation}
\langle f^1,f^2\rangle = \frac{k^4}{18}\, \big( 2 \delta_{T_{2g+}}^2
  + \delta_{T_{2g+}'}^2 \big)
\end{equation}
and
\begin{equation}
\langle f^1,f^1\rangle = \langle f^2,f^2\rangle
  = -\frac{k^4}{9}\, \big( 2 \delta_{T_{2g+}}^2 + \delta_{T_{2g+}'}^2 \big) .
\end{equation}
The Pfaffian is
\begin{align}
P(\mathbf{k}) &= \left[ \langle c,c\rangle - \langle f^1,f^1\rangle - \langle f^2,f^2\rangle \right]^2
  \nonumber \\
&\quad{} + 4 \left[ \langle f^1,f^2\rangle^2 - \langle f^1,f^1\rangle \langle f^2,f^2\rangle \right] ,
\label{3.T1g.line.inf.3x}
\end{align}
wherein the second term evaluates to
\begin{equation}
-\frac{k^8}{27}\, \big( 2 \delta_{T_{2g+}}^2 + \delta_{T_{2g+}'}^2 \big)^2 .
\end{equation}
Since this is generally negative we also expect the nodes to inflate in the $\langle 111\rangle$ directions but they are not attached to the normal-state Fermi surface.

For the equatorial line node, we take $\mathbf{k}=(k_x,k_y,0)$. The superconducting form factors are
\begin{align}
f_0(\mathbf{k}) &= f_3(\mathbf{k}) = f_4(\mathbf{k}) = f_5(\mathbf{k}) = 0 , \\
f_1(\mathbf{k}) &= \delta_{A_{2g+}} (k_x^4 k_y^2-k_y^4 k_x^2)+\delta_{E_{g+}}\, k_y^2
  + i \delta_{T_{2g+}} k_x k_y \nonumber \\
&\quad{} + i \delta_{T_{1g+}'} k_x k_y (k_x^2-k_y^2) \\
f_2(\mathbf{k}) &= i\delta_{A_{2g+}} (k_x^4 k_y^2-k_y^4 k_x^2)-i \delta_{E_{g+}}\, k_x^2
  - \delta_{T_{2g+}}\, k_x k_y \nonumber \\
&\quad{} + \delta_{T_{1g+}'} k_x k_y (k_x^2-k_y^2) .
\end{align}
We find $\langle f^1,f^2\rangle\neq 0$ and $\langle f^1,f^1\rangle \neq \langle f^2,f^2\rangle$. The Pfaffian thus again has the form of Eq.\ (\ref{3.T1g.line.inf.3x}). The second term reads as
\begin{align}
& - 4 k_x^2 k_y^2\, \big[ \delta_{T_{2g+}}^2 - \delta_{E_{g+}}^2 
  - \delta_{A_{2g+}} \delta_{E_{g+}} (k_x^2-k_y^2)^2 \nonumber \\
&\quad{}+ \big( \delta_{A_{2g+}}^2 k_x^2 k_y^2
  - \delta_{T_{1g+}'}^2 \big) (k_x^2 + k_y^2) (k_x^2 - k_y^2) \big]^2
\label{3.T1g.eq.P2}
\end{align}
and is generically negative. We conclude that the equatorial line node is inflated by noninfinitesimal pairing. The resulting BFS is toroidal but pinched on the $k_x$ and $k_y$ axes since Eq.\ (\ref{3.T1g.eq.P2}) gives zero there. Since the first term in Eq.\ (\ref{3.T1g.line.inf.3x}) becomes zero close to but not at the normal-state Fermi surface the BFS generically does not touch the normal-state Fermi surface. This also holds on the $k_x$ and $k_y$ axes. The behavior of the nodes is identical to the case of $T_{1g+}$ pairing for two \textit{s}-orbitals.

\subsection{Orbital doublet}
\label{sub.Eg}

The discussion in Sec.\ \ref{sub.opposite} suggests how to construct further orbital models: the set of orbitals must be closed under the action of the point group. This implies that the orbitals must transform like basis functions of one-dimensional irreps or as complete sets of basis functions of multidimensional irreps. In the previous examples, we have considered two $A_{1g}$ orbitals and one $A_{1g}$ and one $A_{2u}$ orbital. As the simplest example with a multidimensional irrep we here analyze the case of two orbitals transforming like basis functions of $E_g$. This is naturally realized by a doublet of $e_g$ orbitals ($d_{x^2-y^2}$ and $d_{3z^2-r^2}$) per site. Since they are of even parity we have $U_P = \sigma_0 \otimes \sigma_0$. Also, $U_T = \sigma_0 \otimes i\sigma_2$ holds.

The new aspect here is that the orbital part alone can have higher-dimensional irreps. The irreps of Pauli matrices in orbital space are the following: $\eta_0 \equiv \sigma_0$ belongs to $A_{1g+}$. The two matrices $\eta_1 \equiv \sigma_1$ and $\eta_2 \equiv -\sigma_3$ form an $E_{g+}$ doublet. It is easy to check that under rotations $\eta_1$ and $\eta_2$ transform like $k_x^2-k_y^2$ and $(2k_z^2-k_x^2-k_y^2)/\sqrt{3}$, respectively. Finally, $\eta_3 \equiv \sigma_2$ belongs to $A_{2g-}$.

In spin space, $\sigma_0$ of course transforms according to $A_{1g+}$ and $(\sigma_1,\sigma_2,\sigma_3)$ form a $T_{1g-}$ triplet. Combining higher-dimensional irreps from the orbital and spin parts, we obtain reducible product representations. Thus Kronecker products of Pauli matrices in orbital and spin space have to be linearly combined to construct the proper basis matrices. (Such a construction is also implicit in Table \ref{tab.j32.basis} of basis matrices for the $j=3/2$ case above.) Specifically, we require the nontrivial reduction $E_{g+} \otimes T_{1g-} = T_{1g-} \oplus T_{2g-}$. The resulting basis matrices are presented in Table~\ref{tab.eg.basis}.

\begin{table}
\caption{\label{tab.eg.basis}Basis matrices on the internal Hilbert space 
for the case of an $E_g$ doublet of orbitals with spin and point group $O_h$. The basis matrices are irreducible tensor operators of the irreps listed in the second column. For multidimensional irreps, the states transforming into each other under point-group operations are distinguished by the index in the third column.}
\begin{ruledtabular}
\begin{tabular}{@{\hspace{1em}}lcc@{\hspace{1em}}}
$h_\nu$ & Irrep & Component \\ \hline
$\eta_0 \otimes \sigma_0$ & $A_{1g+}$ & \\
$\eta_0 \otimes \sigma_1$ & $T_{1g-}$ & 1 \\
$\eta_0 \otimes \sigma_2$ & & 2 \\
$\eta_0 \otimes \sigma_3$ & & 3 \\
$\eta_1 \otimes \sigma_0$ & $E_{g+}$ & 1 \\
$\eta_2 \otimes \sigma_0$ & & 2 \\
$-\frac{1}{2}\, \eta_1 \otimes \sigma_1 - \frac{\sqrt{3}}{2}\, \eta_2 \otimes \sigma_1$ & $T_{2g-}$ & 1 \\
$-\frac{1}{2}\, \eta_1 \otimes \sigma_2 + \frac{\sqrt{3}}{2}\, \eta_2 \otimes \sigma_2$ & & 2 \\
$\eta_1 \otimes \sigma_3$ & & 3 \\
$\frac{\sqrt{3}}{2}\, \eta_1 \otimes \sigma_1 - \frac{1}{2}\, \eta_2 \otimes \sigma_1$ & $T_{1g-}$ & 1 \\
$-\frac{\sqrt{3}}{2}\, \eta_1 \otimes \sigma_2 - \frac{1}{2}\, \eta_2 \otimes \sigma_2$ & & 2 \\
$\eta_2 \otimes \sigma_3$ & & 3 \\
$\eta_3 \otimes \sigma_0$ & $A_{2g-}$ \\
$\eta_3 \otimes \sigma_1$ & $T_{2g+}$ & 1 \\
$\eta_3 \otimes \sigma_2$ & & 2 \\
$\eta_3 \otimes \sigma_3$ & & 3 \\
\end{tabular}
\end{ruledtabular}
\end{table}

The basis matrices relevant for the normal-state Hamiltonian and for even-parity pairing are
\begin{align}
h_0 &\equiv \eta_0 \otimes \sigma_0 & A_{1g+}, \\
h_1 &\equiv \eta_3 \otimes \sigma_1 & T_{2g+}, \\
h_2 &\equiv \eta_3 \otimes \sigma_2 & T_{2g+}, \\
h_3 &\equiv \eta_3 \otimes \sigma_3 & T_{2g+}, \\
h_4 &\equiv \eta_1 \otimes \sigma_0 & E_{g+}, \\
h_5 &\equiv \eta_2 \otimes \sigma_0 & E_{g+}.
\end{align}
The symmetry properties are thus the same as for the $j=3/2$ case. Hence, the analysis of pairing states is completely analogous to Sec.\ \ref{sub.j32}, except for the different definition of the basis matrices~$h_n$.

\subsection{Orbital triplet}
\label{sub.T2g}

\begin{table}
\caption{\label{tab.GM.basis}Classification of the Gell-Mann matrices acting on orbital space for a $T_{2g}$ orbital triplet and point group $O_h$. The matrices are irreducible tensor operators of the irreps listed in the second column. For multidimensional irreps, the states transforming into each other under point-group operations are distinguished by the index in the third column.}
\begin{ruledtabular}
\begin{tabular}{@{\hspace{3em}}ccc@{\hspace{3em}}}
Matrix & Irrep & Component \\ \hline
$\lambda_0$ & $A_{1g+}$ & \\
$\lambda_3$ & $T_{2g+}$ & 1 \\
$\lambda_2$ & & 2 \\
$\lambda_1$ & & 3 \\
$\lambda_6$ & $T_{1g-}$ & 1 \\
$-\lambda_5$ & & 2 \\
$\lambda_4$ & & 3 \\
$\lambda_7$ & $E_{g+}$ & 1 \\
$-\lambda_8$ & & 2 \\
\end{tabular}
\end{ruledtabular}
\end{table}

We briefly consider the case of three $t_{2g}$ orbitals per site of a cubic lattice, i.e., $d_{yz}$, $d_{zx}$, and $d_{xy}$ orbitals. In this example, the dimension of the internal Hilbert space is $N=6$ and according to Appendix \ref{app.normal} there are 15 basis matrices for the normal-state Hamiltonian and even-parity superconductivity. Their algebra is much more complicated than for $N=4$. In particular, they do not anticommute pairwise. We require a basis of $3\times 3$ matrices that act on the orbital space. We take the Gell-Mann matrices~\cite{SMB20}
\begin{align}
\lambda_0 = &\begin{pmatrix} 1 & 0 & 0 \\ 0 & 1 & 0 \\ 0 & 0 & 1 \end{pmatrix} \!, \\
\lambda_1 = \begin{pmatrix} 0 & 1 & 0 \\ 1 & 0 & 0 \\ 0 & 0 & 0 \end{pmatrix} \!,\:
\lambda_2 = &\begin{pmatrix} 0 & 0 & 1 \\ 0 & 0 & 0 \\ 1 & 0 & 0 \end{pmatrix} \!,\:
\lambda_3 = \begin{pmatrix} 0 & 0 & 0 \\ 0 & 0 & 1 \\ 0 & 1 & 0 \end{pmatrix} \!, \\
\lambda_4 = \begin{pmatrix} 0 & -i & 0 \\ i & 0 & 0 \\ 0 & 0 & 0 \end{pmatrix} \!,\:
\lambda_5 = &\begin{pmatrix} 0 & 0 & -i \\ 0 & 0 & 0 \\ i & 0 & 0 \end{pmatrix} \!,\:
\lambda_6 = \begin{pmatrix} 0 & 0 & 0 \\ 0 & 0 & -i \\ 0 & i & 0 \end{pmatrix} \!, \\
\lambda_7 = \begin{pmatrix} 1 & 0 & 0 \\ 0 & -1 & 0 \\ 0 & 0 & 0 \end{pmatrix} \!,\hspace{0.35em}
&\lambda_8 = \frac{1}{\sqrt{3}} \begin{pmatrix} 1 & 0 & 0 \\ 0 & 1 & 0 \\ 0 & 0 & -2 \end{pmatrix} \!.
\end{align}
The associated irreps are given in Table \ref{tab.GM.basis}. The Gell-Mann matrices have to be combined with the Pauli matrices acting on spin space to form basis matrices for the combined internal degrees of freedom. $\sigma_0$ of course transforms according to $A_{1g+}$ and $(\sigma_1,\sigma_2,\sigma_3)$ to $T_{1g-}$. Possible basis matrices for $H_N(\mathbf{k})$ and even-parity pairing must belong to $g+$ irreps. They can thus be constructed by combining $\lambda_j$ for $j \in \{0,1,2,3,7,8\}$ with $\sigma_0$ and by combining $\lambda_j$ with $j \in \{4,5,6\}$ with $\sigma_1$, $\sigma_2$, or $\sigma_3$. The relevant product representations are either trivial or involve the reduction $T_{1g-} \otimes T_{1g-} = A_{1g+} \oplus E_{g+} \oplus T_{1g+} \oplus T_{2g+}$. The 15 allowed basis matrices and the associated irreps are shown in Table \ref{tab.t2g.basis}. Here, we have not chosen any special normalization, except that the entries belonging to the same multiplets have consistent numerical factors. There are also 21 basis matrices belonging to $g-$ irreps, which are disallowed as pairing matrices and are not listed for simplicity. 

\begin{table}
\caption{\label{tab.t2g.basis}Basis matrices on the internal Hilbert space for the case of an $T_{2g}$ triplet of orbitals with spin and point group $O_h$. Unlike for the previous examples, only the basis matrices $h_n$ allowed in the normal-state Hamiltonian and for even-parity pairing are listed. The basis matrices are irreducible tensor operators of the irreps listed in the second column. For multidimensional irreps, the states transforming into each other under point-group operations are distinguished by the index in the third column.}
\begin{ruledtabular}
\begin{tabular}{lcc}
$h_n$ & Irrep & Component \\ \hline
$h_0 \equiv \lambda_0 \otimes \sigma_0$ & $A_{1g+}$ & \\
$h_1 \equiv \lambda_3 \otimes \sigma_0$ & $T_{2g+}$ & 1 \\
$h_2 \equiv \lambda_2 \otimes \sigma_0$ & & 2 \\
$h_3 \equiv \lambda_1 \otimes \sigma_0$ & & 3 \\
$h_4 \equiv \lambda_7 \otimes \sigma_0$ & $E_{g+}$ & 1 \\
$h_5 \equiv -\lambda_8 \otimes \sigma_0$ & & 2 \\
$h_6 \equiv \lambda_6 \otimes \sigma_1 - \lambda_5 \otimes \sigma_2
  + \lambda_4 \otimes \sigma_3$ & $A_{1g+}$ & \\
$h_7 \equiv \lambda_6 \otimes \sigma_1 + \lambda_5 \otimes \sigma_2$ & $E_{g+}$ & 1 \\
$h_8 \equiv \frac{1}{\sqrt{3}}\, (2\lambda_4 \otimes \sigma_3 - \lambda_6 
\otimes \sigma_1
  + \lambda_5 \otimes \sigma_2)$ & & 2 \\
$h_9 \equiv -\lambda_5 \otimes \sigma_3 - \lambda_4 \otimes \sigma_2$ & $T_{1g+}$ & 1 \\
$h_{10} \equiv \lambda_4 \otimes \sigma_1 - \lambda_6 \otimes \sigma_3$ & 
& 2 \\
$h_{11} \equiv \lambda_6 \otimes \sigma_2 + \lambda_5 \otimes \sigma_1$ & 
& 3 \\
$h_{12} \equiv -\lambda_5 \otimes \sigma_3 + \lambda_4 \otimes \sigma_2$ & $T_{2g+}$ & 1 \\
$h_{13} \equiv \lambda_4 \otimes \sigma_1 + \lambda_6 \otimes \sigma_3$ & 
& 2 \\
$h_{14} \equiv \lambda_6 \otimes \sigma_2 - \lambda_5 \otimes \sigma_1$ & 
& 3 \\
\end{tabular}
\end{ruledtabular}
\end{table}

The normal-state Hamiltonian reads as
\begin{equation}
H_N(\mathbf{k}) = \sum_{n=0}^{14} c_n(\mathbf{k})\, h_n ,
\label{3.t2g.HN.3}
\end{equation}
where $c_n(\mathbf{k})$ transforms like $h_n$. For even-parity superconductivity, we can combine the 15 matrices $h_n$ with form factors $f_n(\mathbf{k})$ belonging to all the $g+$ irreps. This obviously generates pairing states for all $g+$ irreps. Any pairing state can be expressed in terms of a pairing matrix of the form
\begin{equation}
D(\mathbf{k}) = \sum_{n=0}^{14} f_n(\mathbf{k})\, h_n .
\label{3.t2g.D.3}
\end{equation}
Moreover, all even-parity pairing states other than $A_{2g+}$ can occur for purely local pairing, i.e., with constant form factors, because the basis matrices $h_n$ in Table \ref{tab.t2g.basis} cover all $g+$ irreps except $A_{2g+}$. Another interesting observation is that purely local pairing with trivial ($A_{1g+}$) symmetry now allows for an orbitally nontrivial contribution from~$h_6$.

All pairing states including nonlocal contributions can be constructed as described above. For the present case of $N=6$, we typically find a larger number of contributions than for $N=4$. For example, $E_{g+}$ pairing can result from the products (form factor times basis matrix) (a) $A_{1g+} \otimes E_{g+}$, (b) $A_{2g+} \otimes E_{g+}$, (c) $E_{g+} \otimes A_{1g+}$, (d) $E_{g+} \otimes E_{g+}$, (e) $T_{1g+} \otimes T_{1g+}$, (f) $T_{1g+} \otimes T_{2g+}$, (g) $T_{2g+} \otimes T_{1g+}$, and (h) $T_{2g+} \otimes T_{2g+}$.

The leading-order contribution of type (a) to the pairing matrix reads as
\begin{align}
D_{x^2-y^2,A_{1g+}}(\mathbf{k}) &\cong \delta_0\, \lambda_7 \otimes \sigma_0
  + \delta_1\, ( \lambda_6 \otimes \sigma_1 + \lambda_5 \otimes \sigma_2 ) , \\
D_{3z^2-r^2,A_{1g+}}(\mathbf{k}) &\cong -\delta_0\, \lambda_8 \otimes \sigma_0
  + \frac{\delta_1}{\sqrt{3}}\, ( 2\lambda_4 \otimes \sigma_3 \nonumber \\
&\quad{}- \lambda_6 \otimes \sigma_1 + \lambda_5 \otimes \sigma_2 ) ,
\end{align}
where one of the constants $\delta_0$ and $\delta_1$ could be set to unity. We omit the construction of the other contributions.

The expression $F(\mathbf{k}) = \sum_n c_n(\mathbf{k})\, f_n(\mathbf{k})$ transforms like the pairing matrix $D(\mathbf{k})$ and the condition $F(\mathbf{k})=0$ determines the location of IP nodes, as shown in Sec.\ \ref{sec.theory}. In principle, we can now obtain the form factors $c_n(\mathbf{k})$ appearing in Eq.\ (\ref{3.t2g.HN.3}) and $f_n(\mathbf{k})$ in Eq.\ (\ref{3.t2g.D.3}) and thus $F(\mathbf{k})$ and the IP nodal structure. Pairing of not infinitesimal amplitude, in particular the existence of BFSs, could then be analyzed following Sec.\ \ref{sec.theory}. This requires the calculation of the Pfaffian $\Pf \tilde{\mathcal{H}}(\mathbf{k})$. Due to the absence of simple algebraic relations between the basis matrices $h_n$, there is no simple analytical expression in terms of the functions $c_n(\mathbf{k})$ and $f_n(\mathbf{k})$ so that such an analysis would likely require a numerical study of $\Pf \tilde{\mathcal{H}}(\mathbf{k})$. We do not execute this program here. BFSs have been predicted for $E_{g+}$ pairing in an $N=6$ model for $\mathrm{Sr}_2\mathrm{RuO}_4$ \cite{SMB20}, which has the point group $D_{4h}$.

\section{Discussion and conclusions}
\label{sec.conclusions}

We have analyzed possible superconducting pairing symmetries for materials with local degrees of freedom beyond the electronic spin, such as orbital or basis site. Local degrees of freedom can enable unconventional pairing, in particular with BFSs. We find that this is the case even in the simple (and somewhat artificial) case of two \textit{s}-orbitals at each lattice site, where the orbital index appears to be a spectator: Such a model permits orbitally nontrivial pairing of $T_{1g}$ symmetry. The main step taken in this paper is to go beyond local pairing. This implies that not only the normal state is characterized by momentum-dependent form factors $c_n(\mathbf{k})$ but also the superconducting pairing is characterized by momentum-dependent form factors $f_n(\mathbf{k})$. In case of even-parity pairing, the functions $f_n(\mathbf{k})$ have the same symmetry properties as the $c_n(\mathbf{k})$.

There are three distinct types of contributions to pairing with nontrivial symmetry: (a) purely local pairing which is internally anisotropic, (b) internally isotropic pairing with nontrivial momentum-space form factors, and (c) contributions with nontrivial internal and momentum-space structure. Type (a) has been studied in Refs.\ \cite{ABT17,BAM18}, whereas type (b) is exemplified by $d_{x^2-y^2}$ pairing in cuprates. Types (a) and (c) are internally anisotropic, which means that the pairing matrix $\Delta(\mathbf{k}) = D(\mathbf{k})\, U_T$ acts nontrivially on the internal degrees of freedom. $D(\mathbf{k})$ then contains one or more basis matrices on the internal Hilbert space that are not proportional to the identity matrix.

Nonlocal pairing permits symmetries of pairing states that usually cannot all be realized for local pairing. In fact, \emph{all} even-parity ($g$) irreps of the magnetic point group occur for nonlocal pairing since they all have mo\-men\-tum-space basis functions and thus at least permit a contribution of type (b). In particular, there is always at least one basis matrix belonging to the trivial irrep, which is even, namely the identity matrix. Since the constant function of momentum also has full symmetry, local pairing with full, trivial symmetry is always allowed. Of course, it can be energetically suppressed by a local repulsive interaction.

The example of two orbitals of opposite parity reveals that if parity acts nontrivially on the internal degrees of freedom, basis matrices become possible that are odd under parity; see Sec.\ \ref{sub.opposite}. These basis matrices can combine with odd-in-momentum form factors to form even-parity superconducting states~\cite{Ber74,LiB19}.

Odd-parity superconductivity is characterized by $u+$ irreps (odd under inversion, even under time reversal), which requires $g-$ basis matrices and $u-$ form factors $f_\nu(\mathbf{k})$. All $u-$ irreps possess momentum-space basis functions and thus allow to write down such form factors. Moreover, inspection of multiplication tables for irreps shows that for any magnetic point group with inversion, the existence of one $g-$ basis matrix is sufficient to generate pairing states belonging to all $u+$ irreps.

The IP nodes are determined by both the normal-state and the corresponding superconducting form factors, $c_n(\mathbf{k})$ and $f_n(\mathbf{k})$, respectively, not by the superconducting factors alone. Specifically, the simple criterion $\sum_n c_n(\mathbf{k})\, f_n(\mathbf{k})=0$ diagnoses the presence of IP nodes. The position of the IP nodes is generically unaffected by nonlocal contributions to the extent that it is determined by symmetry. It is then only determined by the irrep of the pairing state. As a counterexample, for pairing belonging to the second ($3z^2-r^2$) component of $E_g$ for the point group $O_h$, there are always two line nodes but these do not lie in high-symmetry planes and are only constrained to pass through the $\langle 111\rangle$ directions.

Pairing states belonging to one-dimensional irreps can break TRS if the pairing has more than one contribution, which allows nontrivial phase factors. In the resulting TRS-breaking state, the symmetry-imposed line nodes of the time-reversal-symmetric state persist for infinitesimal pairing. The same mechanism is also possible for multidimensional irreps, in addition to the more natural case of phase factors between different components.

If the pairing strength is not infinitesimal the point or line nodes can be inflated into BFSs. Like for local pairing \cite{ABT17,BAM18}, the BFSs are given by the zeros of the Pfaffian $P(\mathbf{k})$ of the BdG Hamiltonian, unitarily transformed into antisymmetric form. There are three possible cases for the inflated nodes: (1) they are forced to contain the original node and thus remain attached to the normal-state Fermi surface, (2) they do not remain attached to the original node and thus generically shift away from the normal-state Fermi surface with increasing pairing strength, or (3) inflated line nodes remain attached to the original line node only in high-symmetry directions (we have only observed the situation that the inflation also vanishes there). For increasing pairing, the BFSs typically grow and eventually merge. In case (2), the merged pockets can eventually shrink and finally disappear if this does not violate any remaining nonzero topological invariants they possess \cite{BrS17,BAM18}. In real materials, this only seems likely if the BFSs are located \emph{inside} small normal-state Fermi pocket(s) of a poor metal. It is an intriguing open question whether the resulting fully gapped superconducting state is topologically nontrivial.

In principle, new BFSs can emerge at strong coupling, when quasiparticle bands are shifted through the Fermi energy. However, we expect that such BFSs are usually energetically disfavored and that the system can avoid them by developing a suitable momentum-dependent pairing amplitude.

The case of an internal Hilbert space of dimension $N=4$ is special \cite{HeL21}. We here only discuss the standard case where the internal degrees of freedom include the electron spin so that the transformation $P\mathcal{T}$ squares to $-\openone$. Then, there are exactly six basis matrices $h_n$ with simple algebraic properties: One of them, $h_0\propto\openone$, commutes with all others, which anticommute pairwise. This structures allows us to find relatively compact expressions for the Pfaffian $P(\mathbf{k})$ in terms of the form factors $c_n(\mathbf{k})$ and $f_n(\mathbf{k})$. For $N>4$, there is no comparable algebraic structure and the Pfaffian could only be given explicitly in terms of the components of the transformed Hamiltonian. The number of terms in the resulting expression is exponentially large in $N$.

It is useful to review and compare our results for one-dimensional $A_{2g+}$ pairing and multidimensional $T_{1g+}$ pairing, which illustrate some of our general remarks. Note that it is hard to envision local degrees of freedom that realize an operator with $A_{2g+}$ symmetry, ultimately because of the high minimum order $l=6$ of basis functions. Hence, purely local pairing is unlikely to exist and it indeed does not appear in the examples we have considered. For infinitesimal pairing, the $A_{2g+}$ pairing state for any model has six symmetry-imposed line nodes in the $\{110\}$ mirror planes. Since they are present in the real and imaginary parts of the gap function, they persist for TRS-broken states.

We find that beyond infinitesimal pairing, the nodes of the TRS-broken $A_{2g+}$ states either persist as line nodes or are inflated into BFSs. Specifically, the nodes are inflated in the mirror planes for the electrons with effective spin $j=3/2$ but not for the cases of two \textit{s}-orbitals and of two orbitals of opposite parity. The origin of this difference is that for the $j=3/2$ case, in the mirror planes, the two amplitudes $\delta_{Eg+}$ and $\delta_{T2g+}$ contribute to the Pfaffian whenever there is a phase difference between them. For the other two cases, there is a single amplitude in the mirror planes and its phase can be gauged away. Thus there is no inflation. A general insight here is that if only a single amplitude contributes in some high-symmetry direction or plane the breaking of TRS cannot lead to the formation of a BFS or of a gap since the physics is (gauge) invariant under changes of the phase of this amplitude. This argument also applies to multidimensional irreps.

TRS-breaking pairing states are more natural for multidimensional irreps. In our context, $T_{1g+}$ is an interesting pairing symmetry. Whereas it appears for purely local pairing in the case of two \textit{s}-orbitals, it only appears for nonlocal pairing for effective-spin-$3/2$ fermions. Our study suggests that for both cases all point and line nodes appearing for TRS-broken $T_{1g+}$ pairing state with order parameter $(1,i,0)$ are inflated for noninfinitesimal pairing. The point nodes on the $k_z$-axis are the only ones which remains attached to the normal-state Fermi surface, while all other point and line nodes are shifted away from the normal-state Fermi surface and thus could annihilate for strong coupling.

To conclude, nonlocal pairing typically permits a much larger number of possible pairing symmetries. Their nodal structure, including the possibility of BFSs, can be analyzed based on symmetry. For nodes at infinitesimal pairing, there is a simple yet powerful criterion in terms of a scalar product of form factors. The known criterion for the appearance of BFSs in terms of a Pfaffian extends to nonlocal pairing and general internal degrees of freedom and shows that BFSs generically exist if the superconducting state breaks TRS.

\acknowledgments

The authors thank D. F. Agterberg, P. M. R. Brydon, I. F. Herbut, A. Knoll, S. Kobayashi, and J. M. Link for useful discussions. Financial support by the Deut\-sche For\-schungs\-ge\-mein\-schaft through the Collaborative Research Center SFB 1143, Project A04, the Research Training Group GRK 1621, and the Cluster of Excellence on Complexity and Topology in Quantum Matter ct.qmat (EXC~2147) is gratefully acknowledged.

\appendix

\section{Enumeration of basis matrices}
\label{app.normal}

In this Appendix, we obtain the number of basis matrices that can occur and thus generically do occur in the normal-state Hamiltonian $H_N(\mathbf{k})$. The same basis matrices can appear in the pairing matrix $D(\mathbf{k})$ for $s_T=-1$ and even-parity superconductivity as well as for $s_T=+1$ and odd-parity superconductivity, while in the other two cases, only the remaining basis matrices can appear in $D(\mathbf{k})$. The statements can be obtained in a more general framework but it might be useful to present them using representation theory of point groups.

The following theorem is proven: Let $N$ be the dimension of the Hilbert space describing the local degrees of freedom in the normal state. Let $P$ be the unitary parity operator on this Hilbert space and let $\mathcal{T}$ be the antiunitary time-reversal operator. Then the number of Hermitian basis matrices $h_n$ that can appear in a normal-state Hamiltonian that respects $P\mathcal{T}$ symmetry is
\begin{equation}
n_h = \left\{ \begin{array}{ll}
    \displaystyle \frac{N(N+1)}{2} & \mbox{for $(P\mathcal{T})^2 = +1$} , \\[1.5ex]
    \displaystyle \frac{N(N-1)}{2} & \mbox{for $(P\mathcal{T})^2 = -1$} .
  \end{array} \right.
\end{equation}
As we shall see, the case $(P\mathcal{T})^2 = -1$ can only occur for even $N$. Results for small $N$ are given in Table~\ref{tab.numhn}.

\begin{table}
\caption{\label{tab.numhn}Number of basis matrices $h_n$ that appear in a normal-state Hamiltonian $H_N(\mathbf{k})$ with $P\mathcal{T}$ symmetry for the two cases that $P\mathcal{T}$ squares to $\pm 1$. $N$ is the dimension of the internal Hilbert space. For $(P\mathcal{T})^2=-1$, which is 
the standard case, $N$ has to be even.}
\begin{ruledtabular}
\begin{tabular}{ccc}
& \multicolumn{2}{c}{$n_h$} \\
\raisebox{1.4ex}[-1.4ex]{$N$} & $(P\mathcal{T})^2 = +1$ & $(P\mathcal{T})^2 = -1$ \\ \hline \hline
1 & 1 & -- \\
2 & 3 & 1 \\
3 & 6 & -- \\
4 & 10 & 6 \\
5 & 15 & -- \\
6 & 21 & 15 \\
\end{tabular}
\end{ruledtabular}
\end{table}

To show this, the normal-state Hamiltonian is expanded as
\begin{equation}
H_N(\mathbf{k}) = \sum_{n=1}^{n_h} c_n(\mathbf{k})\, h_n .
\label{A.HN.3}
\end{equation}
Under the magnetic point group $M$, the functions $c_n(\mathbf{k})$ must transform like the corresponding $h_n$ to ensure that $H_N(\mathbf{k})$ is invariant. The operation $P\mathcal{T}$ is an element of $M$. The irreps of $M$ can be uniquely and exhaustively divided into irreps that are even or odd under $P\mathcal{T}$. Only the $P\mathcal{T}$-even irreps have momentum-space basis functions since the momentum $\mathbf{k}$ is $P\mathcal{T}$ even. Hence, all $P\mathcal{T}$-even and no $P\mathcal{T}$-odd $h_\nu$ can occur in Eq.\ (\ref{A.HN.3}). The proposition is thus a statement about the number $n_h$ of $N\times N$ basis matrices $h_\nu$ that are even under $P\mathcal{T}$.

Since $P\mathcal{T}$ is antiunitary there exists a unitary $N\times N$ matrix $U_{PT}$ with $P\mathcal{T} = U_{PT} \mathcal{K}$, where $\mathcal{K}$ is the complex conjugation. Then we have
\begin{equation}
(P\mathcal{T})^2 = U_{PT} \mathcal{K} U_{PT} \mathcal{K} = U_{PT} U_{PT}^* .
\end{equation}
Thus $(P\mathcal{T})^2 = \pm 1$ is equivalent to $U_{PT} U_{PT}^* = \pm 1$ and, since $U_{PT}^*$ is unitary, to $U_{PT} = \pm (U_{PT}^*)^{-1} = \pm U_{PT}^T$.

Case 1: $(P\mathcal{T})^2 = +1$ and symmetric $U_{PT}$. For any unitary symmetric $N\times N$ matrix $U_{PT}$, there exists a unitary matrix $Q$ such that
\begin{equation}
U_{PT} = Q Q^T .
\label{A.UPTQ.3}
\end{equation}
$h_\nu$ being even/odd under $P\mathcal{T}$ means $U_{PT} h_\nu^* U_{PT}^\dagger = \pm h_\nu$, which due to the Hermiticity of $h_\nu$ is equivalent to
\begin{equation}
U_{PT} h_\nu^T U_{PT}^\dagger = \pm h_\nu .
\end{equation}
Equation (\ref{A.UPTQ.3}) then gives
\begin{equation}
Q Q^T h_\nu^T Q^* Q^\dagger = \pm h_\nu ,
\end{equation}
which is equivalent to
\begin{equation}
Q^T h_\nu^T Q^* = (Q^\dagger h_\nu Q)^T = \pm Q^\dagger h_\nu Q .
\end{equation}
Note that $k_\nu \equiv Q^\dagger h_\nu Q$ is Hermitian. The dimension of the vector space over $\mathbb{R}$ of Hermitian $N\times N$ matrices that are also symmetric is ${N(N+1)}/{2}$. Hence, the dimension of the vector space of Hermitian $P\mathcal{T}$-even $N\times N$ and thus the number of $P\mathcal{T}$-even basis elements $h_\nu$ also equals ${N(N+1)}/{2}$. Analogously, the dimension of the space of Hermitian $N\times N$ matrices that are antisymmetric and thus the number of $P\mathcal{T}$-odd basis matrices equals ${N(N-1)}/{2}$. Note that the sum of the two numbers is $N^2$, as expected.

Case 2: $(P\mathcal{T})^2 = -1$ and antisymmetric $U_{PT}$. For any unitary antisymmetric $N\times N$ matrix $U_{PT}$, there exists a unitary matrix $Q$ such that
\begin{equation}
U_{PT} = Q \Lambda Q^T ,
\label{A.UPTQ.5}
\end{equation}
with
\begin{equation}
\Lambda = i\sigma_2 \otimes \openone ,
\end{equation}
which clearly means that $N$ must be even. We therefore write $N=2M$. $h_\nu$ being even/odd under $P\mathcal{T}$ means
\begin{equation}
U_{PT} h_\nu^T U_{PT}^\dagger = \pm h_\nu .
\end{equation}
Equation (\ref{A.UPTQ.5}) gives
\begin{equation}
Q \Lambda Q^T h_\nu^T Q^* \Lambda^\dagger Q^\dagger = \pm h_\nu ,
\end{equation}
which is equivalent to
\begin{equation}
\Lambda Q^T h_\nu^T Q^* \Lambda^\dagger = \Lambda (Q^\dagger h_\nu Q)^T 
\Lambda^\dagger = \pm Q^\dagger h_\nu Q .
\label{A.QhQ.7}
\end{equation}
Let $k_\nu \equiv Q^\dagger h_\nu Q$ (Hermitian). We write $\Lambda$ and $k_\nu$ in block form as
\begin{align}
\Lambda &= \left(\begin{array}{cc}
    0 & \openone \\ -\openone & 0
  \end{array}\right) , \\
k_\nu &= \left(\begin{array}{cc}
    \kappa_{11} & \kappa_{12} \\ \kappa_{12}^\dagger & \kappa_{22}
  \end{array}\right) ,
\end{align}
where $\kappa_{11}$ and $\kappa_{22}$ are Hermitian. Equation (\ref{A.QhQ.7}) can then be written as
\begin{equation}
\Lambda k_\nu^T \Lambda^\dagger = \begin{pmatrix}
    \kappa_{22}^T & -\kappa_{12}^T \\ -\kappa_{12}^* & \kappa_{11}^T
  \end{pmatrix} = \begin{pmatrix}
    \pm \kappa_{11} & \pm \kappa_{12} \\ \pm \kappa_{12}^\dagger & \pm \kappa_{22}
  \end{pmatrix} = \pm k_\nu .
\end{equation}
This yields the relations $\kappa_{12}^T = \mp \kappa_{12}$ and $\kappa_{22} = \pm \kappa_{11}^T$, and $k_\nu$ thus assumes the form
\begin{equation}
k_\nu = \begin{pmatrix}
    \kappa_{11} & \kappa_{12} \\ \kappa_{12}^\dagger & \pm \kappa_{11}^T
  \end{pmatrix} ,
\label{A.kqsn.block.3}
\end{equation}
with $\kappa_{11}^\dagger = \kappa_{11}$ and $\kappa_{12}^T = \mp \kappa_{12}$. The blocks are $M\times M$ matrices. The dimension of the vector space spanned by the $k_\nu$ is $M^2 + M(M-1) = M(2M-1)$ for the upper sign and $M^2 + M(M+1) = M(2M+1)$ for the lower sign. The sum is $4M^2 = N^2$, as expected. Hence, the dimension of the vector space of Hermitian $N\times N$ matrices that are even under $P\mathcal{T}$ is ${N(N-1)}/{2}$, whereas the dimension for $P\mathcal{T}$-odd matrices is ${N(N+1)}/{2}$. Note that the two numbers are interchanged compared to the case of $(P\mathcal{T})^2 = +1$. This completes the proof.

We are concerned with systems that satisfy TRS and inversion symmetry separately. Then the $P\mathcal{T}$-even (odd) irreps are the $g+$ and $u-$ ($g-$ and $u+$) irreps. Moreover, the two operations commute \cite{Wig32,Mes14} and $P$ squares to $+1$. Hence, $(P\mathcal{T})^2 = \mathcal{T}^2$. If the internal degrees of freedom include the electron spin we have $\mathcal{T}^2 = -1$ \cite{Wig32} and even dimension $N$. The case $\mathcal{T}^2 = +1$ can only be realized if the spin does not occur explicitly, for example because one spin state is pushed to high energies by a magnetic field. Then $\mathcal{T}$ is not the physical TRS but an effective antiunitary symmetry.

\section{Infinitesimal-pairing nodes}
\label{app.IPnodes}

In this Appendix, we discuss the IP nodes for $s_T=-1$ and odd-parity pairing and for the unconventional sign $s_T=+1$ of time reversal squared. For $s_T=-1$ and odd-parity pairing, it is still true that infinitesimal pairing can be described in a single-band, pseudospin picture. However, it is now in the pseudospin-triplet channel. The pairing matrix in the effective single-band picture thus has the form $D_\mathrm{eff}(\mathbf{k}) = \mathbf{d}(\mathbf{k}) \cdot \bsigma$, where $\bsigma$ is the vector of Pauli matrices representing the pseudospin. Since the pseudospin is even under inversion and odd under time reversal its components belong to one or more $g-$ irreps. One can use representation theory to work out which irreps the components of $\mathbf{d}(\mathbf{k})$ must belong to in order to obtain a pairing state of a certain symmetry. The condition $\mathbf{d}(\mathbf{k})=0$ then gives the symmetry-imposed IP nodes. Nodal gaps are thus less likely than for singlet pairing since they must satisfy three scalar conditions.

We now turn to the nonstandard case $s_T=+1$. According to Appendix \ref{app.normal}, this allows an effective single-band model with Hilbert-space dimension $N=1$. Equation (\ref{1.UTsymm.3}) then implies that $U_T=1$ and thus $\Delta(\mathbf{k})=D(\mathbf{k})$. The only Hermitian basis matrix is $h_0=1$. Hence, for a single-band model, the full symmetry information is carried by the form factor $f_0(\mathbf{k})$. $h_0$ belongs to the trivial irrep, which of course is a $g+$ irrep. Table \ref{tab.pairing.states} then shows that for $N=1$ only odd-parity pairing with $u-$ form factor $f_0(\mathbf{k})$ is possible. The analysis of possible pairing states is analogous to the case with $s_T=-1$ and even parity, except that $g+$ irreps are replaced by $u-$ irreps.

Even-parity pairing states for $s_T=+1$ cannot be described by an effective $N=1$ model but are possible for $N=2$. In fact, there are multiple possibilities to implement this case because Eq.\ (\ref{1.UTsymm.3}) now allows $U_T$ to be any symmetric unitary $2\times 2$ matrix, while $U_P$ can be any unitary $2\times 2$ matrix that squares to $\openone$. The specific $U_T$ and $U_P$ and thus the symmetry properties of the $2\times 2$ basis matrices $h_1$, $h_2$, $h_3$ (which are linear combinations of Pauli matrices) depend on the underlying system. Universal properties are therefore unlikely and we do not pursue this here.

\section{Existence and properties of the Pfaffian}
\label{app.Pfaffian}

In this Appendix, we review the main results for the Pfaffian. A simpler proof than in \cite{ABT17,BAM18} is presented. The BdG Hamiltonian (\ref{1.HBdG.2}) satisfies the charge-conjugation symmetry $\mathcal{U}_C \mathcal{H}^T(-\mathbf{k}) \mathcal{U}_C^\dagger = - \mathcal{H}(\mathbf{k})$, where $\mathcal{U}_C = \sigma_1 \otimes \openone$. For it to also satisfy inversion symmetry, there must exist a unitary matrix $U_P$ such that
\begin{equation}
\mathcal{U}_P\, \mathcal{H}(-\mathbf{k})\, \mathcal{U}_P^\dagger = \mathcal{H}(\mathbf{k}) ,
\end{equation}
where
\begin{equation}
\mathcal{U}_P = \begin{pmatrix}
    U_P & 0 \\ 0 & U_P^*
  \end{pmatrix} .
\end{equation}  
This is a special case of Eqs.\ (\ref{1.genUtrans.3}) and (\ref{1.genUtrans.3a}). The two symmetries imply $\mathcal{C}P$ symmetry,
\begin{equation}
\mathcal{U}_{CP}\, \mathcal{H}^T(\mathbf{k})\, \mathcal{U}_{CP}^\dagger = 
- \mathcal{H}(\mathbf{k}) ,
\label{CC.sup.1}
\end{equation}
with $\mathcal{U}_{CP} = \mathcal{U}_C \mathcal{U}_P^*$. We find that $\mathcal{C}P$ squares to the identity since
\begin{equation}
(\mathcal{U}_{CP} \mathcal{K})^2 = \mathcal{U}_{CP} \mathcal{U}_{CP}^*
  = \begin{pmatrix}
    U_P^2 & 0 \\ 0 & (U_P^*)^2
  \end{pmatrix}
  = \begin{pmatrix}
    \openone & 0 \\ 0 & \openone
  \end{pmatrix} .
\end{equation}
This implies that $\mathcal{U}_{CP} = (\mathcal{U}_{CP}^*)^{-1} = (\mathcal{U}_{CP}^*)^\dagger = \mathcal{U}_{CP}^T$ so that $\mathcal{U}_{CP}$ is symmetric. For any (complex) symmetric matrix $\mathcal{U}_{CP}$, there exists a unitary matrix $\Omega$ such that $\Lambda = \Omega \mathcal{U}_{CP} \Omega^T$ (note the transpose) is a diagonal matrix with real nonnegative components (Autonne-Takagi factorization \cite{Aut15,Tak25}). $\Lambda$ is evidently unitary. A diagonal unitary matrix with nonnegative components must be $\Lambda=\openone$. We thus obtain $\mathcal{U}_{CP} = \Omega^\dagger \Omega^*$ and Eq.\ (\ref{CC.sup.1}) becomes
\begin{equation}
\Omega^\dagger \Omega^*\, \mathcal{H}^T(\mathbf{k})\, \Omega^T \Omega = 
- \mathcal{H}(\mathbf{k}) .
\end{equation}
This implies that
\begin{equation}
\Omega^*\, \mathcal{H}^T(\mathbf{k})\, \Omega^T
  = - \Omega\, \mathcal{H}(\mathbf{k})\, \Omega^\dagger .
\end{equation}
Hence, $\tilde{\mathcal{H}}(\mathbf{k}) \equiv \Omega\, \mathcal{H}(\mathbf{k})\, \Omega^\dagger$ is antisymmetric:
\begin{equation}
\tilde{\mathcal{H}}^T(\mathbf{k}) = - \tilde{\mathcal{H}}(\mathbf{k}) .
\end{equation}
This shows that for systems with inversion symmetry, the BdG Hamiltonian can always be unitarily transformed into antisymmetric form. For the antisymmetric matrix $\tilde{\mathcal{H}}(\mathbf{k})$, the Pfaffian $\Pf \tilde{\mathcal{H}}(\mathbf{k})$ is well defined. Note that $\Omega$ is independent of $\mathbf{k}$ (and is in fact solely determined by $\mathcal{U}_{CP}$) so that the components of $\tilde{\mathcal{H}}(\mathbf{k})$ are \emph{smooth} functions of $\mathbf{k}$. Hence, the Pfaffian, which is a polynomial of these components, is an smooth function of $\mathbf{k}$. The sign of the Pfaffian is inverted under simultaneous interchange of two rows and the corresponding two columns. This is a unitary transformation, which can be absorbed into $\Omega$. Hence, the above derivation only determines the Pfaffian up to a sign.

If the dimension $N$ of the local Hilbert space is even the Pfaffian is real: The dimension $2N$ of the BdG Hamiltonian $\tilde{\mathcal{H}}(\mathbf{k})$ is a multiple of four and thus the Pfaffian is a polynomial of even degree of its the components. Since $\tilde{\mathcal{H}}(\mathbf{k})$ is Hermitian and antisymmetric these components are purely imaginary. Conversely, for odd $N$, the Pfaffian is purely imaginary. We define
\begin{equation}
P(\mathbf{k}) \equiv \left\{ \begin{array}{ll}
    \Pf \tilde{\mathcal{H}}(\mathbf{k}) & \mbox{for $N$ even,} \\
    i\Pf \tilde{\mathcal{H}}(\mathbf{k}) & \mbox{for $N$ odd}
  \end{array} \right.
\end{equation}
so that $P(\mathbf{k})$ is always real.

Due to $\mathcal{C}P$ symmetry, the spectrum of the BdG Hamiltonian is symmetric. We thus have
\begin{equation}
\det \mathcal{H}(\mathbf{k}) = (-1)^N \prod_{j=1}^N E_j^2(\mathbf{k}) ,
\end{equation}
where $\pm E_j(\mathbf{k})$ are the quasiparticle energies. We assume $E_j(\mathbf{k})\ge 0$ without loss of generality. The determinant equals the square of the Pfaffian. This implies that
\begin{equation}
P(\mathbf{k}) = \pm \prod_{j=1}^N E_j(\mathbf{k}) .
\label{CC.P.5}
\end{equation}
At any momentum $\mathbf{k}$ not on a node, $P(\mathbf{k})$ is strictly positive or negative. We now choose $\Omega$ in such a way that in the normal state $P(\mathbf{k}_\infty)$ is positive at some momentum $\mathbf{k}_\infty$ far from the normal-state Fermi surface. Then for not too large superconducting energy scale, the energies $E_j(\mathbf{k}_\infty)$ do not change sign when superconductivity is switched on and $P(\mathbf{k}_\infty)$ remains positive. Hence, at $\mathbf{k}_\infty$, the sign in Eq.\ (\ref{CC.P.5}) is $+$.

Since the Pfaffian and thus $P(\mathbf{k})$ are smooth functions of momentum $P(\mathbf{k})$ can only change sign at nodes. Conversely, if $P(\mathbf{k})$ is negative somewhere in the Brillouin zone there must be a surface of zeros, i.e., a BFS, separating the regions of positive and negative $P(\mathbf{k})$.

To determine under what conditions $P(\mathbf{k})$ can actually become negative, we need to analyze Eq.\ (\ref{CC.P.5}). The eigenenergies $E_j(\mathbf{k})$ are continuous functions and are smooth, except at zeros and potentially at crossing points. If a single $E_j(\mathbf{k})$ approaches zero linearly the smoothness of $P(\mathbf{k})$ and the choice $E_j(\mathbf{k})\ge 0$ requires the explicit sign in Eq.\ (\ref{CC.P.5}) to flip. Hence, $P(\mathbf{k})$ changes sign and there must be a closed BFS.

On the other hand, if two eigenenergies approach zero linearly and simultaneously, for example because they are degenerate, the explicit sign does not flip. In this case, $P(\mathbf{k})$ does not change sign at the zero but has a second-order zero there. The same applies if a single eigenenergy approaches zero quadratically. If the Pfaffian does not change sign the $\mathbb{Z}_2$ topological invariant, which is the relative sign of $P(\mathbf{k})$ \cite{ABT17,BAM18}, exists but is trivial. This makes BFSs unstable since for any low-symmetry momentum $\mathbf{k}$ with $P(\mathbf{k})=0$, an infinitesimal change of parameters can make $P(\mathbf{k})$ strictly positive.

For $s_T = -1$, which implies even $N$, and preserved TRS, Kramers' theorem \cite{Kra30,Wig32,Mes14} shows that the spectrum has twofold degeneracy for all $\mathbf{k}$. Then, the latter case applies, $P(\mathbf{k})$ does not change sign, and BFSs are not stable. On the other hand, for $s_T = +1$ or broken TRS, there is no mechanism that leads to twofold degeneracy everywhere, the $P(\mathbf{k})$ generically changes sign, and BFSs are stable.

\section{The algebra of basis matrices}
\label{app.algebra}

As shown in Appendix \ref{app.normal}, for $N=4$ and $(P\mathcal{T})^2=-1$, six basis matrices $h_0$, \dots, $h_5$ appear in the normal-state Hamiltonian. One of them is (proportional to) the identity matrix, as discussed in Sec.\ \ref{sec.theory}. We choose $h_0 = \openone$. In this Appendix, we show that the basis matrices can \emph{always} be chosen in such a way that they satisfy the generalized commutation relations
\begin{equation}
h_0 h_n = h_n h_0
\end{equation}
for $n = 1,2,3,4,5$ and
\begin{equation}
h_m h_n = -h_n h_m
\end{equation}
for $m, n = 1,2,3,4,5$ and $m\neq n$. It is well known that (several sets of) six $4\times 4$ matrices with these properties exist \cite{AWH12,Mes14}. The point here is that the basis matrices always realize this structure.

The matrices $h_n$ can be written as $k_n = Q^\dagger h_n Q$, with $Q$ unitary, where the $k_n$ satisfy Eq.\ (\ref{A.kqsn.block.3}) with the upper sign and $\kappa_{11}^\dagger = \kappa_{11}$ and $\kappa_{12}^T = - \kappa_{12}$. For $N=4$, $\kappa_{11}$ and $\kappa_{12}$ are $2\times 2$ matrices. Then, $\kappa_{11}$ has to be a linear combination of $\sigma_0$, $\sigma_1$, $\sigma_2$, $\sigma_3$ with real coefficients and $\kappa_{12}$ can be $\sigma_2$ with an arbitrary complex prefactor. A maximal set of linearly independent matrices is then
\begin{align}
k_0 &= \left(\begin{array}{cc}
  \sigma_0 & 0 \\ 0 & \sigma_0
  \end{array}\right) = \sigma_0 \otimes \sigma_0 , \\
k_1 &= \left(\begin{array}{cc}
  \sigma_1 & 0 \\ 0 & \sigma_1
  \end{array}\right) = \sigma_0 \otimes \sigma_1 , \\
k_2 &= \left(\begin{array}{cc}
  \sigma_2 & 0 \\ 0 & -\sigma_2
  \end{array}\right) = \sigma_3 \otimes \sigma_2 , \\
k_3 &= \left(\begin{array}{cc}
  \sigma_3 & 0 \\ 0 & \sigma_3
  \end{array}\right) = \sigma_0 \otimes \sigma_3 , \\
k_4 &= \left(\begin{array}{cc}
  0 & \sigma_2 \\ \sigma_2 & 0
  \end{array}\right) = \sigma_1 \otimes \sigma_2 , \\
k_5 &= \left(\begin{array}{cc}
  0 & -i\sigma_2 \\ i\sigma_2 & 0
  \end{array}\right) = \sigma_2 \otimes \sigma_2 .
\end{align}
These matrices satisfy $k_0 k_n = k_n k_0$ for $n = 1,2,3,4,5$ and $k_m k_n = -k_n k_m$ for $m, n = 1,2,3,4,5$ and $m\neq n$. Moreover, they are orthonormal with respect to the scalar product $\Tr k_m k_n$. The basis matrices $h_n$, $n=0,\ldots,5$, are related to the $k_n$ by a unitary transformation. Since the $k_n$ satisfy the generalized commutation relations so do the $h_n$. The upshot is that while the specific form of the basis matrices $h_n$ depends on the model, their algebra does not. This result extends to general Hilbert-space dimension $N$ but the algebraic properties are more complicated for $N>4$.

\section{Pfaffian for the four-dimensional case}
\label{app.analytical.Pfaffian}

Here, we briefly discuss the analytical expression for the Pfaffian $P(\mathbf{k})$ of the transformed BdG Hamiltonian $\tilde{\mathcal{H}}(\mathbf{k})$ for the case of $s_T=-1$, $N=4$, and even-parity pairing. The Pfaffian exists and can be chosen to be a smooth function of momentum $\mathbf{k}$, as shown in Appendix \ref{app.Pfaffian}. Then the property $P^2(\mathbf{k}) = \det\mathcal{H}(\mathbf{k})$ fixes $P(\mathbf{k})$ up to an overall sign. As discussed in Appendix \ref{app.Pfaffian}, we choose this sign so that $P(\mathbf{k})>0$ far from the normal-state Fermi surface.

If the superconducting energy scale is not too large, the Pfaffian is given in terms of the coefficients in Eq.\ (\ref{2.HBdG.4}) as
\begin{align}
P&(\mathbf{k}) = \langle c,c\rangle^2 + \langle f^1,f^1\rangle^2
  + \langle f^2,f^2\rangle^2 \nonumber \\
&{}+ 4\, \big( \langle c,f^1\rangle^2 + \langle f^1,f^2\rangle^2
  + \langle f^2,c\rangle^2 \big) \nonumber \\
&{}- 2\, \big( \langle c,c\rangle\, \langle f^1,f^1\rangle
  + \langle f^1,f^1\rangle\, \langle f^2,f^2\rangle + \langle f^2,f^2\rangle\, \langle c,c\rangle \big) ,
\label{D.Pfr.5}
\end{align}
with the Minkowski-type scalar product
\begin{equation}
\langle A,B\rangle \equiv A_0 B_0 - \sum_{n=1}^5 A_n B_n .
\end{equation}
This proposition is proved by evaluating $P^2(\mathbf{k})$ and showing that it agrees with the determinant of the BdG Hamiltonian. This cumbersome calculation can be simplified by realizing that the Pfaffian is invariant under simultaneous rotations of the five-vectors
\begin{align}
\vec{c} &= (c_1,c_2,c_3,c_4,c_5) , \\
\vec{f}^{\,1} &= (f^1_1,f^1_2,f^1_3,f^1_4,f^1_5) , \\
\vec{f}^{\,2} &= (f^2_1,f^2_2,f^2_3,f^2_4,f^2_5) .
\end{align}
The sign of $P(\mathbf{k})$ is also correct: The assumption of not too large superconducting energy scale means that the $f^1_n$ and $f^2_n$ for $n=0,\ldots,5$ are small compared to the $c_n$ far from the normal-state Fermi energy. Then, at such momenta we get $P(\mathbf{k}) \cong \langle c,c\rangle^2 > 0$.

For large superconducting energy scale, the whole Brillouin zone is affected by superconductivity and we cannot choose the sign of $P(\mathbf{k})$ by continuity from the normal state. This simply means that there is no useful distinction between the inside and the outside of the BFS. The conclusions of this paper remain valid, though.


\begin{thebibliography}{99}
\bibitem{endnote.TRSBexample}For example, a somewhat artificial TRS-breaking state could be realized by $D(\mathbf{k}) = (\delta_0 + i\delta_1 k^2)\, h_0$, $\delta_0, \delta_1\in \mathbb{R}$.
\bibitem{endnote.rotated.Eg}Also note that the pairing states with $(\tdelta^1_0,\tdelta^2_0) \propto (-1/2,\sqrt{3}/2)$ and $(-1/2,-\sqrt{3}/2)$ are just rotated versions of $(1,0)$ and thus degenerate with it.
\bibitem{endnote.Psmooth}The BFS is not restricted to the high-symmetry direction since the Pfaffian $P(\mathbf{k})$ is a smooth function of $\mathbf{k}$. Hence, if $P(\mathbf{k})<0$ on a segment of a high-symmetry axis the Pfaffian remains negative in a neighborhood.
\bibitem{endnote.opp.C4z}Note that the matrix representation of the rotation $C_{4z}$ is $\sigma_3 \otimes e^{-i \sigma_3 \pi/4}$, for example. Because of the symmetry of the $f_{xyz}$-orbital the rotation leads to an additional sign change compared to the simple rotation of a spin.
\end{thebibliography}
\end{document}